%
%
%
%
%

\input harvmac
\input epsf

\parindent=0pt
\Title{SHEP 00-04}{\vbox{\centerline{A gauge invariant exact 
renormalization group II}}}

\centerline{\bf Tim R. Morris}
\vskip .12in plus .02in
\centerline{\it 
Department of Physics, University of Southampton,}
\centerline{\it Highfield, Southampton SO17 1BJ, UK}
\vskip .7in plus .35in

\centerline{\bf Abstract}
\smallskip 
A manifestly gauge invariant and regularized renormalization group flow
equation is constructed for pure $SU(N)$ gauge theory in the
large $N$ limit. In this way we make precise and concrete the notion
of a non-perturbative gauge invariant
continuum Wilsonian effective action.
Manifestly gauge invariant
calculations may be performed, without gauge fixing,
and receive a natural interpretation in terms of
fluctuating Wilson loops.
Regularization is achieved by covariant higher derivatives and
by embedding in a spontaneously broken $SU(N|N)$ 
supergauge
theory; the resulting heavy fermionic vectors are
Pauli-Villars fields. We prove the finiteness of this method to one
loop and any number of external gauge fields. A duality is uncovered
that changes the sign of the squared coupling constant.
As a test of the basic formalism we compute
the one loop $\beta$ function, for the first time without any
gauge fixing, and prove its universality with respect to
cutoff function.

\vskip -1.5cm
\Date{\vbox{
{hep-th/0006064}
\vskip2pt{June, 2000.}
}
}

\def\ins#1#2#3{\hskip #1cm \hbox{#3}\hskip #2cm}
\def\etc{{\it etc.}\ }
\def\ie{{\it i.e.}\ }
\def\eg{{\it e.g.}\ }
\def\cf{{\it cf.}\ }
\def\viz{{\it viz.}\ }

\def\vv{{\it vice versa}}
\def\aka{{\it a.k.a.}\ }

\def\nonp{non-perturbative}
\def\phi{\varphi}
\def\D{{\cal D}}
\def\p{{ p}}
\def\q{{ q}}
\def\r{{ r}}
\def\s{{ s}}
\def\rt{{\tilde r}}
\def\k{{ k}}
\def\u{{\bf u}} 
\def\v{{\bf v}}
\def\x{{ x}}
\def\y{{ y}}
\def\z{{ z}}

\def\A{{\bf A}}
\def\B{{\bf B}}
\def\C{{\bf C}}
\def\d{{\bf D}}
\def\F{{\bf F}}
\def\M{{\bf M}}
\def\X{{\bf X}}
\def\Y{{\bf Y}}
\def\ZZ{{\bf Z}}
\def\di#1{{\rm d}_{#1}}

\def\FF{{\cal F}}
\def\AA{{\cal A}}
\def\BB{{\cal B}}
\def\Z{{\cal Z}}
\def\Cu{{\cal C}}
\def\Bb{{\bar B}}
\def\ct{{\tilde c}}
\def\ph#1{\phantom{#1}}
\def\hS{{\hat S}}

\def\str{{\rm str}\,}
\def\tr{{\rm tr}\,}
\def\al{\alpha}
\def\be{\beta}
\def\la{\lambda}
\def\si{\sigma}
\def\proof{\noindent{\sl Proof.}}

\parindent=15pt

\newsec{Introduction}
In ref. \ref\ymi{T.R. Morris, Nucl. Phys. B573 (2000) 97, hep-th/9910058.}
we presented a gauge invariant Wilsonian RG
(renormalization 
group) \ref\kogwil{K. Wilson and J. Kogut, Phys. Rep. 12C (1974) 75.},
formulated directly in the continuum.
This formulation was shown to have many attractive features,
in particular the fact that
manifest gauge invariance 
can be maintained at all stages of the calculation
and thus also in the solution for the effective action $S$, 
no gauge fixing or ghosts being required, and the equations 
may be reinterpreted  in terms of fluctuations in 
the natural gauge invariant order parameters, namely Wilson loops.
However, the formulation presented was not sufficient to regularise all 
ultra-violet divergences. In this paper we solve this problem whilst
preserving all these attractive aspects \ref\alg{T.R. Morris, 
in {\it The Exact Renormalization Group}, Eds Krasnitz {\it et al},
World Sci (1999) 1, and hep-th/9810104.}.

Our formulation thus furnishes 
for the first time a precise and concrete realisation of the notion of
a non-perturbative gauge invariant
continuum Wilsonian effective action \alg. 
In recent years there has been 
substantial progress in solving supersymmetric gauge 
theories \ref\sw{See for example the reviews:
K. Intriligator and N. Seiberg, Nucl. Phys. Proc. Suppl.
45BC (1996) 1, hep-th/9509066;\ C. Gomez and R. Hernandez, 
hep-th/9510023;\ L. Alvarez-Gaum\'e and S.F. Hassan, Fortsch. Phys. 45 (1997)
159, hep-th/9701069.} by computing just such an
effective action, even though this object has never been defined.
(Only certain general properties were required.)
Whilst we concentrate here solely on pure Yang-Mills theory, we
see no essential difficulty in generalising the flow equations
to include fermions and scalars and indeed spacetime supersymmetry.
It is clear then that our framework can
underpin these ideas \sw\ref\yosh{S. Arnone, 
C. Fusi and K. Yoshida, J. High Energy Physics 02 (1999) 022.}.

The regularisation employed in ref. \ymi\ arises
essentially from 
an effective cutoff function 
\ref\Pol{J. Polchinski, Nucl. Phys. B231 (1984) 269.}\ref\YKIS{T.R. Morris, 
in {\it Yukawa International Seminar '97} 
Prog. Theor. Phys. Suppl. 131 (1998) 395, hep-th/9802039.}
which is gauge covariantized.
Similarly to gauge covariant higher derivative
regularisation, this is not sufficient to regulate all ultra-violet 
divergences. One loop divergences slip through \ref\oneslip{A.A. Slavnov,
Theor. Math. Phys. 13 (1972) 1064;\ B.W. Lee and J. Zinn-Justin,
Phys. Rev. D5 (1972) 3121.}. 
In standard perturbation theory, this
problem has been cured by supplementing the higher derivative 
regularisation with a system of Pauli-Villars regulator (PV) fields,
the action being bilinear in these fields so that they provide,
on integrating out, the
missing one loop counterterms\foot{and of course other finite
contributions}
\ref\pv{T.D. Bakeyev and A.A. Slavnov, Mod. Phys. Lett. A11 (1996) 1539;\
M. Asorey and F. Falceto, Phys. Rev. D54 (1996) 5290,
Nucl. Phys. B327 (1989) 427;\
C.P. Martin and F. Ruiz Ruiz, Nucl. Phys. B436 (1995) 545.}\ref\warr{B.J. 
Warr, Ann. Phys. 183 (1988) 1.}. 
This solution turns out to be
unwieldy, but worse, here the property of being bilinear in the PV
fields is not preserved by the flow: as the gauge field is 
integrated out higher-point PV interactions are generated. 

Instead, we uncovered a system of
regulating fields that is more natural from the exact RG point of view, 
particularly so in the Wilson loop picture \ymi, 
as we first reported in ref. \alg.
We have gradually realised that hidden in 
this formulation are supermatrices and a
spontaneously broken local
$SU(N|N)$ (in unitary gauge). We use this insight to give a concise
and complete exposition of the formulation sketched in ref. \alg.

As in ref. \alg, we will concentrate on the gauge group $SU(N)$ in the
large $N$ limit. 
All the ideas adapt to finite $N$ and indeed other gauge
groups,  except that the embedding in the
appropriate supergauge group should be formulated in such a way
as to make this connection more manifest. 
The disadvantage of the regularisation
framework reported here and in ref. \alg,  is that it was 
developed intuitively from the bottom up, without us being aware of
the underlying local $SU(N|N)$ structure. Whilst many aspects
fell out correctly nevertheless, the formulation given in ref. \alg\ 
is limited to one loop.
Complete regularisation
should be achieved in a manifestly local $SU(N|N)$ 
framework, for reasons that we will outline later.
The full exposition of this latter formulation is however
left for a future paper. 

Such a framework may
of course be used independently of the Wilsonian RG, and provides
a novel and elegant four dimensional `physical'\foot{in the usual sense
that it directly suppresses higher momentum modes}
regularisation for gauge theory
which, as we have already intimated, appears to generalise straightforwardly
to (chiral) fermions, spacetime supersymmetry and so on. 
Subtleties in its precise definition and 
properties \alg\ are discussed later and in the
conclusions, however a full treatment is left to a later paper 
\ref\sunn{S. Arnone, Yu. A. Kubyshin,
T.R. Morris and J.F. Tighe, in preparation.}.
 
One fascinating property
reported here,\foot{and paranthetically in ref. \alg} 
is a duality that effectively exchanges 
the squared coupling constant $g^2$ with $-g^2$. 
At the moment it is not clear to us whether this duality
survives in a manifestly local $SU(N|N)$ framework.
We comment further in the conclusions. 

Let us stress however that there are two main threads in this
paper. On the one hand we introduce this natural gauge invariant
regularisation, as described above. On the other hand, we go on to use
it to repair the divergences reported in ref. \ymi, and thus develop a
consistent calculational framework in which manifest gauge invariance
can be maintained at all stages.
The fact that the $SU(N)$ gauge invariance is 
in this way 
explicitly 
preserved, thus with no gauge fixing or BRST ghosts,
results in  elegant and highly constrained relations.
One important consequence is that there is no wavefunction renormalization,
the only quantity requiring renormalization being the coupling constant
\ymi. Manifest gauge invariance is 
also a necessary component of our PV regularisation scheme, as we will
show.

Non-perturbative approaches to non-Abelian gauge theory
that proceed by gauge fixing, must face up to the challenging problem
of Gribov copies \ref\Grib{V. Gribov, Nucl. Phys. B139 (1978) 1;\
I. Singer, Comm. Math. Phys. 60 (1978) 7;\
See \eg  P. van Baal, hep-th/9711070;\
M. Asorey and F. Falceto, Ann. Phys. 196 (1989) 209;\
K. Fujikawa, Nucl. Phys. B468 (1996) 355.}. 
Here, these problems are entirely avoided \ref\oGrib{For other approaches
to evade the Gribov problem see \eg 
V.M. Villanueva, J. Govaerts, and J-L. Lucio-Martinez,
J. Phys. A33 (2000) 4183; S.V. Shabanov and J.R.
Klauder, Phys. Lett. B456 (1999) 38.}.
Indeed we may turn the issue around, and
use the present formulation, since it is already well defined without
gauge fixing, to investigate explicitly 
the (quantum) consequences
of gauge fixing and Gribov copies directly
in the continuum.

In view of the novelty of the construction presented here and in
ref. \ymi,
a basic test of the formalism is surely desirable. We compute concrete
expressions for all the elementary vertices for two certain choices of
covariantization but for general cutoff functions.
We then use one of these to derive, directly in the continuum \YKIS, the 
classical values  of the two, three and
four point vertices in $S$ and the one-loop contribution to the
two-point vertex. From this we compute the
one-loop $\beta$ function. Throughout the calculation, we
use entirely general cutoff functions,\foot{up to some basic
criteria on normalisation and ultraviolet decay rates} and
maintain manifest gauge invariance. The fact that we
obtain the result $\beta_1=-{11\over3}{N\over(4\pi)^2}$,
independently of the choice of
cutoff functions is encouraging confirmation that the expected
universality of the continuum limit
has been successfully incorporated. The fact that it agrees with the
usual perturbative  result, demonstrates
for the first time {\sl explicitly}  that the 
one-loop $\beta$ function is free from Gribov 
problems, as expected.

The paper is structured as follows. 
In sec. 2, after preliminary
definitions, we state the flow equation in superfield notation
and show that it is manifestly gauge invariant, leaves the partition
function invariant, and recall the property of
quasilocality and that the flow corresponds to integrating out \ymi. 
This last feature relies on showing that the integrals are indeed 
ultraviolet regularised,
which is established for physical one-loop vertices in sec. 9 (see 
below). In setting up the definitions, we also show that the present
formulation is a reformulation of that sketched in ref. \alg, and 
show why its 
form follows essentially from spontaneously broken $SU(N|N)$. Full
and partial supermatrix differentials, the resulting
trapped $\sigma_3$s, covariantizations, super `wines', cutoff 
functions $c$ and $\ct$, and the renormalization condition are
all introduced here. The essence of the calculation can be followed
by reading this section, sec. 3 (which demonstrates that the 
isolation of perturbative contributions follows in the same way as
for the pure gauge case \ymi), and secs. 7 and 8 that cover respectively
the tree-level and one-loop calculation. 

However, to arrive at the equations in secs. 7 and 8, one needs to
extract the Feynman rules from the flow equation. We do this for
the reader in sec. 4. This also serves to fix the nomenclature 
for the Feynman rules and
to explain carefully the precise relation between the flow equation,
the Wilson loop diagrams, and their Feynman diagrammatic expansion.

Underlying these rules are the wine vertices themselves. Although
these are defined implicitly in sec. 2, see also refs. \ymi\alg,
clearly we need explicit expressions for concrete calculations.
These 
are derived in sec. 5, for two forms of covariantization.
For the form that we will use here for calculation, we also need 
to resolve their values at certain special momenta. In this
section we also work out the wine vertices' large momentum behaviour. 
This is not needed for the explicit tree and one-loop calculations, 
but it is needed at a very rough level to establish finiteness at
one loop (again see below). We pause here to work out their
large momentum behaviour in much more detail than required,
because it is elegant and interesting, falls out with little
effort, and may later be important.

In sec. 6 we uncover and describe the symmetries of the flow equation.
On the one hand this helps to understand the equation and the
novelties of the underlying $SU(N|N)$ regularization at a deeper level,
in particular we show that charge conjugation invariance and fermion
number mix to form a $Z_4$, that the superfields are only pseudoreal,
and demonstrate the existence of a duality that in a sense exchanges
$g^2$ with $-g^2$. On the other hand, 
a precise delineation of all symmetries is
needed to constrain certain `counterterms'
in the tree-level calculations in sec. 7.

Indeed in sec. 7, we will see that the classical vertices suffer a form
of divergence as a result of certain freedoms in the Pauli-Villars 
sector. These in turn lead to the introduction of some new parameters
$\gamma$. We compute only those vertices that we need for the one-loop
calculation in sec. 8. We also streamline the calculation a little,
by borrowing some general results from sec. 9.

Sec. 8 starts by showing how the $\beta$ function is determined, 
in principle non-perturbatively,
following ref. \ymi. We then specialize to the concrete one-loop
calculation. We explain in particular how we handle the calculation
for general cutoff functions by integrating by parts so as to lower
the degree of differentiation, whilst bringing all terms to a 
canonical algebraic form. In this way we reduce the computation
to a set of boundary terms in $D=4$ dimensions which however depend on
the power with which the cutoff functions decay. In fact as we
already demonstrated in refs. \alg\ymi, see also \warr, the result
is ambiguous due to certain total derivative terms that integrate to
finite surface terms. Keeping $D\ne4$ allows all us to discard 
all such terms, and we find that as $D\to4$ we recover the famous 
result for $\beta_1$, as already indicated.

In section 9 we give a proof of finiteness for all one-loop physical
vertices (\ie with no external PV fields). We show at the end of the
section where the difficulties lie in ensuring finiteness for a 
larger set of diagrams in this `unitary gauge' formulation. The
first part -- up to Lemma 3 -- explains in broad outline why our
exact RG has these finiteness properties; it is these considerations
that motivated the form of the exact RG. Although the rest of section
9 stands apart from the paper and can be skipped on first reading, 
they contain the reasons for some finer details in our flow equation. 

Finally in sec. 10, we present our conclusions, some comparisons
with earlier attempts at a manifestly gauge invariant calculations, 
and indicate future directions.



\newsec{SU(N$|$N), Pauli-Villars regularisation, and the exact RG}

We formulate the approach in $D$ Euclidean dimensions, 
specializing to $D=4$ only when required. 

In ref. \alg\ we defined
the Pauli-Villars (PV) regulator fields as follows. 
We took the tensor product of the gauge group with itself,
writing
$SU_1(N)\times SU_2(N)$ to distinguish the two groups. 
Here we write the Hermitian generators of
the two groups similarly as $\tau_1^a$ and $\tau_2^a$,
where each set of generators are orthonormalised as 
$\tr(\tau^a\tau^b)=\half\delta^{ab}$.
Their associated gauge fields are $A^1_\mu$, 
the physical gauge field, and $A^2_\mu$, which is unphysical and part
of the regularisation scheme. (The fields are valued in the Lie algebra
\ie $A^i_\mu\equiv{A^i}^a_\mu \tau_i^a$, $i=1,2$.)
We introduced \alg\ a fermionic Pauli-Villars (PV)
field $(B_\mu)^{i_1}_{j_2}$ and its complex conjugate 
$(\Bb_\mu)^{i_2}_{j_1}$. As indicated, $B$ transforms
as the fundamental of $SU_1(N)$ and as the complex conjugate 
fundamental of $SU_2(N)$. Finally, we introduced two bosonic
real scalar adjoint PV fields $C^1$ and 
$C^2$. The $A^i$'s are massless (of course) while $B$ and the $C^i$ have
masses at the effective cutoff $\Lambda$.
Originally we figured out the statistics, representation content,
and the interactions of these fields intuitively
in such a way as to ensure finiteness of the quantum corrections
in the RG flow equation. We found that there was very little freedom in the
choice of interactions if this was
to be achieved. The majority of the third lecture in ref. \alg\ was
devoted to describing the construction from this point of view,
and will not be repeated here.

We now understand these choices in terms of spontaneously broken $SU(N|N)$.
Let the supergauge field of $SU(N|N)$ be $\AA_\mu$, which we write 
in supermatrix form, \ie as a matrix representation
with bosonic diagonal elements, and fermionic off-diagonal elements:
\eqn\superg{\AA_\mu =\pmatrix{A^1_\mu&B_\mu\cr \Bb_\mu&A^2_\mu\cr}\quad.}
Since $\AA$ is valued in the graded Lie algebra of $SU(N|N)$, we require
that its supertrace vanishes: $\str \AA=0$, 
where the supertrace is defined by
$$\str {\bf X} 
=\str\pmatrix{X^{11} &X^{12}\cr X^{21} &X^{22}\cr}=\tr X^{11}-\tr X^{22}
$$
for any supermatrix ${\bf X}$. (The extra signs incurred on commutation
mean that for such matrices it is only the supertrace
that is cyclically symmetric, and thus in particular $SU(N|N)$ invariant.) 
We also introduce the superscalar 
\eqn\supersc{ \Cu=\pmatrix{C^1&D\cr{\bar D}&C^2\cr}\quad,}
but do not require $\str\Cu=0$. 

Note that the conditions $\tr C^1=\tr C^2=0$ 
are thus not imposed as in ref. \alg. 
Similarly, the conditions $\tr A^1_\mu =\tr A^2_\mu=0$ are not 
directly imposed. Only the constraint
$\tr A^1_\mu -\tr A^2_\mu=0$ has been applied
so far. Actually we can, and here will,
require that the $U(1)$ components of the supermatrix algebra,
$\tr (A^1_\mu+A^2_\mu)$ and $\tr (C^1+C^2)$, are
also absent, by a suitable modification of the
matrix commutator representation of the super-Lie product
in $SU(N|N)$ \ref\bars{I. Bars, in {\it Introduction to supersymmetry
in particle and nuclear physics}, Eds Casta\~nos {\it et al},
Plenum, New York (1984) 107.}. 

(Although it is not required for the rest of the paper, it may help
to pause on these points. Bars' observation
is that the super-Lie product
may be equally well represented by a bilinear (anti)symmetric
``$*$'' bracket
$$[\ ,\ ]^*_{\pm}=[\ ,\ ]_{\pm}-{1\over2N}{\rm tr}[\ ,\ ]_{\pm},$$
where $[\ ,\ ]_{\pm}$ is applied to the supergenerators and is a commutator
or anticommutator as appropriate. This effectively removes the unit
matrix as a representative of the algebra, and thus its bosonic $U(1)$ 
subgroup. In fact we
will handle these $U(1)$ components in a different way in our proof that
spontaneously broken $SU(N|N)$ acts as a regulator \sunn. For the
present paper it is only important to note that these modifications are 
subleading in $N$ and thus vanish in the $N=\infty$ limit
that we will take shortly. Let us stress that
there are in any case no $U(1)$ components in the present formulation \alg.
In this paper, we use $SU(N|N)$ to interpret and systemize our
earlier results \alg, and in this interpretation we may understand
the missing $U(1)$ factors in the way that we have just described.)

We choose the Lagrangian so that $\Cu$ picks up the expectation
value $<\!\Cu\!>\sim{\Lambda}\pmatrix{1&0\cr0&-1\cr}$, 
breaking $SU(N|N)$ spontaneously to $SU(N)\times SU(N)$.
In unitary gauge 
the Goldstone modes $D$ vanish (eaten by the $B_\mu$'s),
leaving the massive `Higgs' $C^i$
and massive vector fermions $B_\mu$.  
In this way, we elegantly recover exactly the
spectrum of fields introduced in ref. \alg. Remarkably, with 
covariant higher derivative regularisation, 
we also recover, up to some small 
details, the 
$\hS$ interactions used in 
ref. \alg\ for the one-loop $\beta$ function calculation!

Note that the only remaining massless fields are the
gauge fields of $SU(N)\times SU(N)$, which 
are of course charge neutral under each others
gauge group. Thus at energies much less than $\Lambda$ 
we are left only with these gauge fields which decouple into
the required physical $SU(N)$ Yang-Mills theory, and
a copy which as we will see has the opposite sign squared coupling
constant and is thus unphysical.

Not surprisingly, the $SU(N|N)$ theory described above has very good
ultra-violet behaviour. Technically this arises 
in the unbroken theory, because any quantum
correction involving $\tr 1=N$ in the $SU(N)$ theory, here involves
$\str 1=0$. (We will refer to this as `the supertrace mechanism'; of
course at the level of component fields it arises through exact
cancellation between bosonic and fermionic degrees of freedom.)
Indeed in the large $N$ limit the symmetric phase of
the supertheory thus has no quantum corrections at all. With covariant
higher derivative regularisation, we expect to be able to ensure
that the remaining corrections even at $N\ne\infty$ are finite. However, 
the full development of these investigations is left for the future. 

Here, and from now on, we will take the $N=\infty$ limit
and work solely with the theory described in ref. \alg, which is closely
related to spontaneously broken $SU(N|N)$
in unitary gauge as described above (and further outlined below).
As we will show, it appears that we cannot
ensure the full regularisation of the theory this way except to one loop
with external gauge fields. We interpret this as symptoms of
the differences described above and of the 
expected poor ultraviolet behaviour of the unitary gauge.

It is helpful to introduce the $2N\times 2N$ `Pauli' matrices
$$\sigma_1=\pmatrix{0&1\cr1&0\cr}\ins11{and}\sigma_3=\pmatrix{1&0\cr0&-1\cr}
$$
(in terms of which for example we have $\str\X=\tr\sigma_3\X$) and to
separate the bosonic and fermionic parts of the superfields, thus
$$\A_\mu=\pmatrix{A^1_\mu &0\cr 0&A^2_\mu\cr},\quad
\B_\mu=\pmatrix{0&B_\mu\cr \Bb_\mu&0\cr},\quad
\C=\pmatrix{C^1&0\cr0&C^2\cr},\quad 
\omega=\pmatrix{\omega^1&0\cr0&\omega^2\cr}\quad,$$
where $\omega$ is the unbroken $SU(N)\times SU(N)$ gauge transformation.
Defining now
$$\d_\mu=\partial_\mu-i\A_\mu\ins11{and}\nabla_\mu=
\partial_\mu-i\AA_\mu\quad,$$
both $\A$, and $\AA=\A+\B$, gauge transform under $SU(N)\times SU(N)$:
\eqn\gatr{\delta\A_\mu=\d_\mu\cdot\omega := [\d_\mu,\omega]\ins11{,}
\delta\AA_\mu=\nabla_\mu\cdot\omega \equiv [\nabla_\mu,\omega]\quad.}

The formalism of ref. \ymi\ can be lifted to the supertheory 
as follows. 
Functional derivatives are defined with respect to
the supertrace, thus
\eqn\pfuncd{\eqalign{{\delta\over\delta \A_\mu} &=
\pmatrix{{\delta/\delta A^1_\mu} &0\cr 0&
-{\delta/\delta A^2_\mu}\cr}\quad,\quad
{\delta\over\delta\B_\mu}=
\pmatrix{0&-{\delta/\delta \Bb_\mu}\cr {\delta/\delta B_\mu}
&0\cr}\cr
{\delta\over\delta\C}& =\pmatrix{{\delta/\delta C^1}&0\cr0&
-{\delta/\delta C^2}\cr}\quad,\quad
{\delta\over\delta\AA}={\delta\over\delta\A}
+{\delta\over\delta\B}\quad,\cr}}
(where the adjoint derivatives may be defined 
by $\delta/\delta A^i=2\tau_i^a\delta/\delta {A^i}^a$, $i=1,2$,
as in ref. \ymi. At large $N$, $\delta/\delta C^i$ will also be
thought of as an adjoint derivative, \ie we take the $C^i$ to be traceless.
Again this restriction makes no difference in the 
large $N$ limit, but it is convenient for the $C^i$ terms to inherit in
this way the conventions associated with the $\tau_i^a$ being normalised
to $1/2$.)
 
These transform homogeneously; their other important properties are
easier to state by ignoring the $x$ dependence
and $\mu$ index (as in ref. \ymi). Let $\M$, $\Y$ and $\ZZ$ also
be supermatrix representations as defined above. If $s(\M)$ is a (bosonic)
function of $\M$ such that
\eqn\pvar{\delta s(\M)=\str\delta \M \,\Y\quad,}
then, in precise analogy with ref. \ymi,
\eqn\pvarel{{\partial s\over\partial \M}=\Y\quad.}
(Here in the adjoint parts we use 
the completeness relation for $SU(N)$, which leads to the same relation
as the unrestricted functional derivative, up to $1/N$ corrections
that ensure the coefficient matrix is projected 
onto its traceless part \ymi.
These corrections can be neglected since,
as stated above, we are only interested from here 
on, in the $N=\infty$ limit.)

This leads to `supersowing' under the supertrace
\eqn\psow{\str \X{\partial s\over\partial \M} =\str \X\Y\quad,}
 and to `supersplitting'
\eqn\psplit{\delta \X=\Y
\delta \M\,\ZZ\qquad\Longrightarrow\qquad\str{\partial
\over\partial \M}\X=\str \Y\str \ZZ\quad.}

If however $\M$ is only 
block diagonal  like $\A$ or $\C$,
equivalently purely bosonic,
then $\Y$  in \pvarel\ and \psow\
is replaced by $\di+\Y$, 
the block diagonal (\aka bosonic)  part of $\Y$. 
On the other hand, if $\M$ is only 
block off-diagonal, like $\B$,
equivalently purely fermionic,
then $\Y$  in \pvarel--\psow\
is replaced by $\di-\Y$, 
the block off-diagonal (\aka fermionic) part of $\Y$. 

The projectors $\di\pm$ may be expressed as:
\eqn\proj{\di\pm\Y =\half\left(\Y\pm\Y^-\right)\quad,}
where $\Y^-=\sigma_3\Y\sigma_3$ has
opposite sign fermionic components compared to $\Y$.
Either from this or directly, we have the identities
\eqn\ids{\str \X\di\pm\Y=\str(\di\pm\X)\Y
=\str\di\pm\X\di\pm\Y\quad.} Summarising,
for a `partial' supermatrix $\M =\di\pm\M$, \psow\ is altered to
\eqn\bpsow{\str \X{\partial s\over\partial \M} =\str \X\di\pm\Y\quad.}
These alterations and the relations \ids\ are obvious
in the bosonic / fermionic language.

Similarly, for a partial supermatrix $\M =\di\pm\M$,
the splitting relation \psplit\ is altered:
\eqn\bpsplit{\str{\partial
\over\partial \M}\X=\half\left(\str \Y\str \ZZ\pm\str\sigma_3\Y\str
\sigma_3\,\ZZ\right)\quad,}
as follows most readily by writing $\delta \X=\Y
\di\pm(\delta \M)\,\ZZ$, after which the `full' supersplitting relation 
\psplit\ may be used. This `broken' supersplitting relation
(which in a local $SU(N|N)$ formulation would be seen to arise from 
insertions of $<\!\Cu\!>\sim\Lambda\sigma_3$) provides the only reason 
why there are any
quantum corrections at all in the large $N$ limit.
As discussed in ref. \ymi, the quantum corrections split
open the traces, and in the large $N$ limit, they survive only in the
terms where all fields vacate one of the two traces so as
to leave $\tr 1=N$. In the present case only the broken terms
survive, through $\str\sigma_3=2N$; the unbroken quantum corrections
have no way to survive because vacating a supertrace leaves 
behind $\str 1$ which vanishes.  Expanding \bpsplit\ in $N\times N$ block
components, in the bosonic case we have
$\tr Y^{11}\tr Z^{11}+\tr Y^{22}\tr Z^{22}$, and in the fermionic
case $-\tr Y^{11}\tr Z^{22}-\tr Y^{22}\tr Z^{11}$. 
Note the minus signs
in the latter, as expected for a fermionic loop.

Recall that a central construction of the gauge invariant flow equation in
ref. \ymi, is the `wine', the covariantization of a (smooth) momentum space
kernel $W_p\equiv W(p^2/\Lambda^2)$. 
In position space we write the kernel as
\eqn\kdef{W_{\x\y} \equiv\int\!\!{d^D\!p\over(2\pi)^D}
\,W(p^2/\Lambda^2)\,{\rm e}^{i\p.(\x-\y)}\quad.}
For two $N\otimes{\bar N}$ representations of the gauge
group $SU(N)$, $v(\y)$ and $u(\x)$, we write \eqnn\wv
$$\displaylines{
u\{W\}v= \hfill \wv\cr 
\sum_{m,n=0}^\infty\int\!\! d^D\!x\,d^D\!y\,
d^D\!x_1\cdots d^D\!x_n\,d^D\!y_1\cdots d^D\!y_m\,
W_{\mu_1\cdots\mu_n,\nu_1\cdots\nu_m}
(\x_1,\cdots,\x_n;\y_1,\cdots,\y_m;\x,\y) \cr
\hfill \tr\left[\, u(\x)\, A_{\mu_1}(\x_1)\cdots A_{\mu_n}(\x_n)\, 
v(\y)\,
A_{\nu_1}(\y_1)\cdots A_{\nu_m}(\y_m)\,\right]\quad,\cr}$$
where without loss of generality we may insist that $\{W\}$
satisfies $u\{W\}v\equiv v\{W\}u$. This equation defines the wine
vertices $W_{\mu_1\cdots\mu_n,\nu_1\cdots\nu_m}$, given the
method of covariantization  \ymi. As in \ymi, we write
the $m=0$ vertices (where there is no second product of gauge
fields), more compactly as 
\eqn\compac{W_{\mu_1\cdots\mu_n}(\x_1,\cdots,\x_n;\x,\y)
\equiv W_{\mu_1\cdots\mu_n,}(\x_1,\cdots,\x_n;;\x,\y)\quad,}
while the $m=n=0$  term is
just the original kernel \kdef, \ie 
\eqn\mno{W_,(;;\x,\y)\equiv W_{\x\y}\quad.}

For two supermatrix representations
${\bf v}(\y)$ and ${\bf u}(\x)$ we define 
in precise analogy, the supergauge invariant
\eqnn\pwv
$$\displaylines{ 
\u\{W\}\v= \hfill \pwv\cr 
\sum_{m,n=0}^\infty\int\!\!d^D\!x\,d^D\!y\,
d^D\!x_1\cdots d^D\!x_n\,d^D\!y_1\cdots d^D\!y_m\,
W_{\mu_1\cdots\mu_n,\nu_1\cdots\nu_m}
(\x_1,\cdots,\x_n;\y_1,\cdots,\y_m;\x,\y) \cr
\hfill \str\left[\, \u(\x)\, \AA_{\mu_1}(\x_1)\cdots \AA_{\mu_n}(\x_n)\, 
\v(\y)\,
\AA_{\nu_1}(\y_1)\cdots \AA_{\nu_m}(\y_m)\,\right]\quad.\cr}$$
Note that
the superwine's vertices $W_{\mu_1\cdots\mu_n,\nu_1\cdots\nu_m}$ are 
the same as those in \wv. Equation \pwv\ corresponds precisely
to the `peppering'  prescription in ref. \alg!

We restrict, as in ref. \ymi, to covariantizations which traverse back
and forth along a coincident Wilson line, \ie such that they may be
represented as 
\eqn\wili{
\u\{W\}\v=\int\!\!\!\!\int\!\!d^D\!x\,d^D\!y\int\!\!\D\Cu_{\x\y}\
\str \u(\x)\,\Phi[\Cu_{\x\y}]\,\v(\y)\,\Phi^{-1}[\Cu_{\x\y}]\quad.}
Here the measure $\D\Cu_{xy}$ over curves $\Cu_{xy}$ from $x$ to $y$,
is normalised by
\eqn\pnorm{\int\!\!\D\Cu_{\x\y}\ 1 = W_{\x\y}\quad,}
and encodes our choice of covariantization. The  (super)
Wilson lines are
defined by the path ordered exponential
\eqn\defWl{\eqalign{\Phi[\Cu_{\x\y}]&
=P\exp-i\int_{\Cu_{\x\y}}\!\!\!\!\!\!dz^\mu
\AA_\mu(z)\quad,\cr
&=1-i\int_0^1\!\!\!\!d\tau\,{\dot z}^\mu\! \AA_\mu(z)
-\int_0^1\!\!\!\!d\tau_2\!\int_0^{\tau_2}\!\!\!\!\!d\tau_1\,
{\dot z}\!\cdot\! \AA(\tau_1)\,
{\dot z}\!\cdot\! \AA(\tau_2)\ +\cdots\quad,}}
where we have parametrized $\Cu_{\x\y}$ by $z^\mu(\tau)$, 
$\tau\in[0,1]$, $\z(0)=\x$, $\z(1)=\y$. Since this definition
is independent of the parametrization of the path, the same is
true of the measure $\D\Cu_{xy}$, without loss of generality.
We also choose the measure to be Lorentz covariant, satisfy the
exchange symmetry below \wv, and to be smooth in momentum space,
\ie to yield vertices that are Taylor expandable to all orders
in momenta \ymi. 

One example that satisfies all these criteria is
to utilise the momentum representation to write \ymi: 
\eqn\pourchoice{\u\{W\}\v={\rm str}\int\!\!d^D\!x\, {\bf u}(\x)
\,W(-\nabla^2/\Lambda^2)\cdot \v(\x)
\quad.}
This is the covariantization that we will use 
in this paper to calculate the one-loop $\beta$ function.
[We will see later that it corresponds to coincident lines \wili.]
Another example is simply to use two straight super Wilson lines \ymi:
\eqn\straw{\u\{W\}\v=\int\!\!\!\!\int\!\!d^D\!x\,d^D\!y\
W_{\x\y}\, \str \u(\x)\Phi[\l_{\x\y}]\v(\y)\Phi^{-1}[\l_{\x\y}]\quad,}
$\l_{\x\y}$ being the straight line between $\x$ and $\y$.
We will compute concrete formulae for the wine vertices in both
covariantizations.

Introducing the superfield strength $\FF_{\mu\nu}=i[\nabla_\mu,\nabla_\nu]$,
we write the `seed' action \alg\ as 
\eqn\pseed{\hS=\half\FF_{\mu\nu}\{c^{-1}\}\FF_{\mu\nu}
+\Lambda^2\,\BB_\mu\{\ct^{-1}\}\BB_\mu
+\sigma\Lambda^4\,\C\{1\}\C\quad.}
The first term, which we will refer to as $\hS_A$, 
is simply the $\hS$ of ref. \ymi\ translated to superfields
(corresponding again to peppering \alg), in particular $c(p^2/\Lambda^2)>0$
is a smooth cutoff function satisfying $c(0)=1$ (in fact
without loss of generality) and $c(x)\to0$ as $x\to\infty$.\foot{As in
refs. \ymi\alg, $x$ used as a 
generic argument for these functions, 
should not be confused with the position $x$ and the position space
kernel $c_{xy}$, defined as in \kdef.}
In addition we have two further terms for $\B$ and $\C$, where
\eqn\bb{\BB_\mu=\B_\mu+\nabla_\mu\cdot\C\quad,\qquad\qquad
\C\{1\}\C\equiv\str\int\!\!d^D\!x\,\C^2(\x)\quad,}
and  $\ct(p^2/\Lambda^2)>0$ is another smooth cutoff profile 
whose properties \alg\ are recalled below. \pseed\ is nothing but the seed 
action of ref. \alg, translated into this language.
Note however that such terms are expected in spontaneously broken $SU(N|N)$; 
thus $\hS_A$ is just the higher-derivative regularised pure
gauge part, in particular the wrong sign $A^2_\mu$ action \alg\ is now
seen to be a consequence of the superfield (in particular 
supertrace) structure; the second term, which we name $\hS_B$, 
collects the remaining kinetic pieces after expanding around 
$<\!\Cu\!>$ and imposing unitary gauge,\foot{apart from one significant 
difference described below} in particular providing the mass term for 
$\B$; the last term, $\hS_C$, is the mass term for the `Higgs'. (Here 
$\sigma>0$ a free parameter. Note that the $\C$ field we discussed earlier
appears here as $\Lambda\C$. This change of variables is required for the
finiteness of the flow equation, as explained in sec. 9.)
Recall that the point of $\C$ was to
cancel the longitudinal divergences from $\B_\mu$ \alg. In the
$SU(N|N)$ language this cancellation is unsurprising because the
longitudinal part of $\B_\mu$ is nothing but ${\bf D}$,
the eaten fermionic partner of $\C$.

Actually for $\hS_B$ to correspond exactly to spontaneously broken
$SU(N|N)$, the $\B$-$\C$ cross-terms should contain an insertion of
$i\sigma_3$ arising from an explicit $<\!\Cu\!>$. This in turn leads
to some differences in the discrete symmetries here, as explained in
sec. 5. Also, the self-interactions of the Higgs are missing: these are
not needed for the regularisation (to one loop with only external gauge 
fields). 

We require $\ct(x)\to0$ as $x\to\infty$, to regularise $\C$ propagation,
while for the $\ct^{-1}$ term not to disturb the high energy behaviour
of the transverse part of $\B_\mu$ we clearly require
\eqn\ctlim{c/ (x\ct)\to0\ins11{as}x\to\infty\quad.}
(We require $\ct_0\equiv\ct(0)>0$ 
in order for it to act as a mass for $\B_\mu$ and
positive kinetic term for $\C$ but we do not
require $\ct_0=1$.) More
precise requirements on the UV asymptotics of 
$c$ and $\ct$ will be needed and are derived in sec 9.
We introduce three new kernels via
\eqn\KLM{K(x)={d\over dx}\left({x\ct c\over x\ct+c}\right),\quad
xL(x)={d\over dx}\left({x^2\ct^2\over x\ct+c}\right)\quad{\rm and}\quad
xM(x)={d\over dx}\left({x^2\ct\over x+\sigma\ct}\right)}
($c\equiv c(x)$, $\ct\equiv\ct(x)$ here). 
Finally we can write our full exact RG equation:\eqnn\pRG
$$\eqalignno{
\Lambda{\partial\over\partial\Lambda}S&[\A,\B,\C]=
{1\over2\Lambda^2}\left({1\over N}{\delta\over\delta \AA_\mu}
-{\delta S\over\delta \AA_\mu}\right)\{c'\}
{\delta\Sigma_g\over\delta \AA_\mu}
+{1\over2\Lambda^4}\left({\l\over N}
-\l S\right)\{L\}\l\Sigma_g\cr
&
+{1\over2\Lambda^2}\left({1\over N}{\delta\over\delta \B_\mu}
-{\delta S\over\delta \B_\mu}\right)\{K-c'\}_\A
{\delta\Sigma_g\over\delta \B_\mu}
+{1\over2\Lambda^4}\left({1\over N}{\delta\over\delta \C}
-{\delta S\over\delta \C}\right)\{M-L\}
{\delta\Sigma_g\over\delta \C},\cr
{\rm where}&\qquad\l={\delta\over\delta\C}
+\nabla_\mu\!\cdot\!{\delta\over\delta\AA_\mu}
\qquad{\rm and}\qquad \Sigma_g=g^2S-2\hS\quad. &\pRG\cr
}$$
In here, the first term on the RHS
is just our pure gauge field flow equation \ymi\
with $A_\mu$ replaced by $\AA_\mu$ and trace by supertrace.
Finiteness considerations motivated the precise form of the other terms
(see sec. 9 and ref. \alg.)
By the wine $\{K-c'\}_\A$ we mean that in \pwv\ only $\A$ is used,
rather than the full peppered $\AA$. In $\l$ we apply the $\delta/\delta
\AA_\mu$ first and then $\nabla_\mu$, \ie we understand this 
$\delta/\delta\AA_\mu$ as not differentiating
the $\AA_\mu$ in the $\nabla_\mu$ of $\nabla_\mu\cdot\delta/\delta\AA_\mu$.
Note that as in ref. \ymi, by prime (as in $c'$) we mean
differentiation with respect to its
argument (here $p^2/\Lambda^2$). 
The position space representation \kdef, the covariantization, and the
resulting vertices \wv, are all labelled by the underlying kernel
[\ie in this case replacing the letter $W$ by $c'$ throughout 
\kdef\ -- \pwv].
\pRG\ yields precisely the Feynman rules chosen earlier for the $\beta$ 
function calculation \alg. 

The coupling $g$ is defined as in ref. \ymi\ via the field strength for
$A^1_\mu$ in the $A^1$ part of the action:
\eqn\defg{S|_{A^1}={1\over2g^2}\,{\rm tr}\!\int\!\!d^D\!x\, 
\left(F^1_{\mu\nu}\right)^2+O(\partial^3)}
(discarding the vacuum energy). 
Note that we can and do impose, as in ref.\ymi, 
the requirement of `quasilocality',
and thus in particular that $S$ has a derivative expansion to all orders.
({\it N.B.} Other than of course its supersymmetry,
this is the crucial fundamental property assumed of the Wilsonian 
effective action in supersymmetric theories, which in turn justifies
its holomorphy \sw.)

Clearly, the flow equation \pRG\ is manifestly
$SU(N)\times SU(N)$ gauge invariant. It also
leaves the partition function
$$\Z=\int\!\!\D[\A,\B,\C]\,\e{-NS}$$ 
invariant. To see this, note that \pRG\ implies that \eqnn\zinv
$$\eqalignno{\Lambda{\partial\over\partial\Lambda}\,\e{-NS}&=
-{1\over2\Lambda^2}{\delta\over\delta\AA_\mu}\{c'\}\left({\delta\Sigma_g
\over\delta\AA_\mu}\e{-NS}\right)
-{\l\over2\Lambda^4}\{L\}\left(\e{-NS}\l\Sigma_g\right) &\zinv\cr
&-{1\over2\Lambda^2}{\delta\over\delta \B_\mu}\{K-c'\}_\A\left(
{\delta\Sigma_g\over\delta\B_\mu}\e{-NS}\right)
-{1\over2\Lambda^4}{\delta\over\delta\C}\{M-L\}\left(
{\delta\Sigma_g\over\delta\C}\e{-NS}\right)\quad, \cr
}$$
and hence is a total functional derivative.
Actually, for this to be true we need the $\delta/\delta\AA_\mu$ in 
the leftmost $\l$ to act on everything in the expression and thus
also on the $\AA_\mu$ in $\nabla_\mu$, in apparent contradiction
with the definition given for \pRG. However,
the difference between the two definitions gives
${\delta\over\delta\AA_\mu}\AA_\mu$, which vanishes
by the supertrace mechanism. [To see this set $\Y=1$ in \psplit.]

We can see indirectly 
that the exact RG equation \pRG\ corresponds to integrating
out by the arguments already given in ref. \ymi.

Similarly from \ymi,
we still have that $S$ may be expanded in traces and products
of traces. We can write these as products of supertraces of the fields
$\A$, $\B$ and $\C$, and if necessary from \bpsow\ or \bpsplit, 
embedded $\sigma_3$'s.
Representing the supertraces by closed loops we have the same diagrammatic
notation for the full RG equations as before:
\nfig\fflow{fig 5.}
\midinsert
$$\Lambda{\partial\over\partial\Lambda}\,
\vcenter{\epsfxsize=0.085\hsize\epsfbox{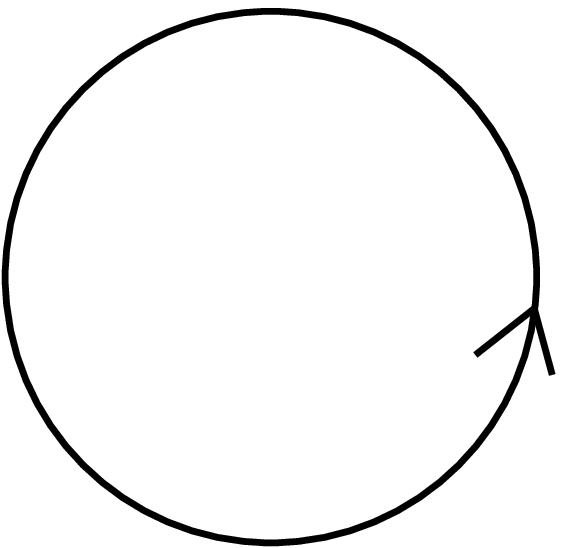}}\,
\ =\ -g^2\!\!\!\!\vcenter{\epsfxsize=0.13\hsize\epsfbox{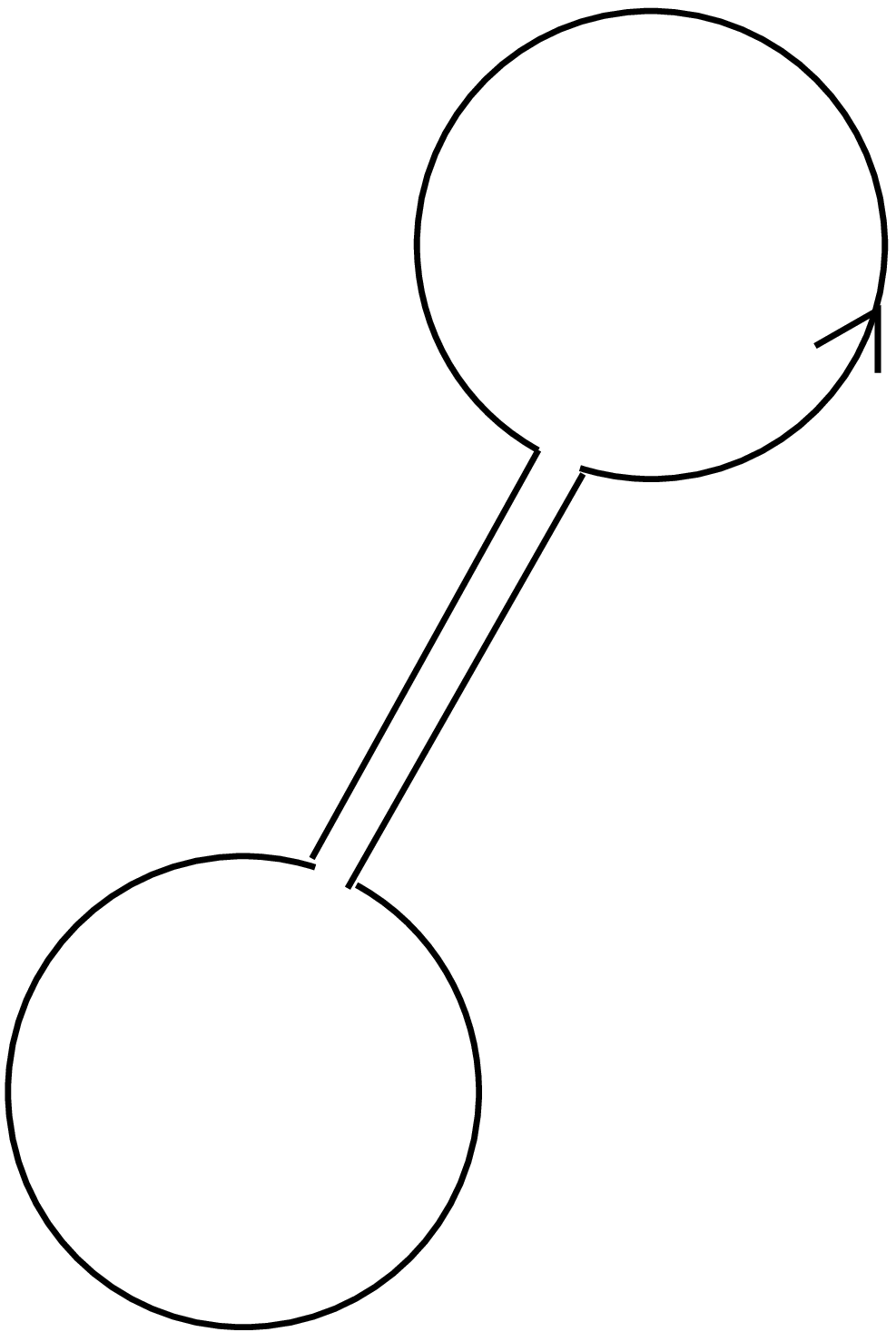}}
\!+2\!\!\!\!\vcenter{\epsfxsize=0.13\hsize\epsfbox{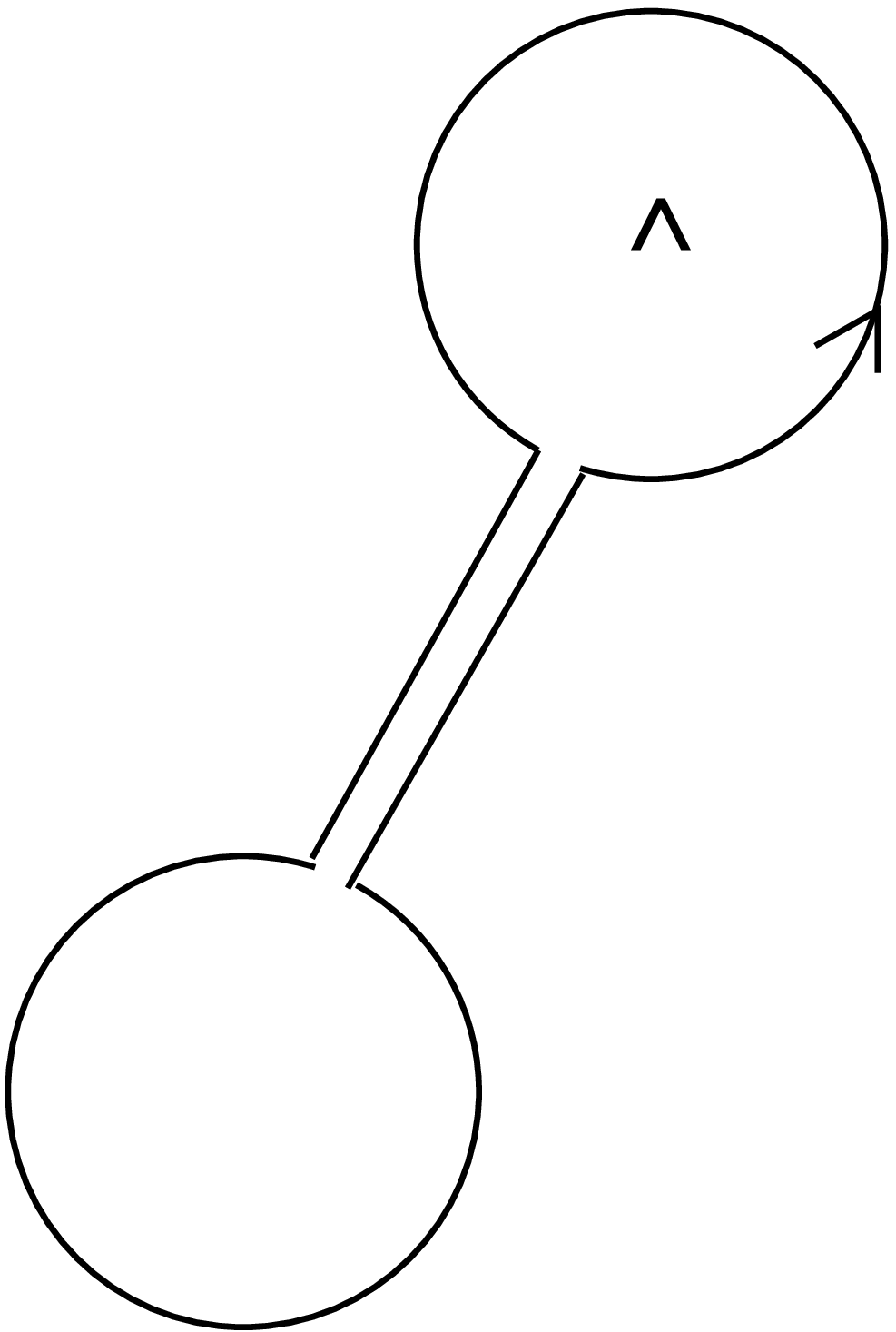}}
\!+{g^2\over N}\ 
\vcenter{\epsfxsize=0.085\hsize\epsfbox{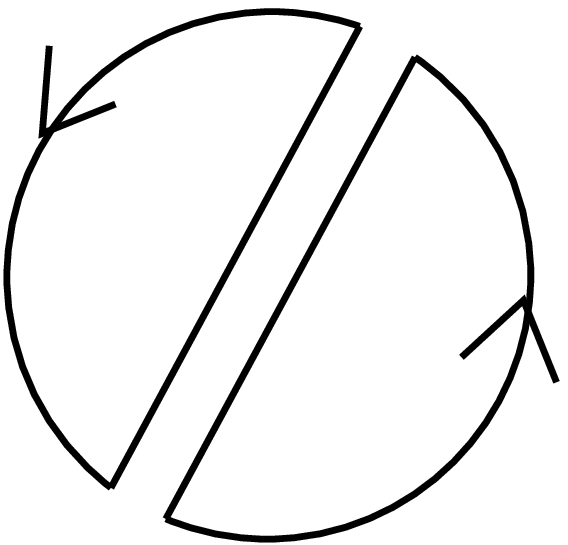}}
\ -{2\over N}\ \vcenter{\epsfxsize=0.085\hsize\epsfbox{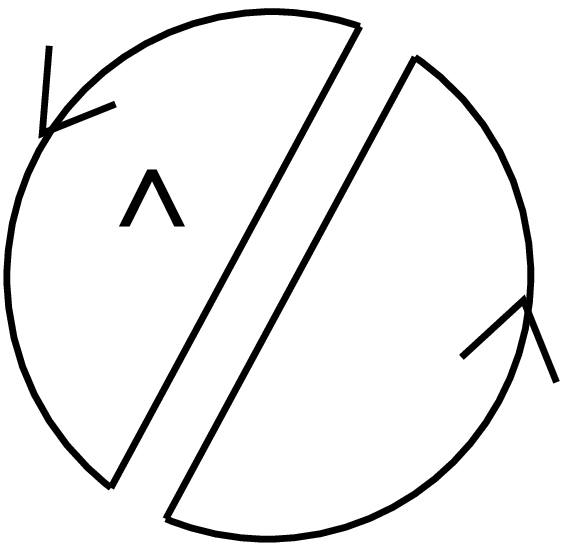}}
$$
{\bf Fig.1.}  Diagrammatic representation of the flow equation.
A circumflex in a circle indicates $\hS$.
\endinsert
\noindent where here we have taken the (na\"\i ve \ymi) large $N$ limit. 
Once again this implies that at most a single (super)trace survives
in the effective action.

The expansion of the loops in powers of the fields yields Feynman
diagrams as before \ymi. Note that the closed loops may no longer 
necessarily be interpreted as integrals over pure gauge (super) Wilson
loops because $\C$s, $\sigma_3$s and individual $\B$s may be inserted.
However, the contributions consist of pure gauge (\viz $\AA$) 
sections joining
isolated $\C$s, $\sigma_3$s and $\B$s, and thus may 
be interpreted as integrals over pure gauge Wilson lines joining a countable
number of such points. 

If we represent diagrammatically
the insertion of a $\sigma_3$ into (a join in) such a superloop,
by a filled arrow pointing to the insertion point, the 
supersowing special cases \bpsow, and the
supersplitting special cases \bpsplit, can be seen to be caused by
the same local process, as shown in 
\fig\fsig{insertion of sigma3's. See diag p14 "Notes on paper".
The insertion of sigma3's appear in the same way, whether or not
the attaching wine results in a tree or loop correction}.
\midinsert
$$
{1\over2}\ \vcenter{\epsfxsize=0.17\hsize\epsfbox{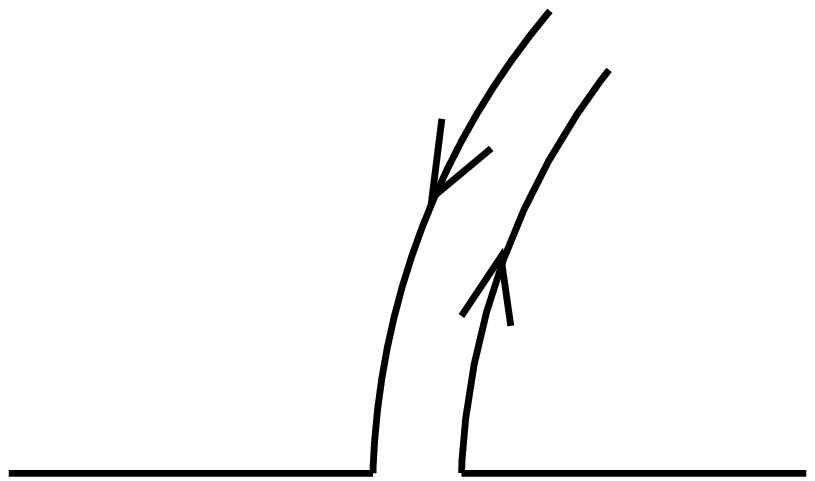}}
\pm{1\over2}\ \vcenter{\epsfxsize=0.17\hsize\epsfbox{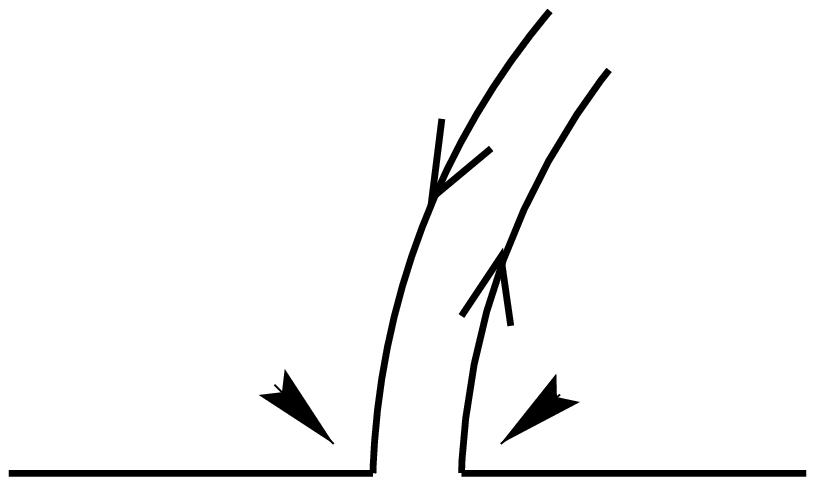}}
$$
{\bf Fig.2.} Insertion of $\sigma_3$s occurs in the same way, whether or not
the attaching wine results in a tree or loop correction.
\endinsert

Contributions to \pRG\ containing terms of form of 
\fig\ftail{wine biting tail} are apparently required
by \zinv\ (where the wine attaches at the top to $S$ or $\hS$).
However, in the large $N$ limit these contributions vanish. 
The reason is as follows. Firstly the $\{K-c'\}_\A$, or $\{M-L\}$,
wines obviously
cannot bite their own tails since they do not contain $\B$, or $\C$,
respectively. Secondly the other end of the wine in \ftail\ must attach
to a vertex with at least two points, because all one-point vertices
vanish by charge conjugation invariance (see sec. 6). 
Since only one superloop
can contain fields in the large $N$ limit, the closed superloop in \ftail\ must
thus be field free. But for the two remaining possibilities $\{c'\}$ and
$\{L\}$, this loop is formed by a `full' supermatrix differential 
$\delta/\delta\AA$ which thus yields $\str 1=0$ by \psplit.
\midinsert
$$
\epsfxsize=0.28\hsize\epsfbox{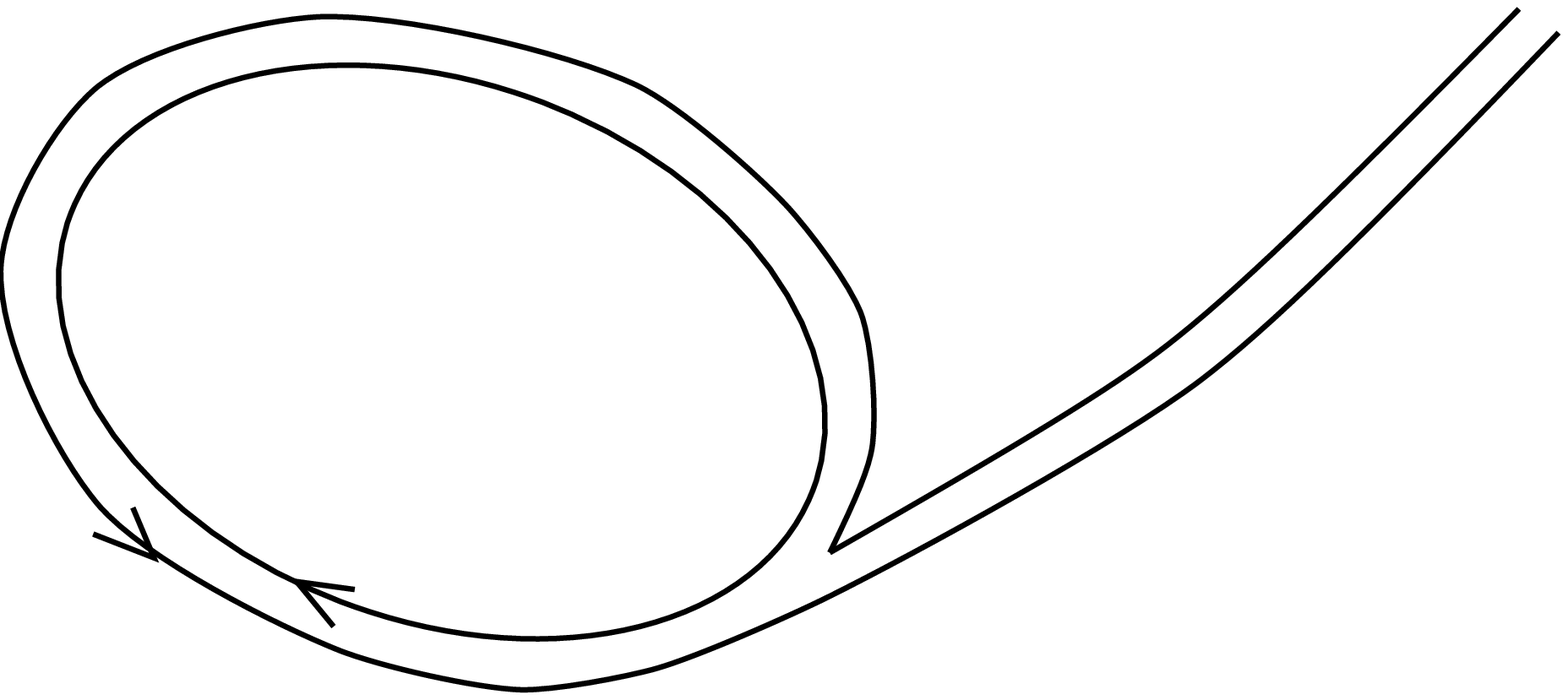}
$$
\centerline{ {\bf Fig.3.} A wine biting its own tail.}
\endinsert

\newsec{Perturbative expansion}

We recall the general structure of the perturbative expansion \ymi.
As in ref. \ymi\
we keep $D$ general at this stage. It will be helpful to access $D=4$
via the limit $D\to4$ \alg.
It will also be helpful to write \pRG\ as
\eqn\ERG{\Lambda{\partial\over\partial\Lambda}S=
-a_0[S,g^2S-2\hS]+a_1[g^2S-2\hS]\quad,}
where we have expanded $\Sigma_g$ and written the classical
terms as the bilinear functional $-a_0$ 
and the quantum terms as the linear functional $a_1$. 
Since this has the same form as the pure gauge case,
the isolation of perturbative contributions from these
equations proceeds as before \ymi.
Thus we see from \ERG, that $S\sim 1/g^2$ at the classical 
level [consistent with \defg], and by iteration, using 
\ERG, that $S$ has as expected the weak coupling expansion 
\eqn\Sloope{S={1\over g^2} S_0+S_1+g^2 S_2 +\cdots\quad.}
Substituting this expansion in \ERG\ and recalling that $g$ will run at
the quantum level, we see that
the $\beta$ function must also take the standard form
\eqn\betafn{\beta:=\Lambda{
\partial g\over\partial\Lambda}=\beta_1g^3+\beta_2g^5+\cdots\quad.}
By \defg, $g^2$ has dimension $4-D$. The $\beta_i$ thus carry
dimensions $(D-4)i$ and (as we will confirm) are not universal except
when $D=4$. From  \Sloope\ and \betafn, we obtain the loopwise expansion
of \ERG:
\eqnn\ergcl\eqnn\ergone\eqnn\ergtwo
$$\eqalignno{\Lambda{\partial\over\partial\Lambda}S_0 &=-a_0[S_0,S_0-2\hS]
&\ergcl\cr
\Lambda{\partial\over\partial\Lambda}S_1&=2\beta_1S_0-2a_0[S_0-\hS,S_1]
+a_1[S_0-2\hS] &\ergone\cr
\Lambda{\partial\over\partial\Lambda}S_2&=2\beta_2S_0-2a_0[S_0-\hS,S_2]
-a_0[S_1,S_1]+a_1[S_1]\quad, &\ergtwo\cr}$$
\etc \ymi.
In a similar way we obtain equations for the weak coupling
expansion of integrated operators which in this notation
are identical to those of ref. \ymi.

\newsec{Feynman rules}
This section sets out the nomenclature we will use for
the Feynman rules and the resulting expressions
as derived by expanding \pRG\ in a power series in the
fields. Along the way we explain, pedantically, 
the precise method for translating \fflow\ to equations for 
individual vertices.
As before \alg\ymi, this is diagrammatically represented 
by replacing the thick black lines of \fflow, 
which represent the exact expressions without
expansion in the fields, by thin lines, and
a series of contributions with increasing numbers of 
points (blobs). The points represent individual fields
and appear in all places on the composite
Wilson loop with equal weight. Where necessary
the points will now be labelled by the flavour of the field that attaches
to it. Specializing to a single supertrace as appropriate for the
large $N$ limit, the field expansion of the effective action,
which is illustrated diagrammatically in \fig\fpsexp{sum of circles with
blobs as in fig.4 ymi but with only one circle and repeated with a sigma3
arrow in a given place}, takes the form:
\eqn\sexp{S =\sum {1\over s_n}\int\!\!d^D\!x_1\cdots d^D\!x_n\,
S_{\ph{\sigma^j}\,a_1\,\cdots \,a_n\,}^{\sigma^j\X_1\cdots \X_n}
(\x_1,\cdots,\x_n)\
\str \sigma_3^j\X_1^{a_1}(\x_1)\cdots \X_n^{a_n}(\x_n)\quad,}
where the superfields $\X_i^{a_i}$ are $\A^{\mu_i}$, $\B^{\mu_i}$ or $\C$,
the indices $a_i=\mu_i$, or null for $\C$, and $j=0$ or $1$. (We will
omit the $\sigma$ superscript for $j=0$ and write it without exponent
when $j=1$.)
Only one cyclic ordering of each list $\X_1\cdots\X_n$ appears in the sum.
Furthermore, if the list $\X_1\cdots\X_n$ is invariant under some
nontrivial cyclic permutations, 
then $s_n$
is the order of the cyclic subgroup, otherwise $s_n=1$. (See sec. 6.1.)
In the process of computing $S$, 
there can arise any number of trapped $\sigma_3$s via \fsig,
however in the field expansion these can always be reduced to at most
one by (anti)commutation through ($\B$), $\A$ or $\C$, and if one remains
it will be placed at the beginning of the supertrace (thus resulting in
a marked superloop as illustrated). 
\midinsert
$$
\epsfxsize=\hsize\epsfbox{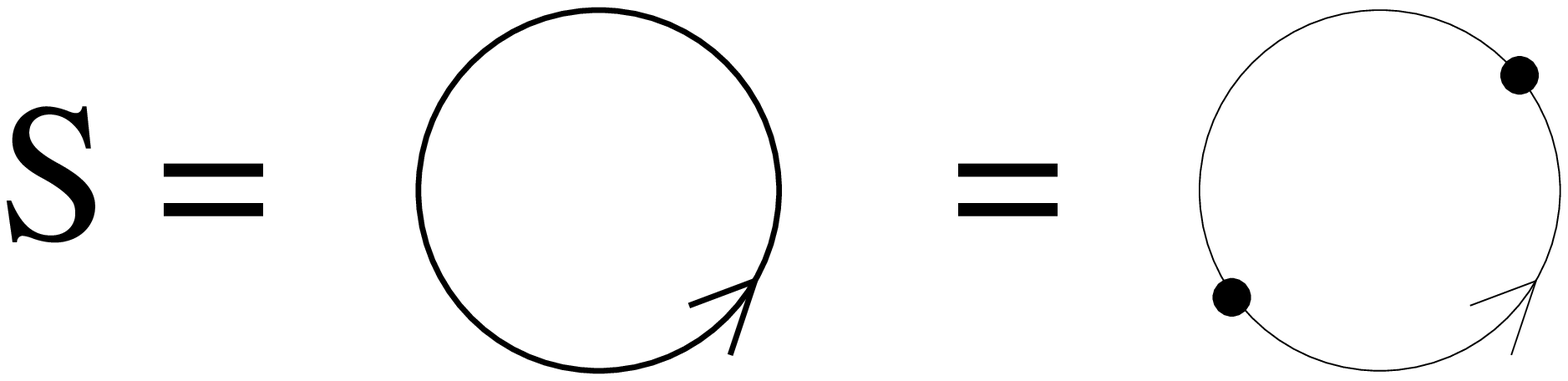}
$$
\centerline{ {\bf Fig.4.} Expansion of the action into supertraces
of fields, with and without a $\sigma_3$ insertion.}
\endinsert

Concentrating on a given vertex, but with $j$ so far undetermined,
and turning to the composite Wilson loops on the RHS of \fflow,
we assign the vertex's
flavours ($\A$, $\B$ or $\C$) together with their associated momenta and 
if appropriate Lorentz indices, to the points, summing over all cyclic
permutations. For each resulting configuration, the flavours of the
component wine and action vertices can then be determined. The position
of any embedded $\sigma_3$s in component action vertices then follow.
Of course some cases already
vanish at this stage due to the absence of the appropriate vertex,
\eg when a $\C$-point is placed on a wine. In other cases there
may be more than one choice for the component vertices, in which
case the corresponding Feynman diagrams are summed over. Attachments
made via partial supermatrices are expanded as in \fsig. 
The full set of $\sigma_3$s can be combined inside the Wilson
loops, (anti)commuting past external fields as necessary, and eliminated
via $\sigma_3^2=1$ and/or $\str\sigma_3=2N$, leaving at most one $\sigma_3$,
which is moved to its canonical position as determined by \sexp.
(Actually, the expansion step \fsig, and all $\sigma_3$s can be ignored for
classical vertices as explained in sec. 7.)
Finally, applying momentum conservation and 
including loop momentum integrals if appropriate, we read off according 
to the Feynman rules set out below, 
the complete expressions for the component vertices,
and multiply the whole by $1/2\Lambda^2$ \ymi.

\midinsert
$$
\epsfxsize=\hsize\epsfbox{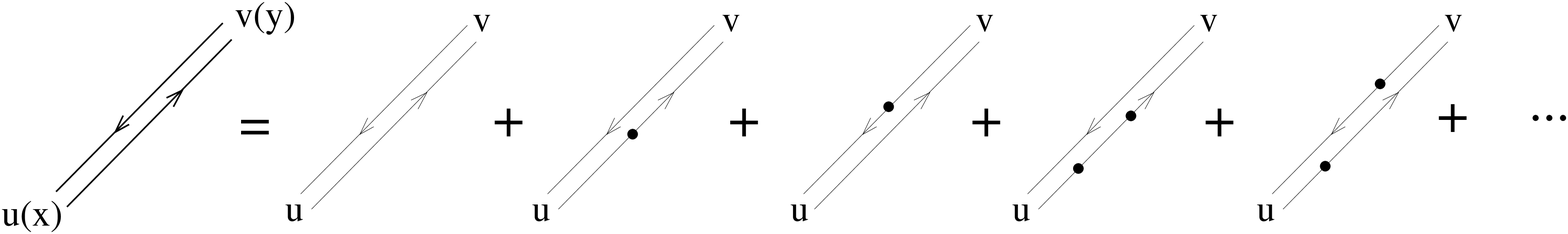}
$$
\centerline{ {\bf Fig.5.} Expansion of the wine in gauge fields.}
\endinsert
The wine
expansion \pwv\ appears as \fig\fwexp{fig 2 of ymi but without the gauge
field labels}. The wine component of the vertices is 
then given by \pwv\ (with $\AA$
replaced by $\A$ in the case of $\{K-c'\}_\A$) and thus in general,
from \wili\ -- \defWl, as \ymi:
 \eqnn\wnpoint 
$$\displaylines{
W_{\mu_1\cdots\mu_n,\nu_1\cdots\nu_m}(p_1,\cdots,p_n;q_1,\cdots,q_m;r,s)
\ (2\pi)^D \delta(\sum_{i=1}^n p_i+\sum_{j=1}^mq_j+r+s) =\hfill\wnpoint\cr
\hfill(-i)^{n+m}\int\!\!\!\!\int\!\!d^D\!u\,d^D\!v\int\!\!\D\Cu_{uv}
\int_u^v\!\!\!\!dx^{\mu_n}_n\!\int_u^{x_n}\!\!\!\!\!\!\!dx^{\mu_{n-1}}_{n-1}
\cdots\int_u^{x_2}\!\!\!\!\!\!\!dx^{\mu_1}_{1}
\int_v^u\!\!\!\!dy^{\nu_m}_m\!\int_v^{y_m}\!\!\!\!\!\!\!dy^{\nu_{m-1}}_{m-1}
\cdots\int_v^{y_2}\!\!\!\!\!\!\!dy^{\nu_1}_{1}\hfill\cr
\hfill \exp{-i\left(r.u+s.v+\sum_i\p_i.x_i+\sum_jq_j.y_j\right)}\quad,\cr
}$$
where the $x_i$ integration is along the curve $\Cu_{uv}$, and the
$y_j$ integration along the same curve but in the opposite direction
\cf \fig\winemom{fig.8 of ymi, except add alpha at r and beta at s,
wine with momentum labels as in (618.5)}. 
(As in ref. \ymi, all momenta are taken to be pointing in to the vertex.)
Explicit expressions up to $O(\AA^2)$, for two covariantizations, 
are given in sec. 5.
\midinsert
$$
\epsfxsize=0.4\hsize\epsfbox{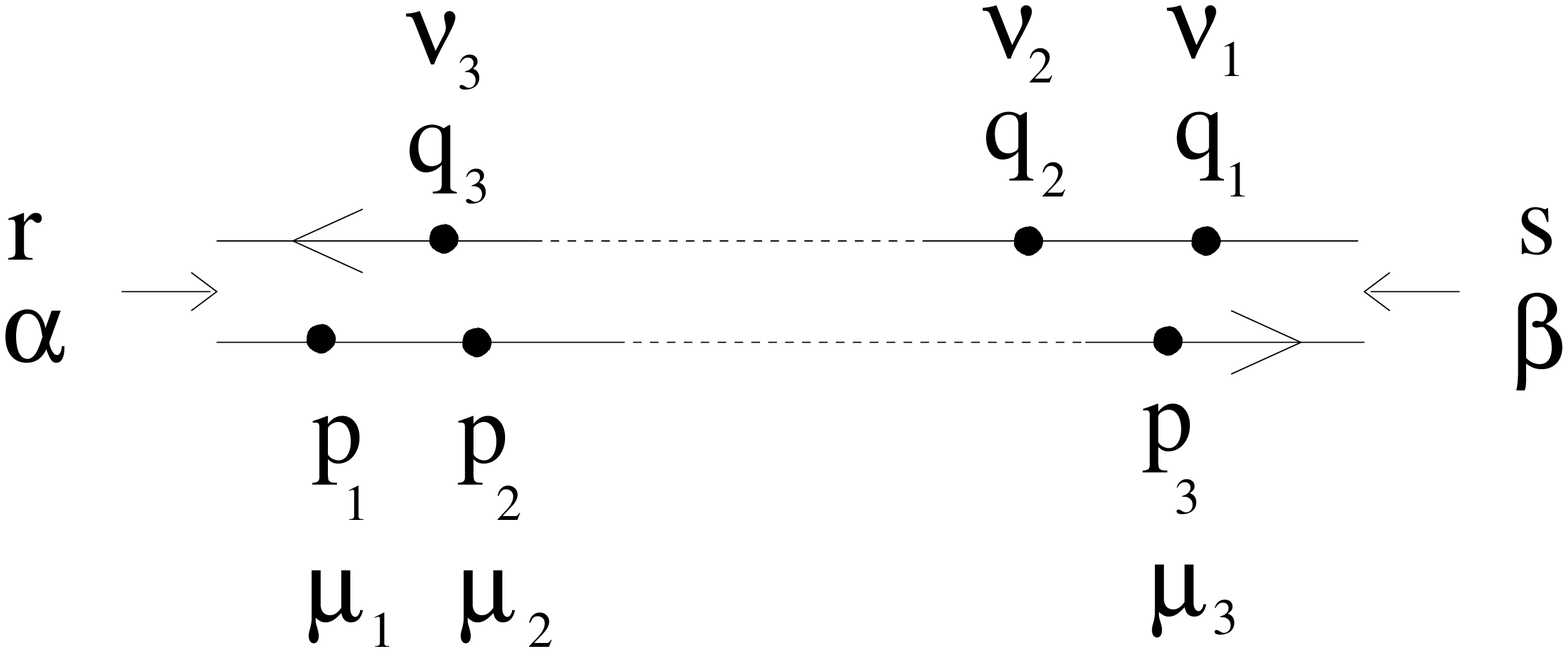}
$$
\centerline{{\bf Fig.6.} 
Feynman rule for the wine, with momentum and Lorentz labels.}
\endinsert

From \pRG, we now have several extra pieces 
to take into account in the full Feynman rule for the wine vertices
compared to ref. \ymi:
extra factors of $1/\Lambda^2$, contributions from $\l$,
the flavour and (if it exists) Lorentz index 
of the functional derivative the wine vertex actually attaches to. (Due
to $\l$, it is no longer the case that one just contracts these Lorentz 
indices.) A complete notation for the wine vertex Feynman rule can 
therefore be given as:
\eqn\wcomp{V^{\X_1\cdots\X_n,\Y_1\cdots\Y_m,\ZZ_1\ZZ_2}_{
\,\mu_1\,\cdots\,\mu_n\,,\,\nu_1\,\cdots\,\nu_m\,,\,a_1\,a_2\,}
(p_1,\cdots,p_n;q_1,\cdots,q_m;r,s)\quad,}
with momenta and indices as represented in \winemom,
where the $\X_i^{\mu_i}(p_i)$ and $\Y_j^{\nu_j}(q_j)$ are $\A$s or $\B$s,
and the functional derivatives are with respect to $\ZZ_1^{a_1}(r)$
and $\ZZ_2^{a_2}(s)$. The index $a_1$ ($a_2$) is the Lorentz index
$\alpha$ ($\beta$) if $\ZZ_1$ ($\ZZ_2$) is $\A$ or $\B$, and null if it is
$\C$.

This notation is convenient for listing the Feynman rules below.
They may all be expressed immediately in terms of the expansion of
$L$ with its extra pieces. For this we write [with the momentum
arguments $(p_1,\cdots,p_n;q_1,\cdots,q_m;r,s)$ here suppressed]:
%
%
\eqn\wab{W_{\mu_1\cdots\mu_n,\nu_1\cdots\nu_m,\al\be}=
\delta_{\al\be}W_{\mu_1\cdots\mu_n,\nu_1\cdots\nu_m}+
L_{\mu_1\cdots\mu_n,\nu_1\cdots\nu_m,\al\be}\quad,}
where shorthands \compac\ and \mno\ apply (see also \kdef\ and above),
and thus in particular at
the zero-point level $W_{p,\alpha\beta}=\delta_{\alpha\beta}W_p
+L_{p,\al\be}$. The first few $\C$-$\A_\be$ and $\A_\al$-$\A_\be$ 
vertices extracted from $\l\{L\}\l$ are:
\eqn\Lvs{\eqalign{L_{p,\be} &\equiv L_{,,\ph{\al}\be}(;;p,-p) = 
ip_\be L_p/\Lambda^2\cr
L_{p,\al\be} &=p_\al p_\be L_p/\Lambda^2\cr
L_{\mu,\beta}(p;q,r) &=-{i}\left\{
r_\beta L_\mu(p;q,r)+\delta_{\mu\beta}L_q\right\}/\Lambda^2\cr
L_{\mu,\alpha\beta}(p;q,r) &=\left\{
r_\beta\delta_{\mu\al}L_r
-q_\al\delta_{\mu\beta}L_q-q_\al r_\beta L_\mu(p;q,r)\right\}/\Lambda^2\cr
L_{\mu\nu,\be}(p,q;r,s) &=-i\left\{s_\be L_{\mu\nu}(p,q;r,s)
+\delta_{\be\nu}L_\mu(p;r,s\!+\!q)\right\}/\Lambda^2\cr
L_{\mu\nu,\al\be}(p,q;r,s) &=\big\{\delta_{\mu\al}\delta_{\nu\be}L_{p+r}
+s_\be\delta_{\mu\al}L_\nu(q;p\!+\!r,s)\cr
&\ph{=\big\{\delta_{\mu\al}\delta_{\nu\be}L_{p+r}}
-r_\al\delta_{\nu\be}L_\mu(p;r,s\!+\!q)
-r_\al s_\be L_{\mu\nu}(p,q;r,s)\big\}/\Lambda^2\quad,
}}
and via the coincident line identities (\cf  sec. 5 and ref. \ymi):
\eqn\Lvvs{\eqalign{L_{\mu,\nu,\be}(p;q;r,s) &= -L_{\mu\nu,\be}(p,q;r,s)
-L_{\nu\mu,\be}(q,p;r,s)\cr
L_{\mu,\nu,\al\be}(p;q;r,s) &=-L_{\mu\nu,\al\be}(p,q;r,s)
-L_{\nu\mu,\al\be}(q,p;r,s)\quad.\cr}}
From \pRG, we thus obtain for the zero-point wine Feynman rules \wcomp:
\eqn\vzero{\eqalign{
V^{,,\A\A}_{\,,,\,\mu\,\nu\,}(;;p,-p) =\ \ \ c'_{p,\al\be}\ &,\qquad
V^{,,\B\B}_{\,,,\,\al\,\be\,}(;;p,-p) =K_{p,\al\be}\ ,\cr
V^{,,\C\C}_{\,,,\ph{\C\C}}(;;p,-p) = M_p/\Lambda^2\ &,\qquad
V^{,,\C\A}_{\,,,\ph{\C}\,\be\,}(;;p,-p) =L_{p,\be}\ ,
}}
and for the one-point wine Feynman rules, using \compac:
\eqn\vone{\eqalign{V^{\A,\A\A}_{\ \mu,\,\al\,\be\,}(p;q,r) &=
V^{\B,\A\B}_{\ \mu,\,\al\,\be\,}(") =
V^{\B,\B\A}_{\ \mu,\,\al\,\be\,}(") = c'_{\mu,\al\be}(p;q,r)\cr
V^{\A,\B\B}_{\ \mu,\,\al\,\be\,}(p;q,r) &=K_{\mu,\al\be}(p;q,r)\cr
V^{\A,\C\C}_{\ \mu,\,\ph{\C\C}}(p;q,r) &={1\over\Lambda^2}M_\mu(p;q,r)\cr
V^{\A,\C\A}_{\ \mu,\ph{\C}\,\be\,}(p;q,r) &=
V^{\B,\C\B}_{\ \mu,\ph{\C}\,\be\,}(")=L_{\mu,\be}(p;q,r)\cr
V^{\A,\A\C}_{\ \mu,\,\al\,\ph{\C}}(p;q,r) &=
V^{\B,\B\C}_{\ \mu,\,\al\,\ph{\C}}(")=-L_{\mu,\al}(p;r,q)\quad.\cr}}
Of course we 
do not list those that vanish due to the absence of an appropriate vertex.
Similarly for the two-point wine Feynman rules:
\eqn\vtwo{\eqalign{
V^{\A\A,\A\A}_{\ \mu\,\nu\,,\,\al\,\be\,}(p,q;r,s) &=
V^{\B\A,\B\A}_{\ \mu\,\nu\,,\,\al\,\be\,}(") =
V^{\B\B,\A\A}_{\ \mu\,\nu\,,\,\al\,\be\,}(") =
V^{\A\B,\A\B}_{\ \mu\,\nu\,,\,\al\,\be\,}(") =
c'_{\mu\nu,\al\be}(p,q;r,s)\cr
V^{\A\A,\B\B}_{\ \mu\,\nu\,,\,\al\,\be\,}(p,q;r,s) &=
K_{\mu\nu,\al\be}(p,q;r,s)\cr
V^{\A,\A,\A\A}_{\ \mu\,,\nu\,,\,\al\,\be\,}(p;q;r,s) &=
V^{\B,\A,\A\B}_{\ \mu\,,\nu\,,\,\al\,\be\,}(") =
V^{\A,\B,\A\B}_{\ \mu\,,\nu\,,\,\al\,\be\,}(") =
c'_{\mu,\nu,\al\be}(p;q;r,s)\cr
V^{\A\A,\C\A}_{\ \mu\,\nu\,,\ph{\C}\be\,}(p,q;r,s) &=
L_{\mu\nu,\be}(p,q;r,s)\cr
V^{\A\A,\A\C}_{\ \mu\,\nu\,,\,\al\ph{\C}}(p,q;r,s) &=
L_{\nu\mu,\al}(q,p;s,r)\cr
V^{\A,\A,\C\A}_{\ \mu\,,\nu\,,\ph{\C}\be\,}(p;q;r,s) &=
L_{\mu,\nu,\be}(p;q;r,s)\cr
V^{\A\A,\C\C}_{\ \mu\,\nu\,,\ph{\C\C}}(p,q;r,s) &=
{1\over\Lambda^2}M_{\mu\nu}(p,q;r,s)\quad.
}}
Here also we only list those that we will need later.
The expansion of $\hS_A$ in \pseed\
is pure $\AA$ and we reserve unlabelled $\hS$ vertices for
this:
\eqn\hsexp{
\hS_A =\sum_{n=2}^\infty{1\over n}\int\!\!d^D\!x_1\cdots d^D\!x_n\,
\hS_{\mu_1\cdots\mu_n}(\x_1,\cdots,\x_n)\
\str \AA_{\mu_1}(\x_1)\cdots \AA_{\mu_n}(\x_n)\quad.}
This way the same explicit expressions as in refs. \ymi\alg\ apply:
\eqn\hSex{\eqalign{\hS_{\mu\nu}(\p) &\equiv\hS_{\mu\nu}(\p,-\p)=
2\Delta_{\mu\nu}(\p)/c_\p\cr
\hS_{\mu\nu\lambda}(\p,\q,\r) &={2\over c_\p}(p_\lambda\delta_{\mu\nu}-
p_\nu\delta_{\lambda\mu})
+2c^{-1}_\nu(\q;\p,\r)(p_\lambda r_\mu-p.r\delta_{\lambda\mu})
+{\rm cycles}\cr
\hS_{\mu\nu\lambda\sigma}(\p,\q,\r,\s) &={1\over c_{\p+\q}}
(\delta_{\sigma\mu}\delta_{\lambda\nu}-
\delta_{\lambda\mu}\delta_{\nu\sigma})+2c^{-1}_\nu(\q;\p,\r\!+\!\s)
(p_\sigma\delta_{\lambda\mu}-p_\lambda\delta_{\sigma\mu})\cr
&\ph{=}+2c^{-1}_\sigma(\s;\p,\r\!+\!\q)
(p_\nu\delta_{\mu\lambda}-p_\lambda\delta_{\mu\nu})
+2c^{-1}_{\nu\lambda}(\q,\r;\p,\s)(p_\sigma s_\mu-p.s\delta_{\sigma\mu})\cr
&\ph{=}+c^{-1}_{\nu,\si}(\q;\s;\p,\r)(p_\lambda r_\mu-p.r\delta_{\lambda\mu})
+{\rm cycles}\cr
}}
\etc,
where in the two-point vertex we set $\p_1=-\p_2=\p$, and introduce
the transverse combination
$\Delta_{\mu\nu}(\p) :=p^2\delta_{\mu\nu}-p_\mu p_\nu$,
in the three-point vertex
we add the two cyclic permutations of $(p_\mu,q_\nu,r_\lambda)$,
and in the four-point vertex the three cyclic permutations
of $(p_\mu,q_\nu,r_\lambda,s_\sigma)$.

\midinsert
$$
\epsfxsize=0.4\hsize\epsfbox{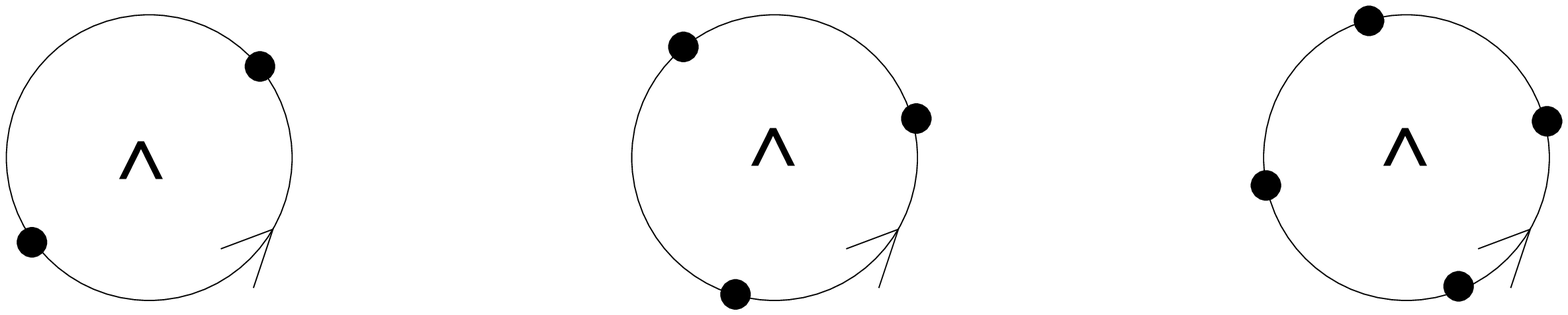}
$$
\centerline{{\bf Fig.7.} 
Feynman rules for the seed vertices.}
\endinsert
The full `seed' vertices that we will need,
are then given as follows, \cf 
\fig\fseed{two, three and four-point seed vertices}.
Two-point vertices (there are no one-point vertices): 
\eqn\hstwo{\eqalign{\hS^{\A\A}_{\,\mu\,\nu\,}(p) &=\hS_{\mu\nu}(p)\cr
\hS^{\B\B}_{\,\mu\,\nu\,}(p) &= 
\hS_{\mu\nu}(p)+2\Lambda^2\delta_{\mu\nu}/\ct_p\cr
\hS^{\C\C}(p) &=2\Lambda^2p^2/\ct_p+2\sigma\Lambda^4\quad,}}
three-point vertices:
\eqn\hsthree{\eqalign{\hS^{\A\A\A}_{\,\mu\,\nu\,\la\,}(p,q,r) &=
\hS_{\mu\nu\la}(p,q,r)\cr
\hS^{\C\C\A}_{\ph{\C\C}\,\la\,}(p,q,r) &=
2\Lambda^2\left\{p_\lambda\,\ct^{-1}_p-q_\la\,\ct^{-1}_q
-p.q\,\ct^{-1}_\la(r;q,p)\right\}\cr
\hS^{\B\B\A}_{\,\mu\,\nu\,\la}(p,q,r) &=
\hS_{\mu\nu\la}(p,q,r)+2\Lambda^2\delta_{\mu\nu}\,\ct^{-1}_\la(r;q,p)\cr
\hS^{\B\B\C}_{\,\mu\,\nu\,\ph{\C}}(p,q,r) &=
2i\Lambda^2\left\{r_\nu\,\ct^{-1}_\mu(p;r,q)
+r_\mu\,\ct^{-1}_\nu(q;p,r)\right\}\quad,
}}
and four-point vertices:
\eqn\hsfour{\eqalign{\hS^{\A\A\A\A}_{\,\mu\,\nu\,\la\,\si\,}(p,q,r,s) &=
\hS_{\mu\nu\la\si}(p,q,r,s)\cr
\hS^{\B\B\A\A}_{\,\mu\,\nu\,\la\,\si\,}(p,q,r,s) &=
\hS_{\mu\nu\la\si}(p,q,r,s)
+2\Lambda^2\delta_{\mu\nu}\,\ct^{-1}_{\la\si}(r,s;q,p)\cr
\hS^{\C\C\A\A}_{\ph{\C\C}\,\la\,\si\,}(") =
2\Lambda^2 &\big\{\delta_{\la\si}\,\ct^{-1}_{p+s}\!+
p_\la\,\ct^{-1}_\si(s;q\!+\!r,p)
-q_\si\,\ct^{-1}_\la(r;q,p\!+\!s)
-p.q\,\ct^{-1}_\la(r,s;q,p)\big\}
}}

\newsec{Explicit expressions for wine vertices}

The wine vertices are defined implicitly
by the choice of covariantization, and
\wv\ or equivalently \pwv. In this section we
derive formulae for the covariantizations
\pourchoice\ and \straw, although in this paper we will only 
use \pourchoice\ for concrete calculations.

\subsec{Straight Wilson Lines}

In the straight line case \straw, the integration over curves 
in \wnpoint, is trivial,
being replaced by its normalisation \pnorm, whilst $\Cu_{uv}$ is just
the straight line which we may parametrize as $x_i=u+(v-u)t_i$
and $y_i=v+(u-v)t'_i$. Translation invariance implies that the integrand
depends on $u$ and $v$ only in the combination $v-u$. 
We replace $v-u$ with derivatives with respect
to the its conjugate momentum
$$p=s+\sum_{i=1}^np_it_i+\sum_{j=1}^mq_j(1-t'_j)\quad.$$
(This expression drops out by substitution in the exponential of \wnpoint\
and is unique up to the use of overall momentum conservation.)
Integrating over the overall position yields the overall momentum conserving
$\delta$ function. Cancelling this from both sides of \wnpoint\ we 
arrive at
\eqnn\wslnpoint\
$$\displaylines{
W_{\mu_1\cdots\mu_n,\nu_1\cdots\nu_m}(p_1,\cdots,p_n;q_1,\cdots,q_m;r,s)
 =\hfill\wslnpoint\cr
\hfill(-1)^{m}\int_0^1\!\!\!\!dt_n\!\int_0^{t_n}\!\!\!\!\!\!\!
dt_{n-1}\cdots\int_0^{t_2}\!\!\!\!\!\!\!dt_{1}
\int_0^1\!\!\!\!dt'_m\!\int_0^{t'_m}\!\!\!\!\!\!\!dt'_{m-1}
\cdots\int_0^{t'_2}\!\!\!\!\!\!\!dt'_{1}
{\partial\over\partial p^{\mu_1}}\cdots{\partial\over\partial p^{\mu_n}}
{\partial\over\partial p^{\nu_1}}\cdots{\partial\over\partial p^{\nu_m}}
W_p\quad.\cr}$$
As an example, the one-point vertex is given by
\eqn\wslone{W^\mu(p;r,s)={2\over\Lambda^2}\int_0^1\!\!\!\!dt\, W'_{tp+s}\,
(tp+s)^\mu\quad.}
We note that these straight
line vertices are completely symmetric on all Lorentz indices.

\subsec{Covariantization via \pourchoice}

We first explain why this covariantization is of the coincident line
form \wili, equivalently \wnpoint. In expanding \pourchoice,
or preferably (since the supermatrices are not necessary here)
\eqn\ourchoice{u\{W\}v={\rm tr}\int\!\!d^D\!x\, u(\x)
\,W(-D^2/\Lambda^2)\cdot v(\x)\quad,}
in powers of the gauge field, we  only need write
$$\displaylines{
u\{W\}v= 
\sum_{m,n=0}^\infty\int\!\!d^D\!x\,d^D\!y\,d^D\!x_1\cdots d^D\!x_n\,
W_{\mu_1\cdots\mu_n}
(\x_1,\cdots,\x_n;\x,\y) \cr
\hfill \tr\left[\, u(\x)\, A_{\mu_1}(\x_1)\cdots A_{\mu_n}(\x_n)\, 
\cdot v(\y)\,\right]\quad,\cr}$$
where these vertices 
are the coefficients of the
ordered product of gauge fields $A_{\mu_1}(\x_1)\cdots$ $A_{\mu_n}(\x_n)$
as before, but moreover, each of
these gauge fields is understood to act by commutation on the expression
to its right.
Expanding the commutators and relabelling the gauge fields that appear
on the right hand side of $v(y)$ in terms of $A_{\nu_i}(y_i)$, as in
\wv, the other vertices are then given in terms of them by
\eqn\coids{
W_{\mu_1\cdots\mu_n,\nu_1\cdots\nu_m}(p_1,\cdots,p_n;q_1,\cdots,q_m;r,s)
=(-)^m\!\!\!\!\!\sum_{interleaves}\!\!\!\!\!
W_{\la_1\cdots\la_{m+n}}(k_1,\cdots,k_{m+n};r,s)\quad,}
where we have (trivially) transferred to momentum space, 
and the sum runs over all interleaves of
the sequences $p_1^{\mu_1},\cdots,p_n^{\mu_n}$ and
$q_m^{\nu_m},\cdots,q_1^{\nu_1}$ \ie combined
sequences $k_1^{\la_1},\cdots,k_{m+n}^{\la_{m+n}}$ in which the $p^\mu$s
remain ordered with respect to each other,
and similarly the $q^\nu$s remain in reverse order.
These are in fact the ``coincident line'' identities of ref. \ymi.
Now we employ a single-Wilson-line representation for the $W$s
on the RHS of \coids, \ie \wnpoint\ with $m=0$. (This
follows from a path integral representation
of the kernel in \ourchoice, expressing the
$A$s in the adjoint representation.)
By isolating the integrals over the coordinates conjugate
to the $q_j$s in the right hand terms of \coids, 
we readily see that they collect together, and on reversing their direction, 
form the $y_j$ integrals of the coincident line representation \wnpoint.

Now we compute the
contribution to $W_{\mu_1\cdots\mu_n}(p_1,\cdots,p_n;r,s)$ from
the $(-D^2)^m$ term in the Taylor expansion of $W(-D^2/\Lambda^2)$.
(This expansion must exist by quasilocality \ymi.) \eqnn\dexp
In this paper we only need explicit expressions to order $A^2$,
but we derive a formula for
the general vertex below, in order to analyse
its large momentum behaviour. Expanding, 
$$\displaylines{(-D^2)^{m}=(-\partial^2)^{m}+
\sum_{\alpha=0}^{m-1}(-\partial^2)^{\alpha}(i\partial.
A+iA.\partial+A^2)(-\partial^2)^{m-1-\alpha}\hfill\dexp\cr
\hfill-\!\!\!\!\!\sum_{{\alpha,\beta,\gamma\ge0\atop
\alpha+\beta+\gamma=m-2}}\!\!\!(-\partial^2)^{\alpha}(\partial.A+A.\partial)
(-\partial^2)^{\beta}(\partial.A+A.\partial)(-\partial^2)^{\gamma}
+O(A^3)\quad.\cr}$$
We already have noted in \mno\ that the zeroth order in $A$ gives 
\eqn\wzero{W(;r,s)=W_s\quad,}
and this is trivially confirmed by \dexp.
Transforming the $O(A)$ terms to momentum space, 
we see that \dexp\ supplies a contribution to $W_\mu(p;r,s)$ of the form
\eqn\wsum{(p+2s)_\mu\sum_{\alpha=0}^{m-1}r^{2\alpha}s^{2m-2-2\alpha}
=(r-s)_\mu{r^{2m}-s^{2m}\over p.(r-s)}}
(where we have used momentum
conservation) and thus resumming the Taylor expansion,
\eqn\wone{W_\mu(p;r,s)=(r-s)_\mu\,{W_r-W_s\over p.(r-s)}\quad.}
Similarly at $O(A^2)$, \dexp\ supplies a contribution to 
$W_{\mu\nu}(p,q;r,s)$ of form
\eqn\wsumm{\delta_{\mu\nu}\sum_{\alpha=0}^{m-1}r^{2\alpha}s^{2m-2-2\alpha}
\ +\ (p+2s+2q)_\mu(q+2s)_\nu\!\!\!\!\!\!
\sum_{{\alpha,\beta,\gamma\ge0\atop
\alpha+\beta+\gamma=m-2}}\!\!r^{2\alpha}(s+q)^{2\beta}s^{2\gamma}\quad.}
The latter sum may readily be evaluated \eg by noting that it is equal
to
$${1\over2\pi i}\oint\!\! {dz\over z^{m-1}}\, 
{1\over (1-r^2z)(1-[s+q]^2z)(1-s^2z)}$$
for a contour of infinitessimal radius encircling the origin (which we close
onto the other poles), and thus after resumming the
expansion of $W$, \eqnn\wtwo
$$\displaylines{W_{\mu\nu}(p,q;r,s)=\delta_{\mu\nu}{W_s-W_r\over s^2-r^2}
\hfill\wtwo\cr
\hfill-(p+2r)_\mu(q+2s)_\nu\left\{{W_{s+q}\over q.(q+2s)\,p.(p+2r)}
+{1\over s^2-r^2}\left[{W_r\over p.(p+2r)}-{W_s\over q.(q+2s)}\right]
\right\}}$$ 
The case where the gauge fields appear on either side of the wine
is then given in terms of this by \coids:
\eqn\wtwop{
W_{\mu,\nu}(p;q;r,s)=-W_{\mu\nu}(p,q;r,s)-W_{\nu\mu}(q,p;r,s)\quad.}

\subsec{Special momenta and covariantization \pourchoice}

The formulae \wone\ and \wtwo\ are ambiguous at certain special 
momenta. One way to determine the correct value is simply to
return to first principles and resum
\wsum\ and \wsumm\ at the special values. However it is comforting to
find that these results also appear uniquely by recalling
that these formulae are only valid when the total momentum flowing into
the vertex vanishes (\ie is conserved), and
taking the limit as the special configuration is approached.
Thus \wone\ needs care at the point $p=0$, \ie $r=-s$,
however by momentum conservation
we can replace $p.(r-s)$ by $s^2-r^2$ 
after which the limit $r\to-s$ is trivial in \wone, giving
\eqn\wonesp{W_\mu(0;-s,s)={2s_\mu\over\Lambda^2}W'_s\quad.}
(This also follows from resumming \wsum\ with $p=0$ and $r=-s$, 
or from the gauge transformation relations \cf (6.2) or ref. \ymi:
$$p^\mu W_\mu(p;-s-p,s)=W_{s+p}-W_s$$
by expanding to first order in $p$ after which $p^\mu$ may be removed
uniquely from both sides. The longitudinal parts of higher
powers in $p$ are also determined uniquely by gauge invariance.
See appendix A of ref. \ymi.)
Note that the case $r=s$ ($p=-2s$) in \wone\ trivially gives zero 
[by a limit or from \wsum].

\wtwo\ needs care at the point $p=-q$ corresponding to $r=-s$:
\eqnn\wtwosp\ 
$$\eqalignno{W_{\mu\nu}(p,-p;r,-r) &=\lim_{\epsilon\to0}
W_{\mu\nu}(p,-p-\epsilon;r+\epsilon,-r) &\wtwosp\cr
&={\delta_{\mu\nu}\over\Lambda^2}W'_r+(p+2r)_\mu(p+2r)_\nu\left\{
{W_{r+p}-W_r\over\left[p.(p+2r)\right]^2}-{W'_r\over\Lambda^2p.(p+2r)}
\right\}\quad. \cr}$$

There are other special momentum configurations that need careful definition
in \wtwo, but we will need only \wtwosp\ here.

\subsec{The general-point vertex in covariantization \pourchoice}
The order $A^n$ term in $(-D^2)^m$ supplies a
contribution to $W_{\mu_1\cdots\mu_n}(p_1,\cdots,p_n;r,s)$
which is a sum over $a$ insertions of $A^2$ and $b$ insertions of 
$\partial.A+A.\partial$ and over all permutations of these factors,
\cf \dexp. In momentum space this reads
\eqn\wnsum{\sum_{a,b\atop 2a+b=n}\sum_{\rm perms}T_1^{i_1}\cdots
T_{a+b}^{i_{a+b}}\!\!\sum_{\al_0,\cdots,\al_{a+b}\atop\sum\al_j=m-a-b}
\prod_{k=0}^{a+b}P^{2\al_k}_{I_k}\quad,}
where the $i_k=$ 1 or 2, according to whether a $\partial.A+A.\partial$
or $A^2$ term is taken from $D^2$ respectively. They yield respectively 
the tensors
$$T^1_k=p_{I_k}^{\mu_{I_k}}+2P_{I_k}^{\mu_{I_k}}\ins11{and}
T^2_k=\delta^{\mu_{I_k-1}\mu_{I_k}}\quad.$$ 
$I_k=\sum_{j=1}^k i_j$ keeps track of the number of gauge fields 
accounted for, with $I_0:=0$ and $I_{a+b}=n$, and
the total momentum flow (directed from $s$)
into the $k^{\rm th}$ insertion is given by
$$P_{I_k}=s+\sum^n_{j=I_k+1}p_j\quad.$$
These definitions are 
illustrated in \fig\fasy{mom flow for gen vertex: see p15}.
Note that $P_{I_0}=P_0=-r$ and $P_{I_{a+b}}=P_n=s$.
\midinsert
$$
\epsfxsize=0.8\hsize\epsfbox{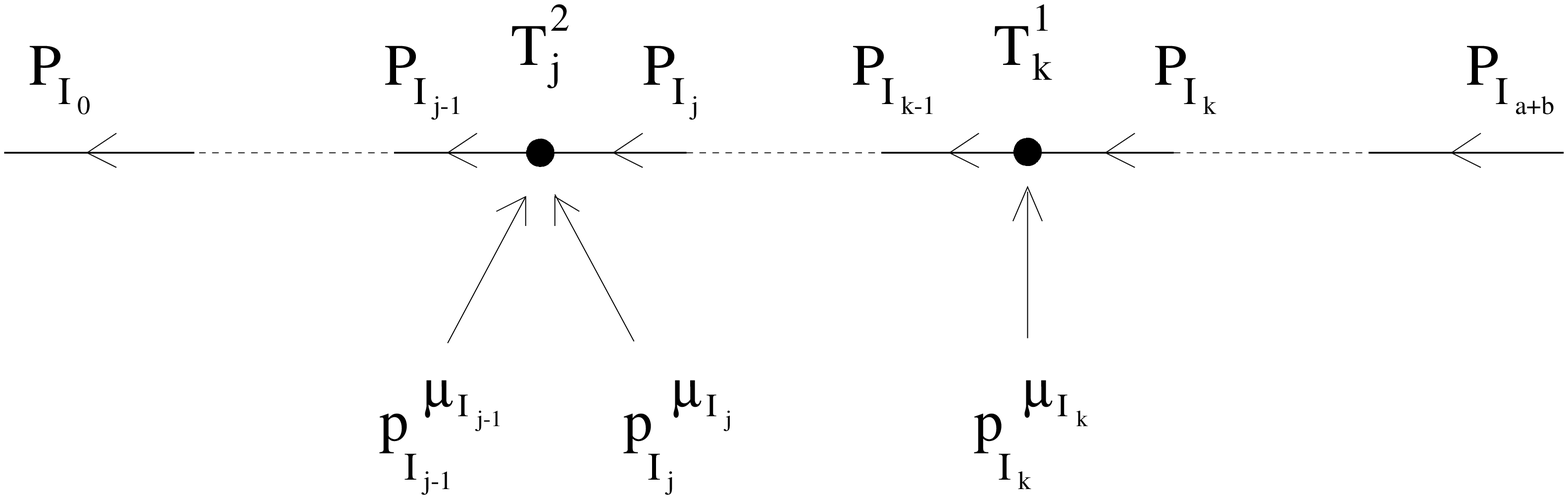}
$$
\centerline{{\bf Fig.8.} 
Tensor structure and momentum flow in the general vertex.} 
\endinsert

As in \wsumm, the RH sum in \wnsum\ may be evaluated readily by 
writing first
$$\sum_{a,b\atop 2a+b=n}\sum_{\rm perms}T_1^{i_1}\cdots
T_{a+b}^{i_{a+b}}
{1\over2\pi i}\oint\!\! {dz\over z^{m+1-a-b}}\, 
\prod^{a+b}_{k=0}{1\over1-z P^2_{I_k}}$$
for a contour of infinitessimal radius encircling the origin. Closing
on to the other poles and resumming the Taylor expansion of $W$, we 
thus obtain
\eqn\wgen{W_{\mu_1\cdots\mu_n}(p_1,\cdots,p_n;r,s)
=\sum_{a,b\atop 2a+b=n}\sum_{\rm perms}T_1^{i_1}\cdots
T_{a+b}^{i_{a+b}}\sum^{a+b}_{j=0}W_{P_{I_j}}\prod_{k=0\atop k\ne j}^{a+b}
{1\over P^2_{I_j}-P^2_{I_k}}\quad.}

\subsec{Large momentum behaviour of covariantization \pourchoice}
We determine the behaviour of such a vertex when a given momentum flow 
through the vertex becomes large. This is needed to study the finiteness 
properties of the exact RG and is used for the proof in sec. 9. We have
already derived in appendix A of ref. \ymi, lower bounds on the divergences,
by gauge invariance considerations. Simply by power counting, we can
now obtain from \wgen, upper bounds on these divergences. In some cases,
where no cancellations can occur or where the lower and upper bound agree,
we can thus be confident that the large momentum behaviour is then precisely
known. Furthermore, we can readily
furnish simple expressions for the coefficients
of the leading large momentum behaviour. 

It is helpful in this section to adopt the following convention. By 
$\approx$ we will mean equal
up to corrections that decay relative to the stated term as 
the large momentum $k\to\infty$;
by $\sim$ we mean also that the stated term is determined only up
to a coefficient of $O(k^0)$.

In fact this level of precision is much more than
we need here, since the proof in sec. 9 only relies on the
following properties:
when the large momentum flow (the loop momentum $k$)
is from end to end ($s\approx k$, $r\approx-k$, 
$k\to\infty$) the behaviour of the $n$ point vertex is no worse than that of
the zero point vertex $W_k$, and when the large momentum instead
enters via the side (one $p_j\approx \pm k$) the `covariantized 
differentiated propagators' $\{c'\}$, $\{K\}$,
$\{L\}$, $\{M\}$ do not diverge.

Consider first the case where the large momentum flow is from end to end
($s\approx k$, $r\approx-k$, $k\to\infty$, all the $p_j$ finite). In this
case all the $P^2_{I_j}\approx k^2$, so that $P^2_{I_j}-P^2_{I_k}\sim k$.
Including all the $T^1\sim k$ ($T^2\sim k^0$ of 
course), one may imagine from \wgen\ that we simply have
$W_{\mu_1\cdots\mu_n}(p_1,\cdots,p_n;r,s)\sim W_k$,
where we have used the fact that the leading behaviour 
apparently arises from having $a=0$. Whilst good enough for our
purposes the ultraviolet behaviour is clearly better than this,
since from \wnsum, each term in the 
Taylor series goes as $k^{2m-n}$ (independent of $a$ and $b$) and 
thus as a whole:
\eqn\endtoend{
W_{\mu_1\cdots\mu_n}(p_1,\cdots,p_n;r,s)\approx W_k/k^n\!\!\!
\sum_{a,b\atop 2a+b=n}\!\!\!2^b\!\!\sum_{\rm perms}{\tilde T}_1^{i_1}\cdots
{\tilde T}_{a+b}^{i_{a+b}}\qquad(s\approx-r\approx k)
\quad,}
where the tensor sum is $O(k^0)$, with 
${\tilde T}^2=T^2$, ${\tilde T}^1=k^{\mu_{I_k}}/k$.
Furthermore we note that this expression agrees with the lower bound
from ref. \ymi, so we are confident that \endtoend\ is correct. 
The problem with the analysis of \wgen\ is that when many terms
are involved as here, there can be many cancellations in \wgen\ of the
leading behaviours as determined by power counting. But of course such
power counting always yields an upper bound on the large momentum 
behaviour.

From \coids\ we readily find the large momentum behaviour for the
$n+m$ point vertex in the case the large momentum $s\approx-r\approx k$
flows from end to end, in particular we see that it is $\sim W_k/k^{m+n}$ 
in agreement with the lower bound from \ymi.

For the more complex cases, the results depend on whether the
kernel $W_k$ decays or grows for large momentum. In the growing cases,
corresponding in this paper to $W=c^{-1}$ or $\ct^{-1}$, we can again
obtain a too pessimistic upper bound from power counting \wgen. The exact
result, agreeing with the lower bound from \ymi, again follows 
most straightforwardly from
\wnsum. Thus, from the maximum power of $k$ in \wnsum, we see in
particular that an $n+m$ vertex with large momentum $k$ entering and
leaving from any two points is again $\sim W_k/k^{m+n}$.

When the kernel $W_k$ decays for large $k$ 
(corresponding in this paper to $c'_k$, $K_k$, $L_k$ and $M_k$)
the leading large $k$ behaviour depends on the precise configuration.
Thus we will show that
\eqnn\endton
\eqnn\endtonn
\eqnn\endtopj
$$\displaylines{
W_{\mu_1\cdots\mu_n,\nu_1\cdots\nu_m}(p_1,\cdots,p_n;
q_1,\cdots,q_m;r,s)\hfill\cr
\approx
-{k^{\mu_n}\over k^2}W_{\mu_1\cdots\mu_{n-1},\nu_1\cdots\nu_m}(p_1,
\cdots,p_{n-1};q_1,\cdots,q_m;r,s\!+\!p_n)
\hfill(s\approx-p_n\approx k)\quad\endton\cr
\approx\left(2{k^{\mu_{n-1}}k^{\mu_n}\over k^4}-
{\delta^{\mu_{n-1}\mu_n}\over k^2}
\right)W_{\mu_1\cdots\mu_{n-2},\nu_1\cdots\nu_m}(p_1,\cdots,p_{n-2};
q_1\cdots q_m;r,s\!+\!p_{n-1}\!+\!p_n)\hfill\cr
\hfill(s\approx-p_{n-1}\approx k)\quad\endtonn
\cr
\sim {1\over k^{n-j+1}} 
W_{\mu_1\cdots\mu_{j-1},\nu_1\cdots\nu_m}(p_1,\cdots,p_{j-1};
q_1,\cdots,q_m;r,s\!+\!\!\sum_{i=j}^np_i)\hfill
(s\approx-p_j\approx k)\quad\endtopj
}$$
Closely similar identities then follow easily for $k$ leaving via $q_1$,
$q_2$ or any $q_i$,
from the charge conjugation invariance identity \eqnn\ccw
$$\displaylines{W_{\mu_1\cdots\mu_n,\nu_1\cdots\nu_m}(p_1,\cdots,p_n;
q_1,\cdots,q_m;r,s)=\hfill\cr
\hfill(-)^{n+m}W_{\nu_m\cdots\nu_1,\mu_n\cdots\mu_1}
(q_m\cdots q_1;p_n\cdots p_1;r,s)\quad,\qquad\ccw\cr}$$
equivalently reversing the
direction of the Wilson lines in \wnpoint, \ymi.

For the cases where the large momentum both enters and leaves through the
side, say $p_{j+1}\approx -p_j\approx k$, we will show that
\eqn\byside{
W_{\mu_1\cdots\mu_n,\nu_1\cdots\nu_m}(p_1,\cdots,p_n;
q_1,\cdots,q_m;r,s)\approx W_{\mu_1\cdots\mu_n,\nu_1\cdots\nu_m}\left[
\delta^{\mu_j\mu_{j+1}}\mapsto{\Delta^{\mu_j\mu_{j+1}}(k)\over k^2}\right],
}
where here the notation means to keep only those terms in
$W_{\mu_1\cdots\mu_n,\nu_1\cdots\nu_m}$ which contain 
$\delta^{\mu_j\mu_{j+1}}$ and for these make the replacement indicated
(no other changes required). In particular this allows us to read off
from \wtwo,
$$W_{\mu\nu}(k,-k-r-s;r,s)\approx {\Delta_{\mu\nu}(k)\over k^2}
{W_s-W_r\over s^2-r^2}$$
and from \wtwosp,
$$W_{\mu\nu}(k,-k;p,-p)\approx {\Delta_{\mu\nu}(k)\over k^2\Lambda^2}W'_p
\quad.$$
For general separation, on one side:
\eqn\bysidee{W_{\mu_1\cdots\mu_n,\nu_1\cdots\nu_m}(p_1,\cdots,p_n;
q_1,\cdots,q_m;r,s)\sim 1/k^{v-1}\qquad (p_{j+v}\approx-p_j\approx k).}
Clearly these conclusions hold also for the $q$ side
(\eg from \ccw, or exchange or Lorentz symmetry, see sec. 6 or \ymi). 
If $k$ enters and leaves by different sides, say $p_i\approx-q_j\approx k$,
then by \coids, \bysidee\ and \byside, we see that the vertex behaves
as $\sim\Delta^{\mu_i\nu_j}(k)/k^2$, irrespective of the values of $i$
and $j$ (reflecting the fact that in \wnpoint\ these points can always
get close to each other).

These properties are arrived at, as follows.
Setting $s\approx k$, $p_n\approx-k$ in \wgen, all other
momenta fixed and finite, the leading term arises when
$i_{a+b}=1$, through $T^1_{a+b}\approx k^{\mu_n}$. Since we can neglect 
the $j=a+b$ term, and factor out $T^1_{a+b}$ and the denominator 
$P^2_{I_j}-P^2_{a+b}\approx-k^2$ in the $j\ne a+b$ terms,
we are left precisely with the expression for the $n-1$ point vertex,
giving \endton\ for $m=0$.
Now if $k$ leaves through $p_{n-1}\approx-k$, then the 
leading terms are furnished by $T^1_{a+b}T^1_{a+b-1}$
and $T^2_{a+b}$. Factoring out the now two divergent denominators 
gives the $m=0$ cases of \endtonn. Proceeding similarly with the
$k$ leaving point further down the line, we readily find \endtopj\
for $m=0$. The large momentum behaviour of these
three cases is confirmed by the fact that 
they saturate the lower bounds derived
in ref. \ymi\ (and in addition the first two clearly involve no
cancellations). Now using \coids, we readily see that they hold also
for $m>0$ (the terms on the RHS of \coids\ only contributing 
at leading order when
no $q$ momenta get between the divergent pair, $p_j$ and $s$).

The cases where both large momenta are on the side, follow similarly.
Thus with $p_{j+1}\approx-p_j\approx k$ in \wgen, we see that the
leading terms come from $T^2=\delta^{\mu_j\mu_{j+1}}$ in which case
no $P_{I_k}$ diverges, and $(T^1)^2\approx k^{\mu_j} k^{\mu_{j+1}}$
with $-1/k^2$ from factoring out a divergent denominator.
Since there is no opportunity for cancellation we can be confident
that \byside\ is then correct, despite the fact that the gauge 
invariance analysis gives a lower bound at $\sim1/k$ \ymi.
Indeed this disagreement between upper and lower bounds
is allowed precisely because the leading
behaviour \byside\ is transverse in $k$ and thus `escapes' the Ward
identities. \bysidee\ follows from a similar analysis and agrees
with gauge invariance analysis expectations after taking into account
the more divergent initial case \byside. Finally the $m>0$ cases
are established as before, by use of \coids.

We note that it is quite straightforward to go further
and establish the order of the next-to-leading
terms that we have been neglecting (typically down by $1/k$),
and even their precise form.

\subsec{Large momentum behaviour of straight line vertices}
We finish this section with a brief remark about  vertices \wslnpoint. 
Broadly speaking,
the straight line vertices have a comparable behaviour.
However, let $W_k$ decay for large $k$, and consider
as an example \wslone, where the large momentum leaves
through the side:
$$W^\mu(-k\!-\!r;r,k)={2\over\Lambda^2}\int_0^1\!\!\!\!dt\, 
W'_{tk+(t-1)r}\,
[tk+(t-1)r]^\mu\quad.$$
We cannot just take the leading term from the
$t$ integrand because this leads to an integral that does not converge
at $t=0$. Substituting $t=\Lambda\sqrt{x}/k$, we obtain the leading
behaviour:
$$W^\mu(-k\!-\!r;r,k)={1\over k}\int_0^\infty{dx\over\sqrt{x}}\, 
W'(\zeta^2)\,\zeta^\mu\quad,$$
where $\zeta={\hat k}\sqrt{x}-r/\Lambda$ (so $\zeta^2=
x-2r.{\hat k}/\Lambda+r^2/\Lambda^2$), with ${\hat k}$ being the unit 
vector in direction $k$, and corrections to the above being $O(1/k^2)$.
We see that as in \endton, $W_\mu\sim1/k$, however unlike \endton\ and
the other cases above, it cannot be expressed as an inverse power of
$k^2$ with a coefficient which is analytic in its momenta (here $k$ and
$r$).  Although mostly a matter of taste, it is the more regular large
momentum behaviour of the vertices following from \pourchoice, that led
us to use this covariantization for the concrete calculations reported
in this paper.

\newsec{Symmetries}
As well as cyclic and exchange symmetry, and the
symmetries of gauge invariance and
charge conjugation, that are inherited and preserved from 
the formulation in ref. \ymi\ (and may be interpreted geometrically
in terms of Wilson loops \ymi\alg), 
some new symmetries appear: fermion
number, which is the $U(1)$ remainder of the original global $U(N|N)$
symmetry, and an interesting $Z_2$ symmetry that exchanges the two
$SU(N)$ subgroups whilst effectively changing the sign of $g^2$. 
This latter symmetry is thus a ``theory space'' symmetry, a symmetry of 
the flow equation \pRG\ but not of the action $S$. 

These symmetries provide the key to understanding the formulation at
a deeper level. We comment on them below, providing definitions where
necessary. We also comment on reality, and include for later some
comments on
Poincar\'e invariance and dimensional assignments.

\subsec{Cyclicity}
Some action vertices inherit symmetries from cyclicity of the
supertrace (for supermatrices). We have already mentioned this in
sec. 4 where these vertices were defined divided by the order of
the symmetry group. Thus the vertices in \hsexp\ are fully cyclically
symmetric:
$$\hS_{\mu_1\cdots\mu_n}(p_1,\cdots,p_n)=\hS_{\mu_2\cdots\mu_n\mu_1}
(p_2,\cdots,p_n,p_1)\quad.$$
Similarly, the vertices \hstwo\ appear in $\hS$ with a factor $1/2$, 
and the $\A\A$ and $\B\B$ vertices are consequently
symmetric under $\mu\leftrightarrow\nu$,
while only the pure $\A$ vertices in the selections presented in
\hsthree\ and \hsfour, have any cyclic symmetry. 

As mentioned below
\sexp, and explained later, odd-loop contributions carry an insertion
of $\sigma_3$. Since $\A$ and $\C$ commute with $\sigma_3$ but $\B$
anticommutes with $\sigma_3$, odd-loop vertices with $\B$ are 
antisymmetric under cyclic permutations that result in the 
original order of flavours but cycle an odd number of $\B$s.
As an example, we see that the odd-loop contributions to 
$S^{\B\B}_{\,\mu\,\nu\,}(p)$ vanish, because these 
must be antisymmetric under 
$\mu\leftrightarrow\nu$, but no such tensor can be constructed.

\subsec{Exchange Symmetry}
From the comment below \wv, we have 
$$W_{\mu_1\cdots\mu_n,\nu_1\cdots\nu_m}(p_1,\cdots,p_n;q_1,\cdots,q_m;r,s)
=W_{\nu_1\cdots\nu_m,\mu_1\cdots\mu_n}(q_1,\cdots,q_m;p_1,\cdots,p_n;s,r),$$
and similarly for the full vertices,
$$\displaylines{V^{\X_1\cdots\X_n,\Y_1\cdots\Y_m,\ZZ_1\ZZ_2}_{
\,\mu_1\,\cdots\,\mu_n\,,\,\nu_1\,\cdots\,\nu_m\,,\,a_1\,a_2\,}
(p_1,\cdots,p_n;q_1,\cdots,q_m;r,s)=\hfill\cr
\hfill V^{\Y_1\cdots\Y_m,\X_1\cdots\X_n,\ZZ_2\ZZ_1}_{
\,\,\nu_1\,\cdots\,\,\nu_m,\,\mu_1\,\cdots\,\mu_n,\,a_2\,a_1\,}
(q_1,\cdots,q_m;q_1,\cdots,q_n;s,r),}$$
as is clear from \winemom, with again similar identities for the 
$\alpha\beta$ vertices of \wab\ -- \Lvvs.

\subsec{Poincar\'e invariance}
Note that (as usual) all vertices accompany $\delta$ functions over
the sum of their momentum arguments. The vertices are thus meaningful
only when momentum is conserved at the vertex (\cf in particular
sec. 5). 

As in ref. \ymi\ we will use the fact that Lorentz
invariance implies that changing the sign of all momentum arguments in
any vertex, changes the sign of those with an odd number of Lorentz
indices
and has no effect on those with an even number. (Of course
this applies to any even dimension $D$. In odd dimensions we need also
parity, which is a symmetry realised straightforwardly here, to rule
out the appearance of $\varepsilon_{\mu_1\cdots\mu_D}$.)

\subsec{Dimensions}
There is of course a scale invariance corresponding to the na\"\i ve,
or engineering, dimensions. However for a number of reasons
(see secs. 2, 3 and 9), especially
in general dimension $D$, the assignments are a little novel. Of course
$[S]=0$ but we also have
$$[g^2]=4-D,\qquad[\beta_i]=(D-4)i,\qquad
[{\cal L}_i]=(D-4)i+4,\qquad[{\hat{\cal L}}]=4,$$
where the ${\cal L}_i$ and ${\hat{\cal L}}$
are the Lagrangians corresponding to $S_i$ and $\hS$. And we have
$$[\A]=[\B]=1\ins11{and}[\C]=0.$$

\subsec{Gauge invariance}
\midinsert
$$
\epsfxsize=0.63\hsize\epsfbox{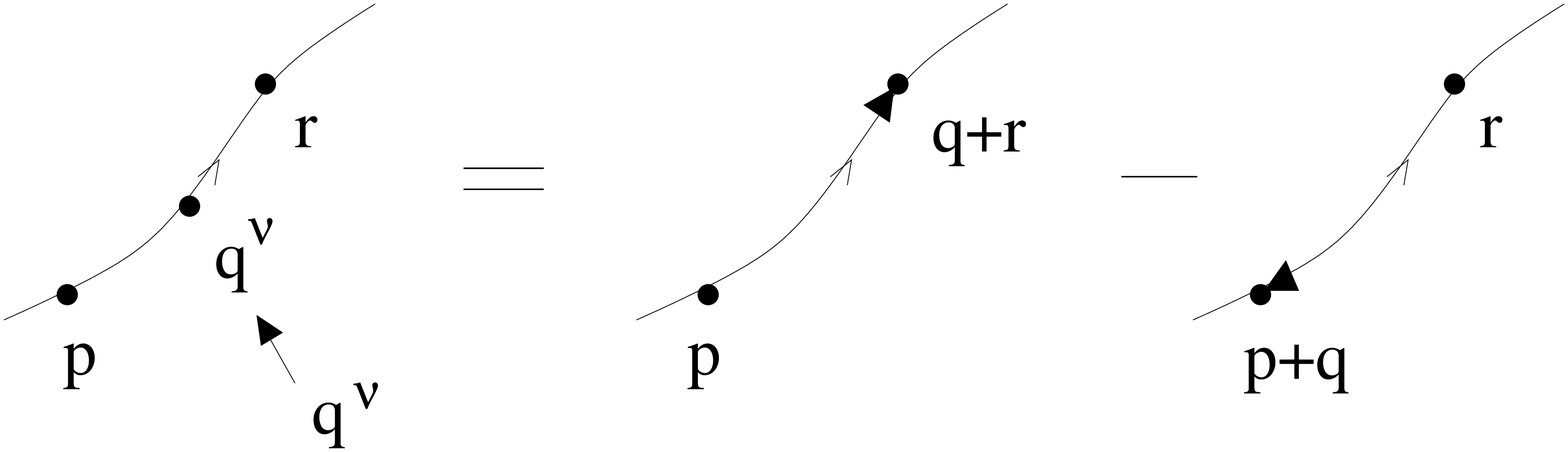}
$$
\centerline{{\bf Fig.9.} 
Graphical representation of gauge invariance identities.} 
\endinsert
These trivial Ward identities follow from the gauge invariance relations
\gatr\  or from Wilson line representations as in \wnpoint, just as they 
did in ref. \ymi. They apply to any pure gauge section (\ie that came 
from pure $\AA$ or pure $A$) and diagrammatically appear as in
\fig\fgatr{Fig. 9. of ymi: Graphical representation of gauge invariance
identities.}.  
Thus,
\eqn\gid{q^\nu U^{\cdots\X\A\Y\cdots}_{\,\cdots\,a\,\nu\,b\,\cdots}
(\cdots,p,q,r,\cdots)=
U^{\cdots\X\Y\cdots}_{\,\cdots\,a\,b\,\cdots}
(\cdots,p,q\!+\!r,\cdots)-
U^{\cdots\X\Y\cdots}_{\,\cdots\,a\,b\,\cdots}
(\cdots,p\!+\!q,r,\cdots)\quad,}
where $U$ is some vertex, and $a$ and $b$ are Lorentz indices or null as 
appropriate. Refs. \ymi\alg\ give as an example the
relations for pure gauge (seed) action vertices.
For wine vertices where the point is at
the end of the line \ymi:\eqnn\wga\
$$\displaylines{p_1^{\mu_1}
W_{\mu_1\cdots\mu_n,\nu_1\cdots\nu_m}(p_1,\cdots,p_n;q_1,\cdots,q_m;r,s)=
\hfill\wga\cr
\hfill W_{\mu_2\cdots\mu_n,\nu_1\cdots\nu_m}(p_1\!+\!p_2,p_3,
\cdots,p_n;q_1,\cdots,q_m;r,s)\quad\cr
\hfill-W_{\mu_2\cdots\mu_n,\nu_1\cdots\nu_m}
(p_2,\cdots,p_n;q_1,\cdots,q_m;r\!+\!p_1,s)\quad,\cr
}$$
with similar identities for contraction with $p_n^{\mu_n}$, $q_1^{\nu_1}$
and $q_m^{\nu_m}$, as is clear from \winemom. In particular, note the
signed momentum that appears for $L_{\mu,\be}$ of \Lvs:
$$p^\mu L_{\mu,\be}(p;q,r)=L_{q,\be}-L_{-r,\be}\quad.$$

\subsec{Charge conjugation invariance}
Recall that the action of charge conjugation on the gauge fields
$A^i_\mu\mapsto -{A^i_\mu}^T$, corresponds to reversal of the sign of the 
underlying Wilson loops \alg\ymi. This means that
the action is invariant under replacing all supertraces of $n$ gauge
fields by the signed reversed order:
\eqn\cstring{\str\A_{\mu_1}\A_{\mu_2}\cdots\A_{\mu_n}\mapsto
(-)^n\str\A_{\mu_n}\cdots\A_{\mu_2}\A_{\mu_1}\quad.}
The `peppering' prescription \alg\ which as we have seen amounts to
replacing $\A$ by $\AA=\A+\B$, means that symmetry \cstring\
must extend to $\B$ fields also. From the form of $\BB$ in \bb, we then
see that $\C$ must also be odd under charge conjugation. 

However, to implement this symmetry on the fields,
we need to include an extra sign for every pair of $B$ or $\Bb$
to compensate for the anticommutation on rearranging 
the order as in \cstring. 
Since by fermion number conservation we know there are as many $\Bb$s as
$B$s (see below), we can incorporate this by an extra sign in the 
definition of the supermatrix transpose; thus we define
\eqn\strans{\X^T:=\pmatrix{{X^{11}}^T &-{X^{21}}^T\cr {X^{12}}^T 
&{X^{22}}^T\cr}\quad.}
(We could instead place the sign on $X^{12}$, or an $i$ [or $-i$] in front
of both fermionic parts.) The action of charge conjugation symmetry 
on the fields is then simply summarised as
\eqn\ccinv{\AA_\mu\mapsto -\AA^T_\mu\quad,\qquad \C\mapsto-\C^T\quad.}
Note that the extension to $SU(N|N)$ thus forces charge conjugation
to be no longer $Z_2$, rather it closes on the $Z_2$ in fermion number 
conservation: $\B\leftrightarrow-\B$, making a $Z_4$ in all.

At the `Wilson loop' level, the symmetry \cstring\ thus holds for all the
fields $\A$, $\B$ and $\C$. Of course charge conjugation \ccinv\ does not 
act on embedded $\sigma_3$s. Instead they maintain their 
relative position in the Wilson loop, \ie flip order together with the
fields in \cstring\ but with $(-)^n$ counting only the fields,
as is clear from \ccinv. This brings us to an important difference
with spontaneously broken $SU(N|N)$: in the present 
formulation the embedded $\sigma_3$s transform differently from $\C$,
and therefore cannot be directly regarded as arising from $<\!\Cu\!>$.
This difference is entirely due to the `missing' $\sigma_3$ in $\hS_B$
discussed below \bb, which would replace the lone $\B$ in $\BB$ by
$i\sigma_3\B$, and thus determine $\C$ to be even under charge conjugation.

\subsec{Reality} 
Note that
of course, we also have the requirement of reality of the Euclidean action
$\hS$ and the flow equation
(equivalent, in a time-reversal invariant theory such as this, to unitarity
of the Minkowksi theory \ref\os{K. Osterwalder
and R. Schrader, Helv. Phys. Acta 46 (1973) 277.}.
Instantons break 
time-reversal invariance and lead to a complex Euclidean $S$, but the
underlying $\hS$ and exact RG equation must still be real. Instanton
contributions will be considered in more detail elsewhere.) For the gauge
fields, the constraints of reality on the vertices follow after 
Hermitian conjugation and the substitution (or identification)
${A^i_\mu}\mapsto {A^i_\mu}^\dagger$. This extends to a change of 
variables on the superfields thus, $\AA_\mu\mapsto\AA_\mu^\dagger$ and 
$\C\mapsto\C^\dagger$. But note that for the same reasons as before, the
transpose part must be defined as in \strans, and thus the superfields are
with this definition only ``pseudo-real'': Hermitian conjugation,
twice-performed, closes on $\B\leftrightarrow-\B$.

Combining reality, charge conjugation and the comments on Poincar\'e
invariance, one readily shows that in momentum space the (wine or action)
vertices associated with an odd number of $\C$ fields
(and any number of $\A$s or $\B$s) are pure imaginary,
whereas those associated with an even number of $\C$ fields are real.
As with the other symmetries outlined in this section, this may be readily
verified on the Feynman rules of  sec. 4.

\subsec{Fermion Number}
In any vertex there are always as many $B$s as $\Bb$s. This has to be so
by the remaining $SU(N)\times SU(N)$ invariance (\cf sec. 2 and ref. \alg),
but it also implies the existence of a fermion number $U(1)$ symmetry:
$B\mapsto B\,\e{i\vartheta}$ and $\Bb\mapsto \Bb\,\e{-i\vartheta}$
(and similarly with $B$ replaced by $D$, when not eaten).
This $U(1)$ is generated by $\sigma_3$:
\eqn\fermin{
\X\mapsto\,\e{i\vartheta\sigma_3/2}\,\X\,\e{-i\vartheta\sigma_3/2}\quad,}
where $\X$ runs over the fields, and thus
extends the global $SU(N|N)$ to $U(N|N)$, appearing in this sense in
the usual way. Note however there is of course
also an essential difference compared to the usual (\ie bosonic) groups,
in that this extra $U(1)$ acts non-trivially in the adjoint representation.
(Also note that the 
fermion number $Z_2$ invariance mentioned above and whose explicit
representation appears below \proj, is just the subgroup generated by
$\vartheta=\pi$.)

\subsec{Duality}
While the above symmetries are interesting to explore in their own right
and as we will see, are required in practical calculation, by far the
most intriguing symmetry we uncovered is a $Z_2$ duality symmetry
which exchanges the r\^ole of the two groups in $SU(N)\times SU(N)$
and at the same time, in a sense that we make explicit below,
changes the sign of the squared coupling constant. At the level of
the flow equation \pRG, it is implemented by
\eqn\dua{g^2\mapsto-g^2\quad,\qquad
\AA\mapsto\AA^e\ins11{and}\C\mapsto\C^e\quad,}
where $\X^e:=\sigma_1\X\,\sigma_1$ and thus for example
\eqn\duaex{\AA^e=\pmatrix{A^2 &\Bb\cr B &A^1\cr}\quad.}
($\sigma_1$ was introduced in sec. 2.)
This transformation is of course not part of $U(N|N)$, not the
least because $\sigma_1$ has bosonic off-diagonal elements.
Note also that $\left(\sigma_3\right)^e=-\sigma_3$. 
Thus from the identity $\str\X=\tr\sigma_3\X$ [or explicitly from
\duaex]
the supertrace of a string of fields is antisymmetric under \dua.
Since from \pseed, $\hS$ is such a single supertrace of superfields, 
it changes sign.
Similarly the RHS of the flow equation \pRG\ picks up a sign via
the single supertrace in \pwv.\foot{Note that the change of variables 
implied
by \duaex, and similarly $C^1\leftrightarrow C^2$, means that the functional
derivatives \pfuncd\ are minus their duals.}
Changing also the sign of
$g^2$ in $\Sigma_g$ of \pRG, we see that the exact RG is 
indeed invariant. It follows that if $S[\A,\B,\C](g^2)$ is a solution,
then so is $S[\A^e,\B^e,\C^e](-g^2)$. 

If 
we imagine $g$ in \pRG\ to be a fixed (\ie independent 
of $\Lambda$) expansion parameter, for example the classical or
bare coupling, $g=g_0$ at $\Lambda=\Lambda_0$, then it is easy to see
that we can take $S$ to be self-dual:
\eqn\sdua{S[\A,\B,\C](g_0^2)=S[\A^e,\B^e,\C^e](-g_0^2)\quad.}
Indeed this follows immediately if as will be the case,
the `initial' condition, $S$ at $\Lambda=\Lambda_0$, is taken to be 
self-dual. 
As a corollary we find
from \Sloope, that all even (odd) order
loop corrections $S_n$ are even (odd) under duality.
In the large $N$ limit, since $S$ is a single supertrace, an
even (odd) loop $S_n$ must thus contain an even (odd)
number of embedded $\sigma_3$s, and this fact can
readily be confirmed explicitly by considering \fsig.

The self-duality at the very least is obscured when we come to
renormalize however. We see immediately that the $\beta$ function
\betafn\ cannot be invariant
under \dua\ unless all the odd-loop $\beta_{2n+1}$ vanish, which is not the 
case. From the expansion \ergone,
\ergtwo, \etc, the non-zero $2\beta_{2n+1}S_0$ terms
mix together terms with even and odd numbers of embedded $\sigma_3$s,
so that the corollary above no longer holds. The underlying reason for
these complications is as follows.
By the above analysis, at one loop (using \defg\ and solving $\beta$)
\eqn\bareone{S[\A,0,0]
={1\over2}\,\str\!\int\!\!d^D\!x\,\left({1\over g_0^2}+2\beta_1\sigma_3
\ln{\Lambda_0\over\Lambda}\right) \F_{\mu\nu}^2 
+O(\partial^3)\quad.}
Defining $1/g^2=1/g^2_0+2\beta_1\ln(\Lambda_0/\Lambda)$ only absorbs
the divergence for the $A^1$ part of the action.
We see that in the continuum limit we are forced to 
introduce two renormalised couplings: $g=g_1$ for $A^1$, and
$g_2$ for $A^2$. Under duality we map to a solution of \pRG, for which
the renormalization condition
\defg\ now insists $g=g_2$. Let us call the couplings for such a solution
${\tilde g}_2={\tilde g}$ and ${\tilde g}_1$. Then by the
duality of \pRG\ and the initial bare action,
\eqn\defgii{\eqalign{
{\tilde g}_2^2(g_0^2,\Lambda_0/\Lambda) &=-g^2(-g_0^2,\Lambda_0/\Lambda)\cr
g_2^2(g_0^2,\Lambda_0/\Lambda) &=-{\tilde g}^2_1(-g_0^2,\Lambda_0/\Lambda)
\quad.}}
We comment further in the conclusions.

\newsec{Classical vertices without gauge fixing}

As we will see, even here there are surprises.
{\sl Classical} solutions will turn out to suffer 
a form of divergence, arising from integration over $\Lambda$,
which is regularised 
by careful choice of the $\Lambda=\Lambda_0\to\infty$ `boundary conditions', 
and will require the introduction of some new `renormalised' 
parameters \alg. 
This has nothing to do with unbounded momentum integrals, and nothing 
to do with gauge invariance {\it per se}: it arises in the Pauli-Villars
sector from the existence of positive powers of the cutoff and some
freedom to add extra interactions. However, the implementation
of a Pauli-Villars exact RG scheme is in itself one of the novel 
developments we report in this paper.

The dictionary for translating the Feynman diagrams, which themselves
follow from expanding (the relevant parts of) \fflow, has already been
given in sec. 4. In fact the expanded Wilson loops look identical
to those in refs. \ymi\alg, a consequence of the equality
of form with that of the pure gauge case
of both \fflow\ and the perturbative development in sec. 3.

As already noted in sec. 2, the tree-level insertions of $\sigma_3$
in \bpsow\ serve only to ensure that $\Y$ and $\X$, 
the remainders of the Feynman diagram on either side, are bosonic or
fermionic as appropriate.\foot{more strictly block diagonal or 
block off-diagonal as appropriate}
But these restrictions are automatically incorporated in 
the explicit Feynman rules, \cf sec. 4. Constructing the tree-level 
vertices out of them, expanding using \fsig\ and
(anti)commuting the $\sigma_3$s together, they all combine to
no overall effect. Thus we omit this step in this section.

Solving the flow equations for the vertices introduces integration
constants, \ie terms independent of $\Lambda$. Of course these must 
be chosen to satisfy all the symmetries of the theory, and we will
signal which symmetries provide non-trivial constraints. Moreover
at the classical level, since $g$ does not run, there is no difficulty
in preserving the self-duality \sdua, which follows here because
the integration constants will be chosen to be single supertraces without
embedded $\sigma_3$s. Thus in this formulation, there are no classical 
vertices with an embedded $\sigma_3$:
\eqn\scsiz{S_{\ph{0\sigma}\,a_1\,\cdots \,a_n\,}^{0\sigma\X_1\cdots \X_n}
=0\quad.}
The reader can find in sec. 6, the list of symmetries together with
relevant comments and definitions.

It is helpful also, to borrow the conclusions on
`drifting' from sec. 9, in particular those
summarised in Lemmas 3 and 4 and Corollary 4. Thus we already know 
that pure-$\A$ classical vertices
are constructed only out of $\hS_A$ vertices, such as in \hSex, and $c'$
vertices \ie the non-$L$ part of 
$c'_{\mu_1\cdots\mu_n,\nu_1\cdots\nu_m,\alpha\beta}$ [\cf \wab]. 
In other words, the pure-$\A$ $S_0$ vertices 
are unchanged from ref. \ymi. Similarly to \hsexp, let us reserve the
unlabelled $S^0_{\mu_1\cdots\mu_n}$ vertices for these, which thus have
the same explicit expressions as in refs. \ymi\alg.
We also know there are no $S_0$ vertices with just one $\C$.

Contributions that seem at variance with the above conclusions,
vanish as a consequence
of drifting, which itself is a consequence of the exact preservation
of gauge invariance. Although these statements are readily verified,
in the interests of compactness we omit the explicit computations.
\vfill\eject

\subsec{Two point vertices}
From \ergcl,  we thus have \ymi\alg:
$$
\Lambda{\partial\over\partial\Lambda}\,
\mathop{\vcenter{\epsfxsize=0.085\hsize\epsfbox{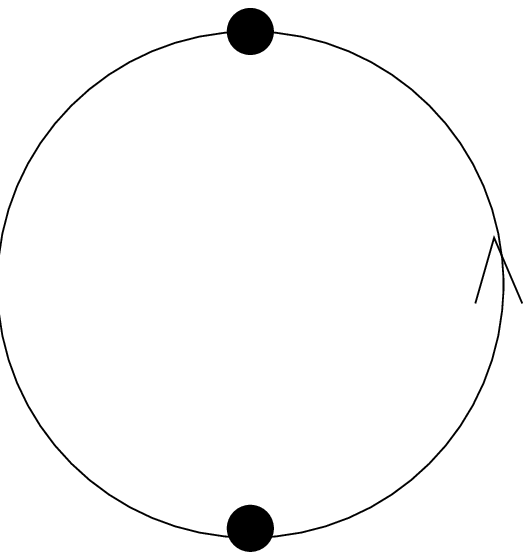}}^{\phantom{p}}
}^{\displaystyle
-p^\nu}_{\displaystyle p^\mu}=\ 2\!\!\!\!\!\!
\mathop{\vcenter{\epsfxsize=0.13\hsize\epsfbox{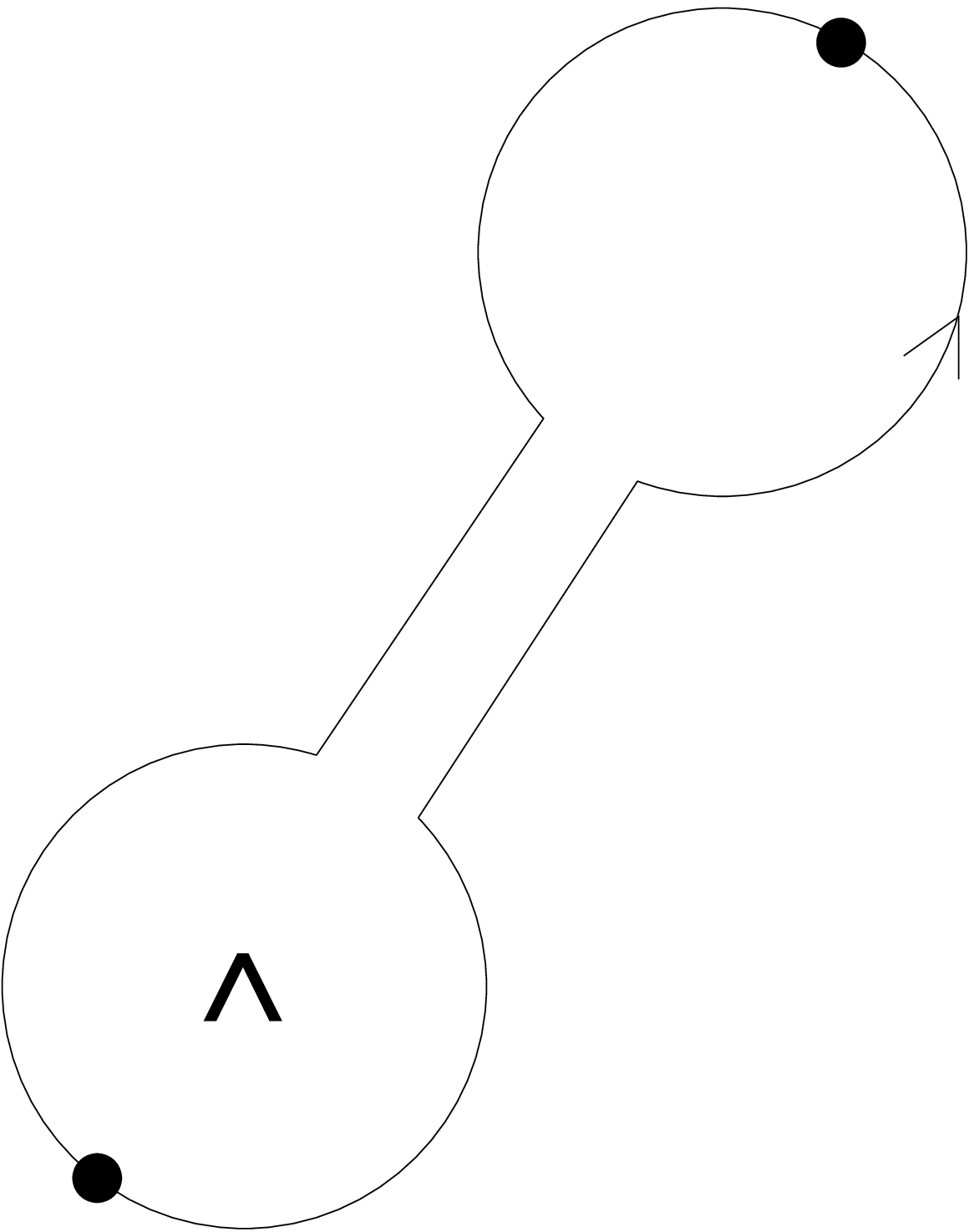}}}_{
\!\!\!\!\!\!\!\!\!\!\!\!\!\!\!\!\displaystyle p^\mu}
^{\displaystyle\;\;\;\;\;\;\;\;\;\;\;\;\;\;\;\;\;\;\;\;-p^\nu}
-\!\!\!\!\!\!\mathop{\vcenter{\epsfxsize=0.13\hsize\epsfbox{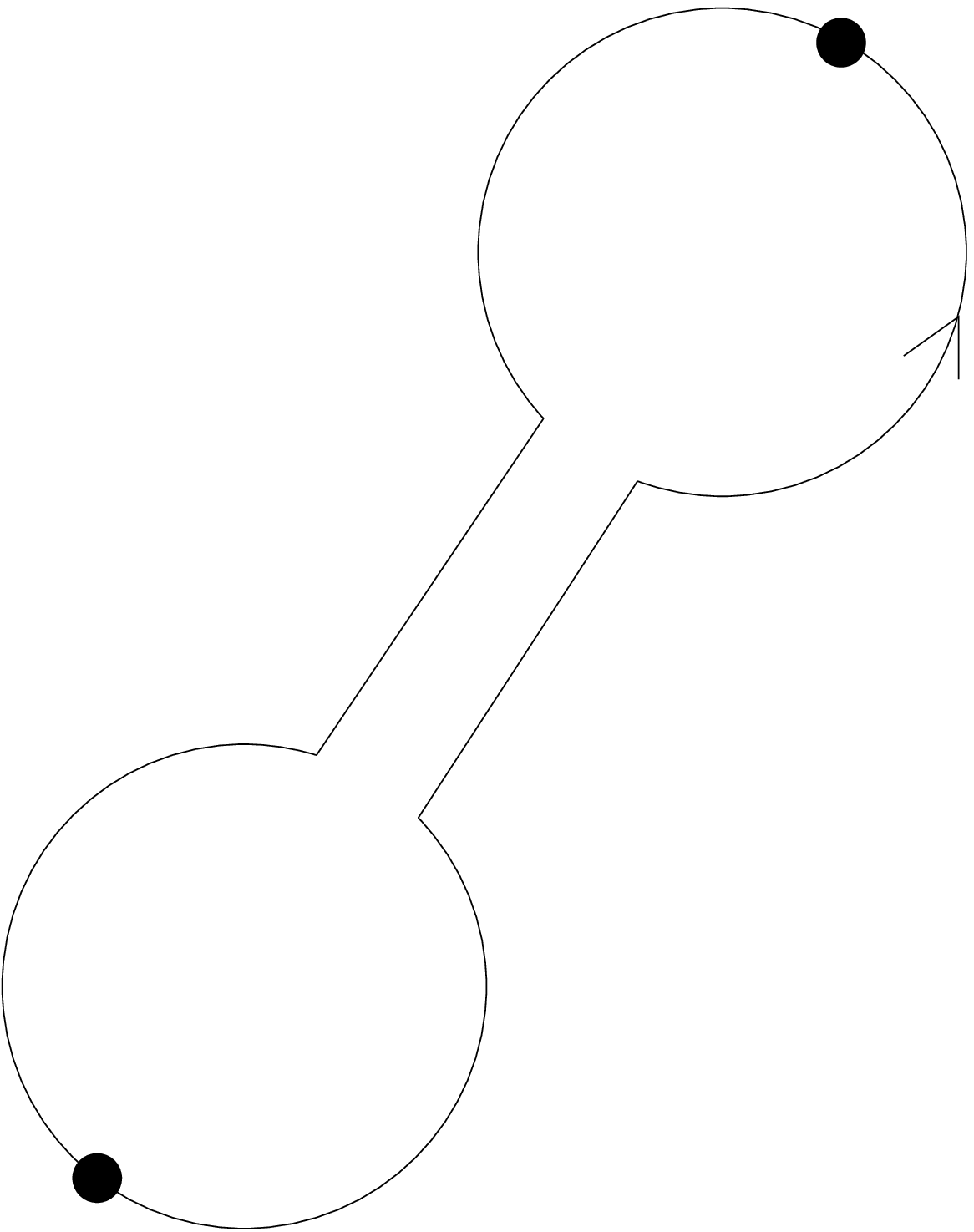}}}_{
\!\!\!\!\!\!\!\!\!\!\!\!\!\!\!\!\displaystyle p^\mu}
^{\displaystyle\;\;\;\;\;\;\;\;\;\;\;\;\;\;\;\;\;\;\;\;-p^\nu}
+(p^\mu\leftrightarrow -p^\nu)$$
\centerline{ {\bf Fig.10.} Feynman diagrams for the two-point vertex.}
\nfig\ftwo{Fig.10 of \ymi: classical two-point flow}
As in refs \alg\ymi, we now adopt the convention that
the empty circle corresponds to $S_0$, not $S$ as in \fflow,
and we have noted that once again since actions' one-point 
vertices vanish (for example by charge conjugation invariance)
we must have at least one blob per lobe. Here however,
we must also assign flavours $\A$, $\B$ or $\C$ to the two points.

From \hstwo\ and \vzero, at first sight we appear to
generate a mixed $\C$-$\A$ vertex. Actually, the required 
$\C$-$\A$ zero-point wine vertex anihilates the $\A$-$\A$
lobe by gauge invariance: $p^\mu\hS_{\mu\nu}(p)=0$. 
This is nothing but a perturbative verification
of the consequences of `drifting' as
discussed above. It is consistent to set to zero 
all $S_0$ vertices for which the RHS of the flow equation vanishes,
as here, and we will do so in this paper.
(In fact in this case, it may be verified that the requirements of
gauge invariance and Lorentz invariance already disallow a
$\C$-$\A$ vertex.)

We are left with:\eqna\dtwop
$$\eqalignno{\Lambda{\partial\over\partial\Lambda}
S^0_{\mu\nu}(p) &={1\over2\Lambda^2}c'_p
\left[2\hS_{\mu\al}(p)-S^0_{\mu\al}(p)\right]
S^0_{\al\nu}(p)+(p_\mu\leftrightarrow -p_\nu), &\dtwop A\cr
\Lambda{\partial\over\partial\Lambda}
S^{0\B\B}_{\ph{0}\,\mu\,\nu}(p) &={1\over2\Lambda^2}
\left[2\hS^{\B\B}_{\,\mu\,\al}(p)-S^{0\B\B}_{\ph{0}\,\mu\,\al}(p)\right]
K_{p,\al\be}\,S^{0\B\B}_{\ph{0}\,\be\,\nu}(p)
+(p_\mu\leftrightarrow -p_\nu), &\dtwop B\cr
\Lambda{\partial\over\partial\Lambda}
S^{0\C\C}(p) &={1\over\Lambda^4}M_p
\left[2\hS^{\C\C}(p)-S^{0\C\C}(p)\right]
S^{0\C\C}(p)\quad, &\dtwop C\cr
}$$
where we have used \vzero, and on \dtwop A,
the drifting simplifications mentioned above.

For completeness we recall how \dtwop A\ is solved \ymi.
By gauge invariance and dimensions,
\eqn\partwo{S^0_{\mu\nu}(p)=2\Delta_{\mu\nu}(p)/f(p^2/\Lambda^2)\quad.}
From \Sloope, we require $f(0)=1$ so
as to be consistent with \defg\ in the $g\to0$ limit. 
Substituting \hSex, one readily finds the unique solution to be $f=c$.

Similarly,
substituting \hstwo, \Lvs, \wab\ and \KLM, one readily verifies that
the two-point classical and seed vertices for $\B$ and $\C$ 
may also be taken to be equal. Thus in total:
\eqn\twop{S^0_{\mu\nu}(p)=\hS_{\mu\nu}(p)\quad,\qquad
S^{0\B\B}_{\ph{0}\,\mu\,\nu}(p)=\hS^{\B\B}_{\,\mu\,\nu}(p)\quad,\qquad 
S^{0\C\C}(p)=\hS^{\C\C}(p)\quad.}
In the case of $\B$ and $\C$ however, these are not the most general 
solutions, presumably reflecting the freedom of reparametrization 
invariance in the non-gauge sector 
\YKIS\ref\reparam{J. Comellas, Nucl. Phys. B509 (1998) 662.}, 
but we will specialize to these equalities since they
simplify the higher-point vertices as in ref. \ymi. In fact the expressions
for $K$, $L$ and $M$ in \KLM\ were determined to make these equalities 
possible. (See also sec. 9.)

\subsec{Three-point vertices}
Similarly, from \ergcl\ and the top two lines of \fflow, we obtain the
following diagrams for the three-point vertex:
$$\Lambda{\partial\over\partial\Lambda}\
\vcenter{\epsfxsize=0.085\hsize\epsfbox{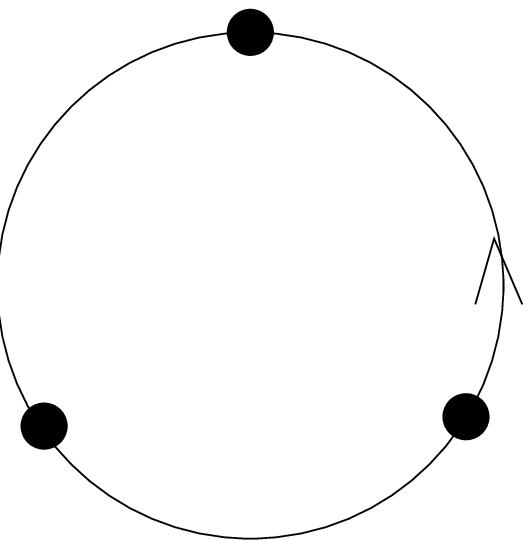}}\,
=\,\, 2\vcenter{\epsfxsize=0.13\hsize\epsfbox{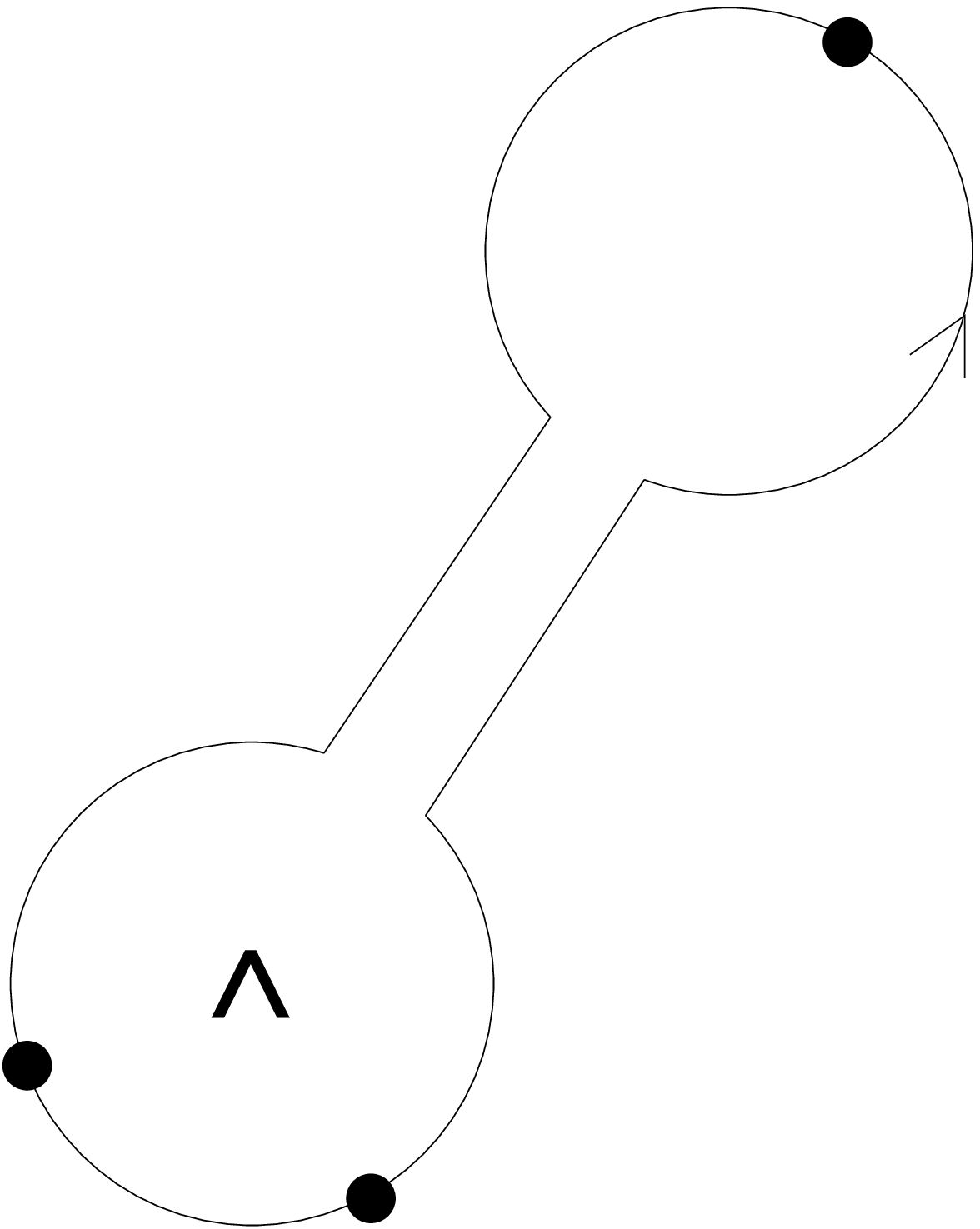}}
+2\vcenter{\epsfxsize=0.13\hsize\epsfbox{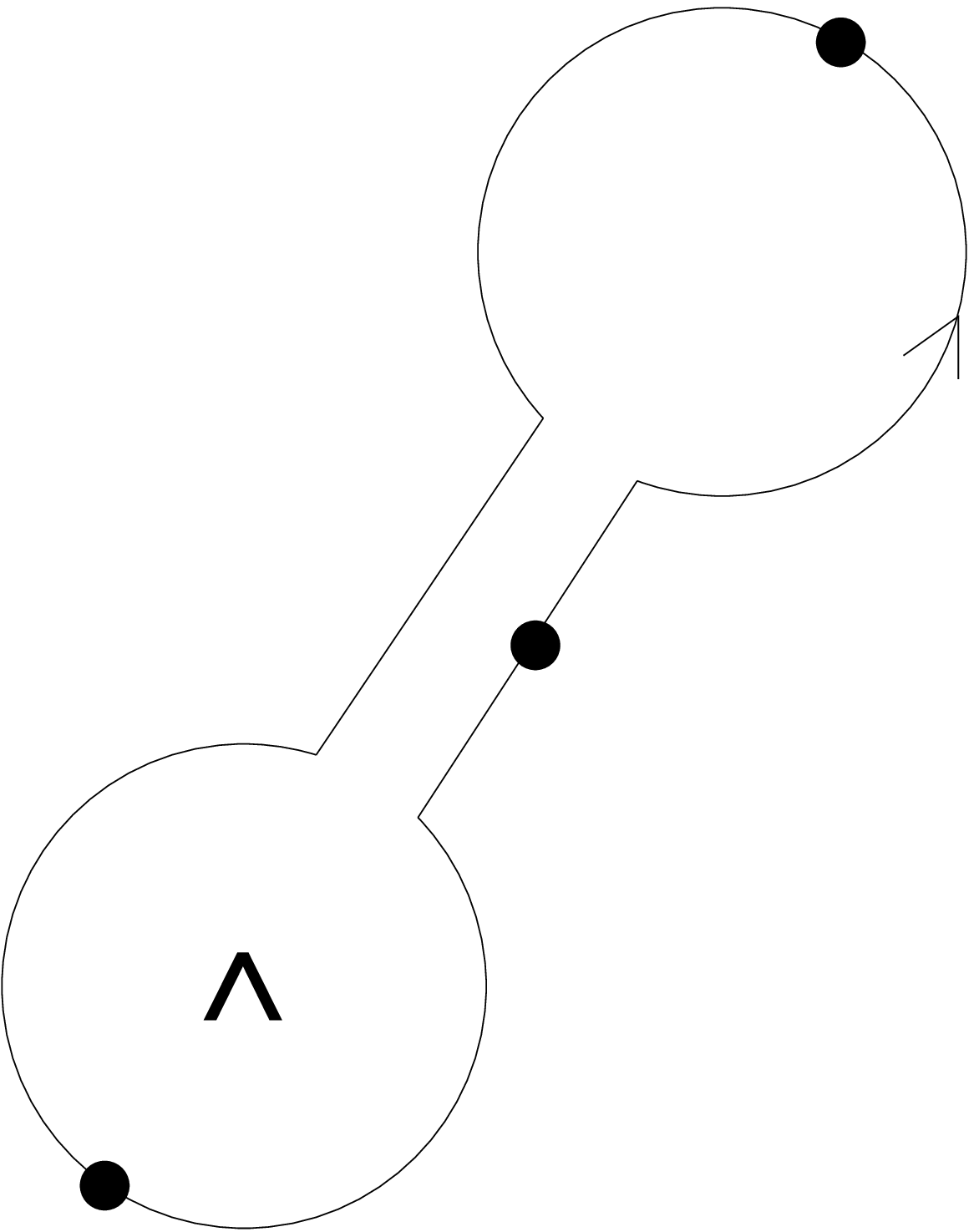}}$$
\centerline{ 
{\bf Fig.11.} Feynman diagrams for the three-point vertex.}
\nfig\fthree{Fig11 ymi three-point fl}
Here we have already simplified with \twop{}, which works precisely
in the same way as in ref. \ymi. We are thus again \ymi\ left on the RHS
with terms that are already determined, allowing the differential 
equation to be integrated immediately. Again this simplification,
provided by the
equalities in \twop{}, persists to all higher point $S_0$ vertices. The
proof is identical to that in ref. \ymi.

Up to cyclicity,
the possible three-point flavours are $\A\A\A$, $\B\B\A$, $\B\B\C$,
$\C\C\A$, and $\C\C\C$. Note that odd numbers of $\B$s are ruled out
by fermion number (the supertrace of such an odd number
vanishes identically) and the classical $\C\A\A$ vertex is ruled out
by drifting (\cf discussion above).

For the pure $\A$ vertex, all the $L$ terms vanish by drifting (again as
outlined above) and we thus obtain the same flow equation as in ref. \ymi,
with the same result:
\eqn\threep{\eqalign{
S^0_{\mu\nu\lambda}(\p,\q,\r)= &-\int_\Lambda^\infty\!\!{d\Lambda_1\over
\Lambda_1^3}\left\{c'_r\hS_{\mu\nu\alpha}(\p,\q,\r)\hS_{\alpha\lambda}(r)
+c'_\nu(\q;\p,\r)\hS_{\mu\alpha}(p)\hS_{\alpha\lambda}(r)\right\}\cr
&+2(r_\nu\delta_{\mu\lambda}-r_\mu\delta_{\nu\lambda})\qquad\qquad\qquad
+{\rm cycles}\quad.}}
As in refs. \ymi\alg, we adopt the convention that the terms in a
particular $\Lambda$-integral are understood to
have $\Lambda$ replaced by the integration variable (here $\Lambda_1$).
The term ``cycles'' means that we add to the 
expression the two cyclic
permutations of $(p_\mu,q_\nu,r_\lambda)$. Recall from ref. \ymi\ (see also
ref. \alg), that the continuum
limit, corresponding to the upper limit $\Lambda_1=\infty$, trivially
exists, and that the integration constant is fixed by gauge invariance
to be the unique covariantization of $\Delta_{\mu\nu}$, \ie the usual
`bare' three-point vertex,
as is also clear from the $\Lambda\to\infty$ limit of \threep, 
since from \defg, \sdua, and by dimensions
\eqn\defcl{S_0[\A,0,0]={1\over2}\,\str\!\int\!\!d^D\!x\, \F_{\mu\nu}^2
+O(\partial^3/\Lambda)}
(or simply by restriction to $A^1$).

In a similar way we obtain for the $\B\B\A$ vertex:
\eqn\pBBA{\eqalign{
&S^{0\B\B\A}_{\ph{0}\,\mu\,\nu\,\la}(p,q,r)=
-\int_\Lambda^\infty\!\!{d\Lambda_1\over\Lambda_1^3}\Big\{
\hS^{\B\B\A}_{\,\mu\,\nu\,\al}(p,q,r)c'_r\hS_{\al\la}(r)
+\hS^{\B\B\A}_{\,\mu\,\al\,\la}(p,q,r)K_{q,\al\be}
\hS^{\B\B}_{\,\be\,\nu}(q)\cr
&+\hS^{\B\B\A}_{\,\al\,\nu\,\la}(p,q,r)K_{p,\al\be}
\hS^{\B\B}_{\,\be\,\mu}(p)
+\hS^{\B\B}_{\,\mu\,\al}(p)c'_{\nu,\al\be}(q;p,r)\hS_{\be\la}(r)
+\hS_{\la\al}(r)c'_{\mu,\al\be}(p;r,q)\hS^{\B\B}_{\,\be\,\nu}(q)\cr
&+\hS^{\B\B}_{\,\nu\,\al}(q)K_{\la,\al\be}(r;q,p)\hS^{\B\B}_{\,\be\,\mu}(p)
\Big\}\cr
&+2\Big\{
(r_\nu\delta_{\mu\la}-r_\mu\delta_{\nu\la})(1+\gamma^{BBA}/2)
+p_\la\delta_{\nu\mu}-p_\nu\delta_{\la\mu}
+q_\mu\delta_{\la\nu}-q_\la\delta_{\mu\nu}
+{\ct'_0\over\ct_0^2}\delta_{\mu\nu}(q-p)_\la\Big\}
}}
where $\gamma^{BBA}$ is a dimensionless real free parameter, and 
the integral needs interpreting with some care -- as we explain
below. The first term has been simplified, once again by drifting -- 
\ie gauge invariance [as above \dtwop{}].
In fact in this way, expanding the wine vertices by \wab, all the terms
in \pBBA\ containing $L$, either vanish or considerably simplify. We omit
the details.

To justify \pBBA, we replace the top limit in the $\Lambda_1$ 
integral by $\Lambda_0$, in which case clearly the integration constant
is identified with the `bare' value of 
$S^{0\B\B\A}_{\ph{0}\,\mu\,\nu\,\la}(p,q,r)$, \ie its value
at $\Lambda=\Lambda_0$.
Unlike the pure-$\A$ case \viz \defg, 
we do not have a renormalization condition to fix this.
(Actually, after identifying all relevant and marginal directions,
these could be introduced of course, but as part of the regulating
structure they are not 
needed -- nor do they simplify the calculations  -- 
at least in this paper.) However by quasilocality \ymi\ and dimensions,
since we may discard all terms that
vanish in the limit $\Lambda_0\to\infty$, 
finding the most general 
integration constant reduces to looking for the usual bare-action type 
terms\foot{\ie  polynomial in momenta, balanced as required 
by non-negative powers of $\Lambda_0$}
consistent with the symmetries of the theory. The exact preservation 
of $SU(N)\times SU(N)$
gauge invariance makes this process simple and elegant.

Thus the most general
integration constant is the $\B\B\A$ part of
\eqn\dBBA{{\cal L}_0|_{\Lambda=\Lambda_0} =\,\str\left\{
{1\over2}\FF_{\mu\nu}^2+{\ct'_0\over\ct_0^2}\B_\mu\nabla^2\B_\mu+
i\gamma^{BBA}\B_\mu\F_{\mu\nu}\B_\nu+\cdots\right\},}
where the first two terms are fixed by gauge invariance from \twop.\foot{
\viz gauge covariantizing \hstwo, or from \pseed, or directly by expanding
\hsthree; here and later the solution may also be readily derived
directly in momentum space, using \gid.}
The ellipses refer to gauge invariant terms that do not
contain the $\B\B\A$ vertex  (and note by \pseed, include some that
diverge in the $\Lambda\to\infty$ limit).
Note that the last term has the right reality and charge conjugation 
properties.

As we will verify, the universality of the continuum limit means that
physical quantities will be independent of $\gamma^{BBA}$. 
But we cannot quite just set it to zero: by dimensions we see that 
if the integral in \pBBA\ yields such a term, then it is
logarithmically divergent (\viz $\sim \int^{\Lambda_0}_\Lambda 
d\Lambda_1/\Lambda_1$). However, also note that
since \dBBA\ contains the most general possible 
such $\Lambda$-independent terms, 
the $\Lambda_1$ integral can {\sl only} diverge this way. A
straightforward calculation confirms that the integral in \pBBA\ 
indeed diverges, as $-16\ln(\Lambda_0/\Lambda)\left(
r_\nu\delta_{\mu\la}-r_\mu\delta_{\nu\la}\right)$, and thus
for a finite continuum limit we set the constant to
\eqn\gaBBA{\gamma^{BBA}=16\ln(\Lambda_0/\mu)+{\rm finite}\quad,}
where $\mu$ is a finite mass scale, and
$\Lambda_0\to\infty$. 

At a more sophisticated level, we may simply impose a definite prescription
for discarding the infinities in  
$\Lambda$ integrals arising in finite continuum solutions such as \pBBA,
for example minimal subtraction of $\Lambda_0$ divergences,
safe in the knowledge that in reality these divergences are actually
cancelled by opposite divergences in parameters in the most general
integration constant. We could go further with this prescription,
and discard these parameters,
since this just amounts to choosing them to be precisely the opposing 
divergences. We will keep them however, as an extra test of universality,
but for simplicity report from now on this more sophisticated approach.
(Actually, we also checked the calculations the dumb way, as in \gaBBA\
and above. We omit the details.)

In this way, the $\B\B\C$ vertex is found to be
\eqn\pBBC{\eqalign{&S^{0\B\B\C}_{\ph{0}\,\mu\,\nu}(p,q,r)=
-\int_\Lambda^\infty\!\!{d\Lambda_1\over\Lambda_1^3}\Big\{
{1\over\Lambda_1^2}\hS^{\B\B\C}_{\,\mu\,\nu}(p,q,r)M_r\hS^{\C\C}(r)
+\hS^{\B\B\A}_{\,\mu\,\nu\,\al}(p,q,r)L_{r,\al}\hS^{\C\C}(r)\cr
&+\hS^{\B\B\C}_{\,\mu\,\al}(p,q,r)K_{q,\al\be}\hS^{\B\B}_{\be\nu}(q)
+\hS^{\B\B\C}_{\,\al\,\nu}(p,q,r)K_{p,\al\be}\hS^{\B\B}_{\be\mu}(p)
-\hS^{\B\B}_{\mu\al}(p)L_{\nu,\al}(q;r,p)\hS^{\C\C}(r)\cr
&+\hS^{\C\C}(r)L_{\mu,\be}(p;r,q)\hS^{\B\B}_{\be\nu}(q)\Big\}
+ i\gamma^{BBC}_1\delta_{\mu\nu}(p^2-q^2)+i\gamma^{BBC}_2(p_\mu p_\nu
-q_\mu q_\nu),
}}
where the two dimensionless real $\gamma^{BBC}_i$ parametrise the
most general integration constant. Once again, the $L$ parts 
considerably simplify on using the gauge invariance relations \gid.
Note that charge
conjugation invariance requires the integration constants to be odd under 
$p_\mu\leftrightarrow q_\nu$. Dimensions suggest and explicit calculation
confirms that the $\gamma^{BBC}_i$ mop up logarithmic divergences
in the $\Lambda_1$ integral.

The $\C\C\A$ vertex is found to be:
\eqn\pCCA{\eqalign{S^{0\C\C\A}_{\ph{0\C\C}\,\la}(p,q,r) &=
-\int_\Lambda^\infty\!\!{d\Lambda_1\over\Lambda_1^5}\Big\{
\hS^{\C\C\A}_{\ph{\C\C}\,\la}(p,q,r)\left[M_pS^{\C\C}(p)+M_qS^{\C\C}(q)
\right]\cr
&+\Lambda_1^2\hS^{\C\C\A}_{\ph{\C\C}\,\al}(p,q,r)c'_r\hS_{\al\la}(r)
+\hS^{\C\C}(q)M_\la(r;q,p)\hS^{\C\C}(p)\Big\}\cr
&+2{\ct'_0\over\ct_0^2}(p^2+q^2)(q-p)_\la+\gamma^{CCA}(r^2p_\la-r_\la r.p),
}}
where $\gamma^{CCA}$ is another dimensionless real free parameter.
Here the integration constant is constrained by gauge invariance, 
\twop\ and \pseed\ to be the $\C\C\A$ vertex in
\eqn\dCCA{{\cal L}_0|_{\Lambda=\Lambda_0} =\,\str\nabla_\mu\cdot\C\left(
\Lambda_0^2\delta_{\mu\nu}+{\ct'_0\over\ct_0^2}\delta_{\mu\nu}\nabla^2+
i\gamma^{CCA}\F_{\mu\nu}\right)\nabla_\nu\cdot\C+\cdots,}
where the ellipses do not contain $\C\C\A$ vertices.
The first term precisely cancels an equal and opposite quadratic 
divergence in the $\Lambda_1$ integral, as it must, and thus 
by the prescription below \gaBBA, is not displayed in \pCCA. The last term,
which verifies the reality and charge conjugation symmetries,
mops up a logarithmic divergence in the $\Lambda_1$
integral.

Finally, the three-point classical $\C$ vertex is:
\eqn\pCCC{S^{0\C\C\C}(p,q,r)=
i\int_\Lambda^\infty\!\!{d\Lambda_1\over\Lambda_1^5}\left\{
\left[\hS^{\C\C}(p)-\hS^{\C\C}(q)\right]L_r\hS^{\C\C}(r)\ +{\rm cycles}
\right\},}
where `cycles' stands for the two cyclic permutations of $(p,q,r)$.
Here we have `drifted'
$\hS^{\C\C\A}_{\ph{\C\C}\,\la}L_{r,\la}$ terms by using \Lvs\ and \gid.
Remarkably, the integration constant must vanish. Charge conjugation and 
cyclic symmetry require that it be antisymmetric under exchange of any pair
of momenta. In this case, any such expression
can be generated by writing down Lorentz invariant terms antisymmetric under 
$p\leftrightarrow q$ and then adding the cyclic iterands.
Using momentum conservation, it is then straightforward
to show that all such polynomials up to dimension 4 vanish on adding the
cycles. Similarly one checks that the superficially
quartically divergent integral in \pCCC\ is actually
finite, as consistency requires.

\subsec{Four-point vertices}
As already discussed, the diagrams are the same as in ref. \ymi:
$$\eqalign{
\Lambda{\partial\over\partial\Lambda}\
\vcenter{\epsfxsize=0.085\hsize\epsfbox{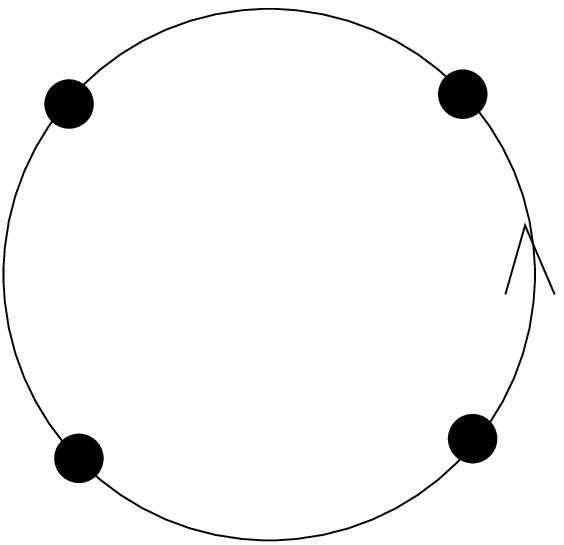}}\,
=\, -\vcenter{\epsfxsize=0.13\hsize\epsfbox{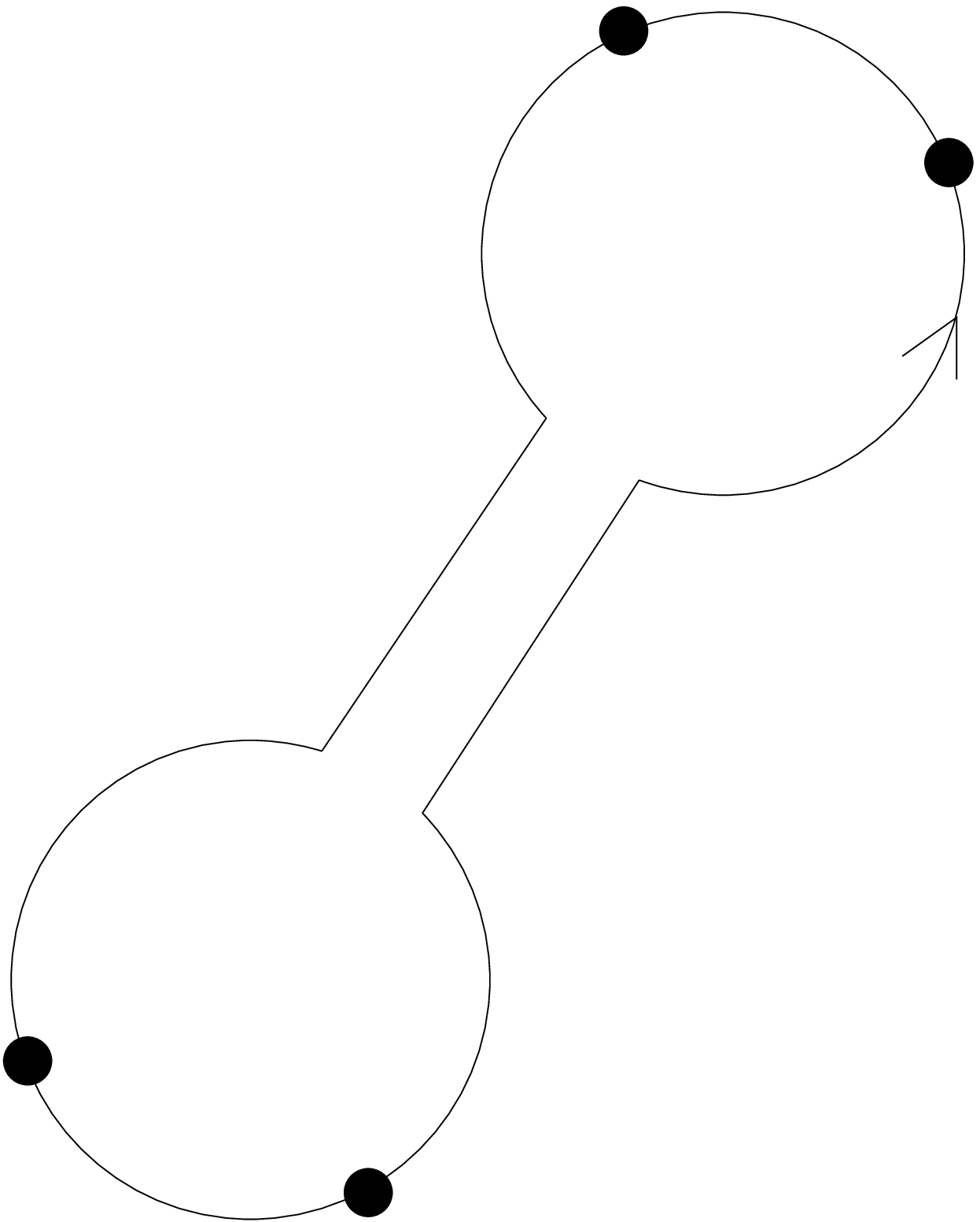}}
&+2\vcenter{\epsfxsize=0.13\hsize\epsfbox{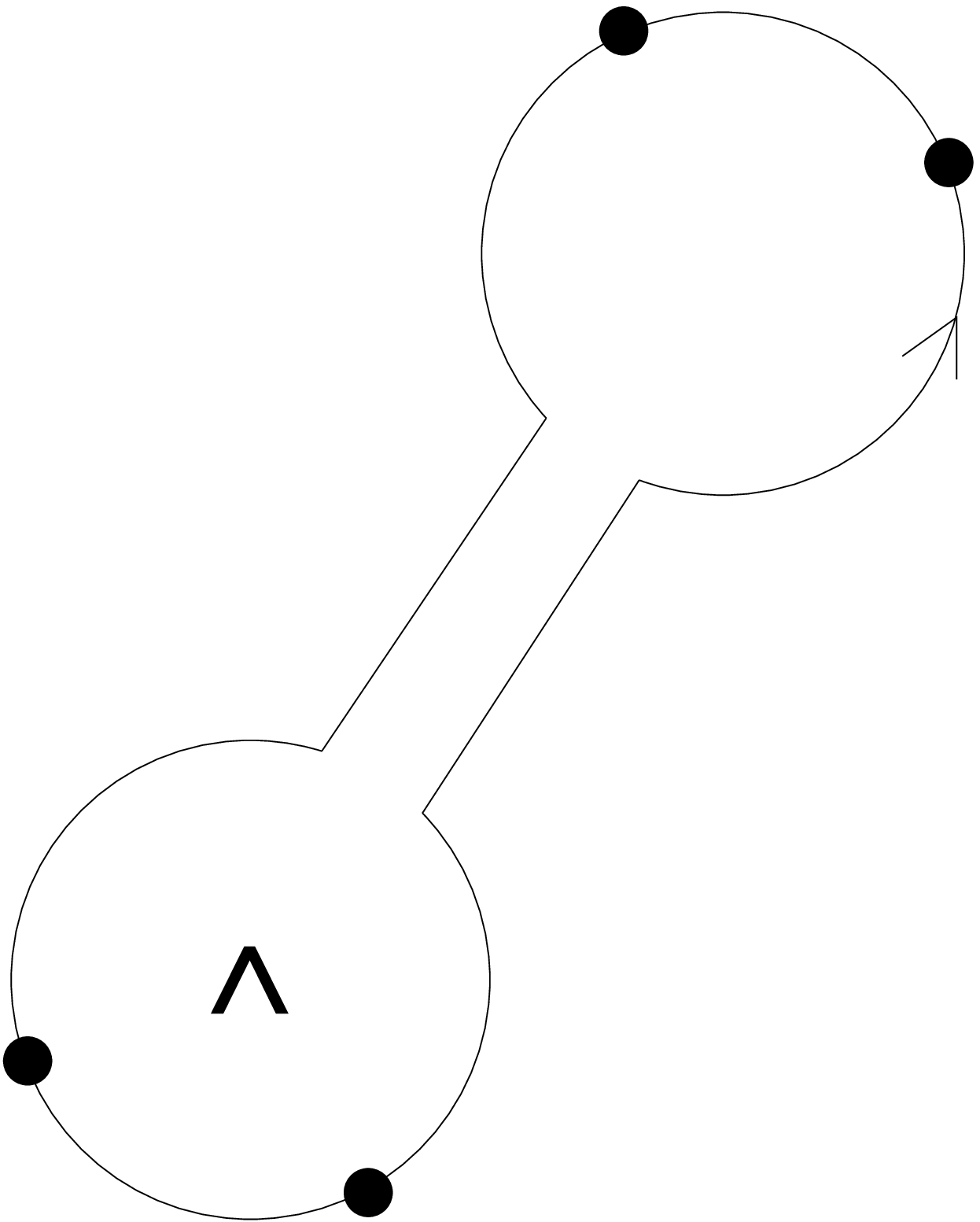}}
+2\vcenter{\epsfxsize=0.13\hsize\epsfbox{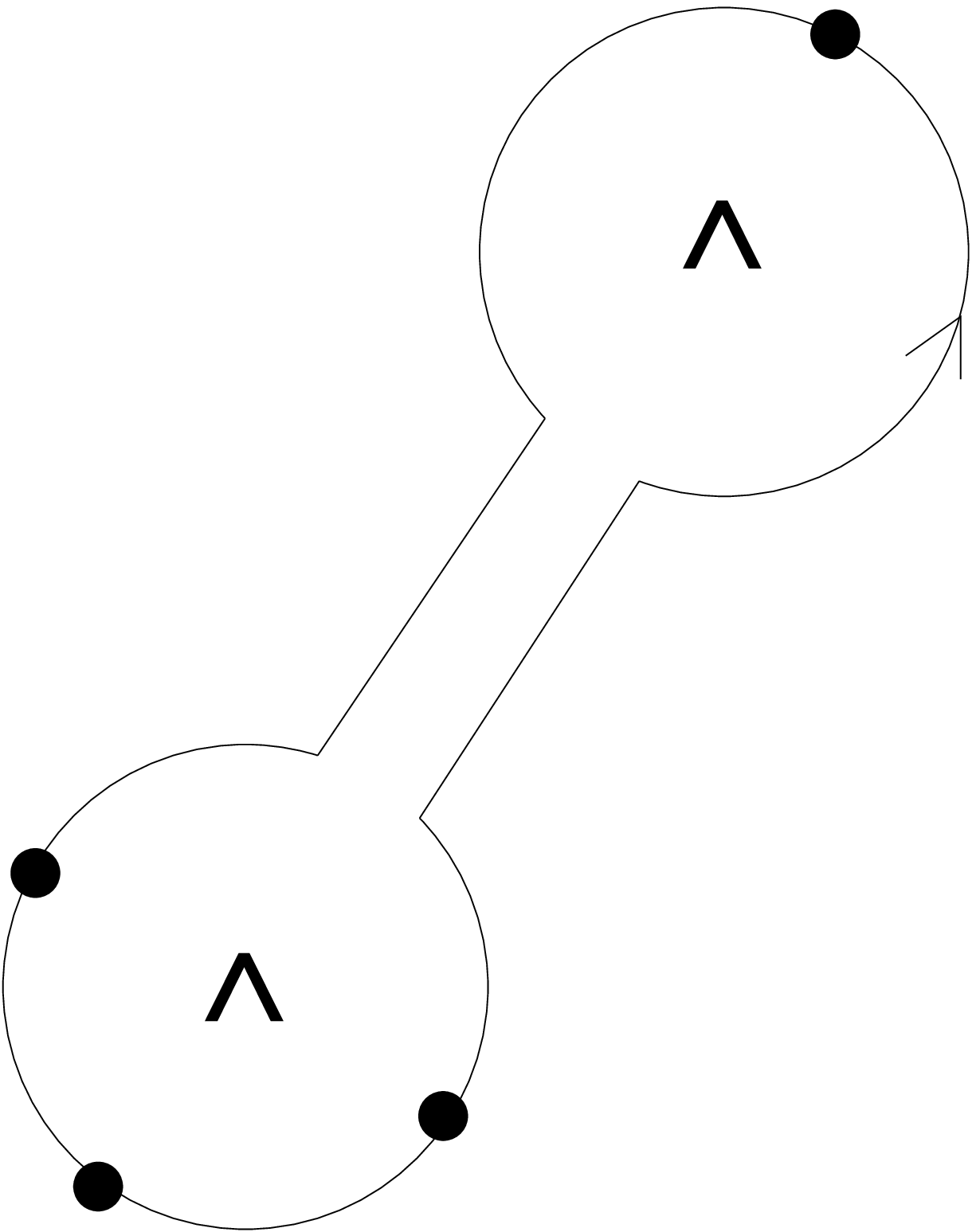}}
+2\vcenter{\epsfxsize=0.13\hsize\epsfbox{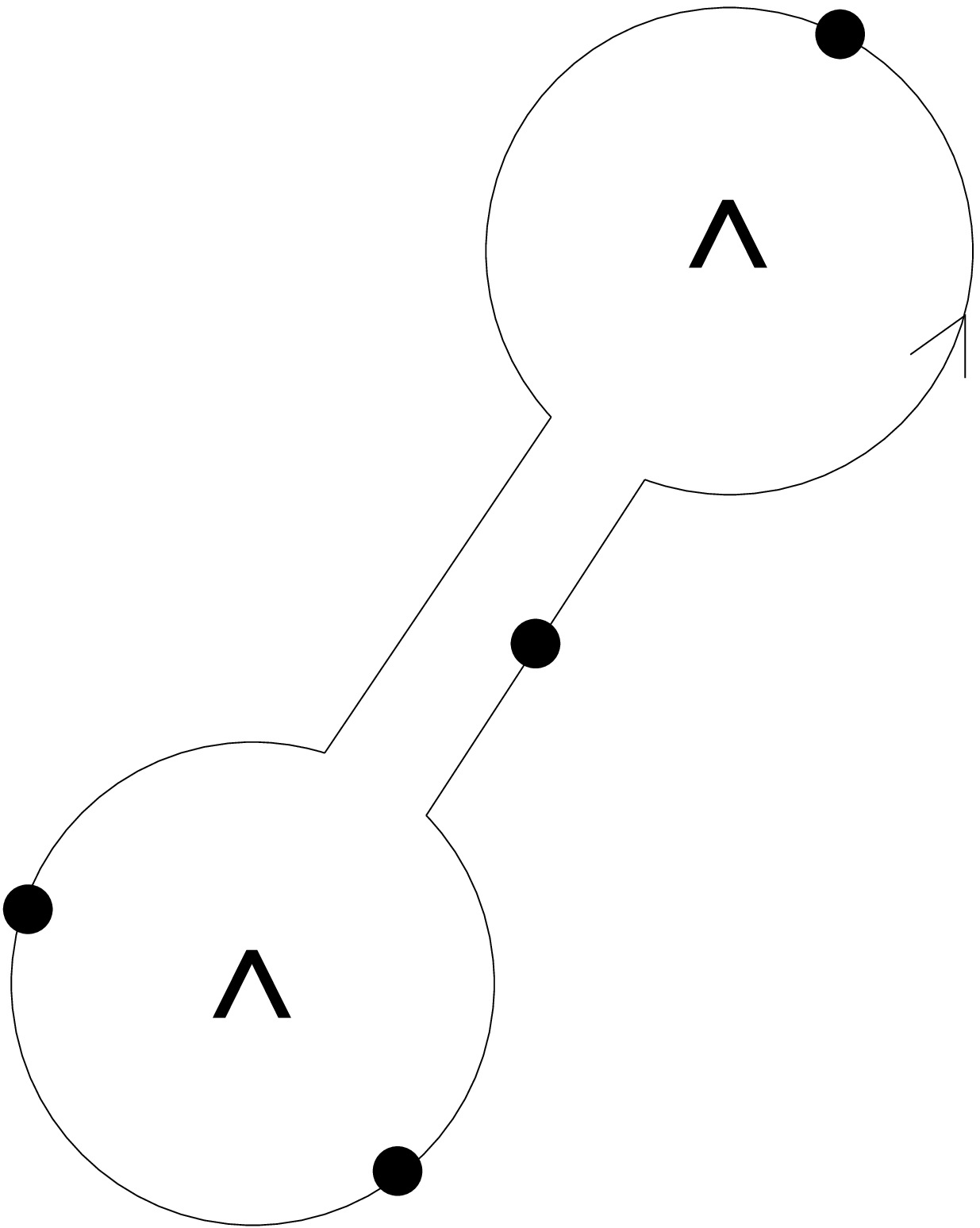}}\cr
&+2\vcenter{\epsfxsize=0.13\hsize\epsfbox{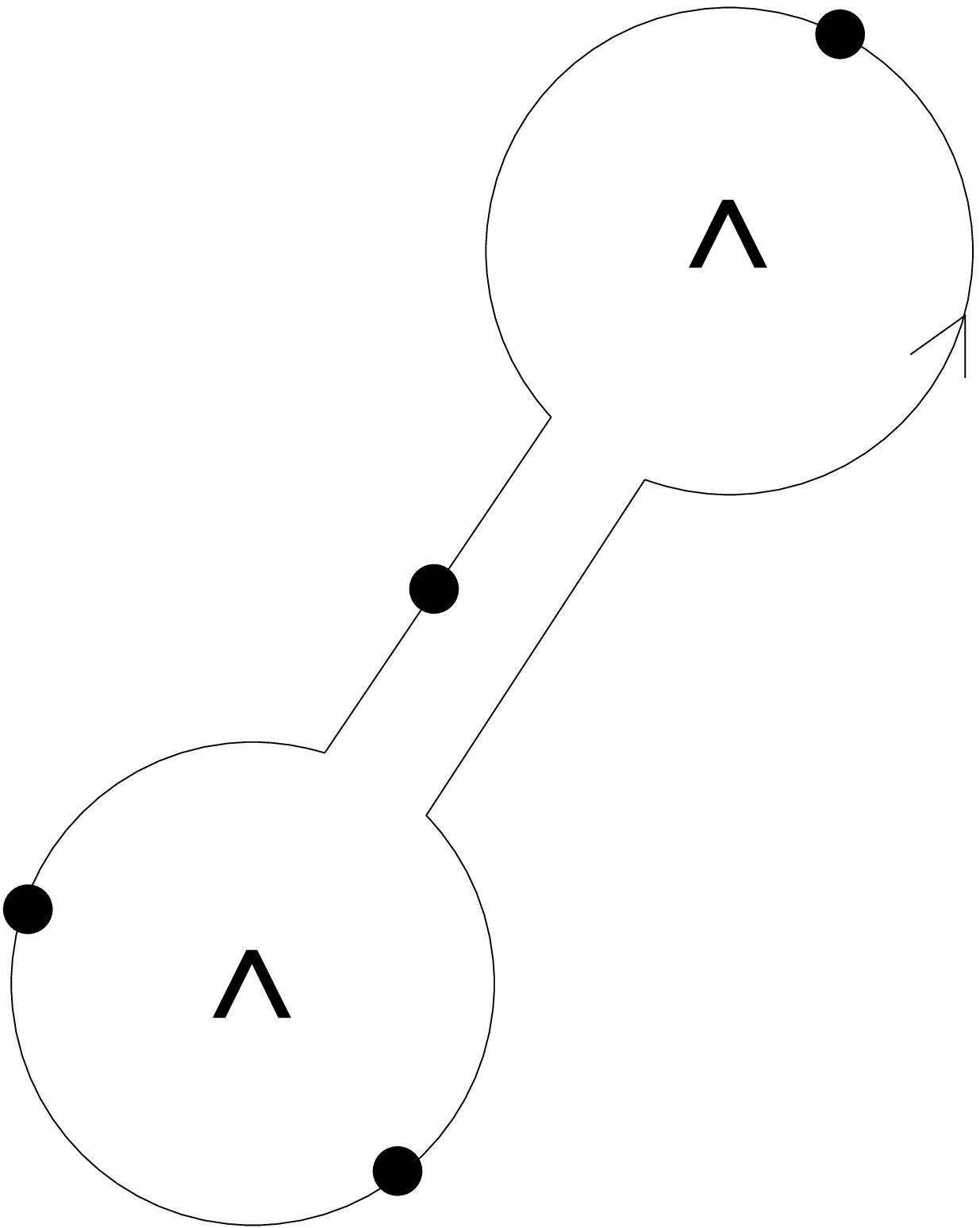}}
+2\vcenter{\epsfxsize=0.13\hsize\epsfbox{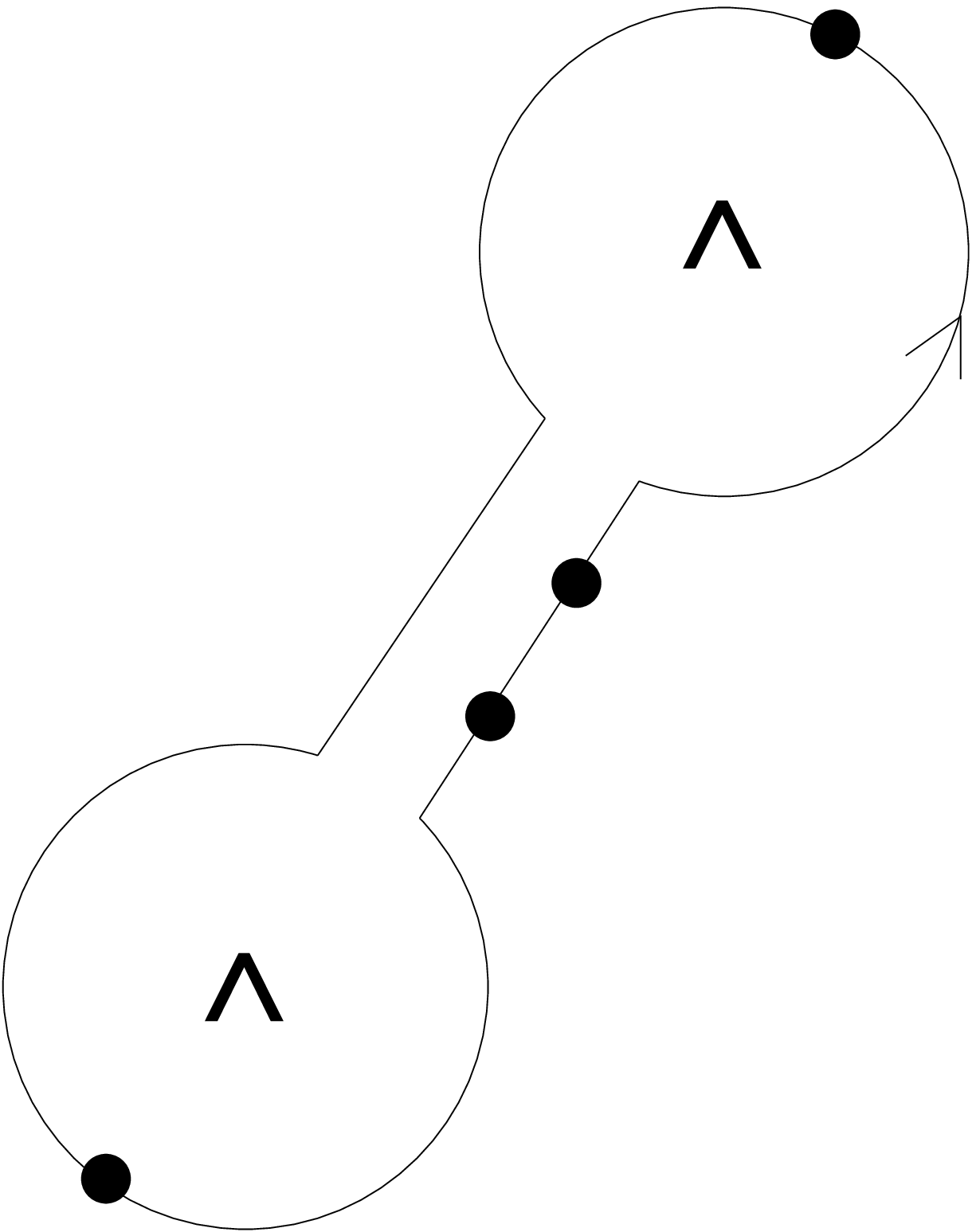}}
+\vcenter{\epsfxsize=0.13\hsize\epsfbox{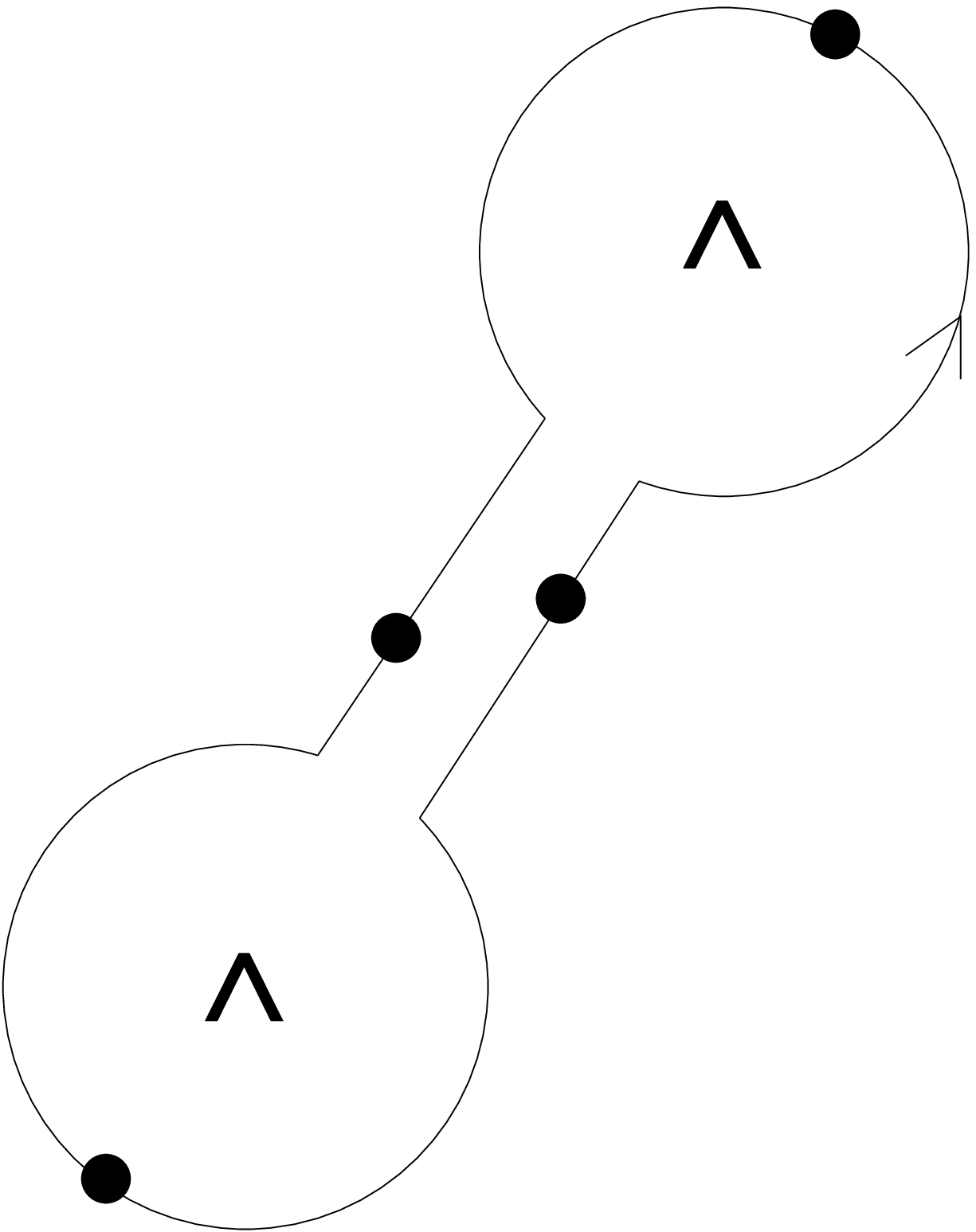}}
}
$$
\centerline{ 
{\bf Fig.12.} Feynman diagrams for the four-point vertex.}
\nfig\ffour{Fig12 ymi four-point fl}

We will concentrate on the vertices that we will need for the
$\beta_1$ computation: $\A\A\A\A$, $\B\B\A\A$ and
$\C\C\A\A$. (All the three-point $S_0$
vertices are however needed, either to derive these four-point vertices
or directly.) The $\C\A\A\A$ vertex need not be considered since it
vanishes by drifting. 

For the same reasons as given for the three-point vertex \threep,
we obtain the same four-$\A$ vertex as in ref. \ymi:
\eqn\fourp{\eqalign{S^0_{\mu\nu\la\sigma}(p, &q,r,s) =
-\int_\Lambda^\infty\!\!{d\Lambda_1\over
\Lambda_1^3}\Big\{
c'_{p+q}\left(
\hS_{\mu\nu\alpha}(p,q,r\!+\!s)
-\half S^0_{\mu\nu\alpha}(p,q,r\!+\!s)
\right)S^0_{\alpha\lambda\sigma}(p\!+\!q,r,s)\cr
&+\hS_{\sigma\alpha}(s)\hS_{\mu\nu\alpha}(p,q,r\!+\!s)c'_\lambda(r;p\!+\!q,s)
+\hS_{\lambda\alpha}(r)\hS_{\mu\nu\alpha}(p,q,r\!+\!s)c'_\sigma(s;r,p\!+\!q)\cr
&+\hS_{\mu\alpha}(p)\hS_{\alpha\sigma}(s)c'_{\nu\la}(q,r;p,s)
+\half\hS_{\mu\alpha}(p)\hS_{\alpha\lambda}(r)c'_{\nu,\sigma}(q;s;p,r)\cr
&+c'_s\hS_{\sigma\alpha}(s)\hS_{\mu\nu\la\alpha}(p,q,r,s)
+ {\rm cycles} \Big\}
+2\delta_{\mu\sigma}\delta_{\nu\lambda}
-4\delta_{\mu\lambda}\delta_{\nu\sigma}+2\delta_{\mu\nu}\delta_{\lambda\sigma}
\quad,}}
where `cycles' stands for the three cyclic permutations of $(p_\mu,q_\nu,
r_\la,s_\si)$.

The $\B\B\A\A$ vertex is:
\eqn\pBBAA{\eqalign{&S^{0\B\B\A\A}_{\ph{0}\,\mu\,\nu\,\la\,\sigma}(p,q,r,s)=
-\int_\Lambda^\infty\!\!{d\Lambda_1\over\Lambda_1^3}\Big\{
-S^{0\B\B\A}_{\ph{0}\,\mu\,\nu\,\al}(p,q,r\!+\!s)c'_{p+q}
 S^0_{\al\la\si}(p\!+\!q,r,s)\cr
&-S^{0\B\B\A}_{\ph{0}\,\mu\,\al\,\si}(p,q\!+\!r,s)K_{p+s,\al\be} 
 S^{0\B\B\A}_{\ph{0}\,\be\,\nu\,\la}(p\!+\!s,q,r)
+\hS^{\B\B\A}_{\,\mu\,\nu\,\al}(p,q,r\!+\!s)c'_{p+q}
 S^0_{\al\la\si}(p\!+\!q,r,s)\cr
&+S^{0\B\B\A}_{\ph{0}\,\mu\,\nu\,\al}(p,q,r\!+\!s)c'_{p+q}
 \hS_{\al\la\si}(p\!+\!q,r,s)
+\hS^{\B\B\A}_{\,\mu\,\al\,\si}(p,q\!+\!r,s)K_{p+s,\al\be}
 S^{0\B\B\A}_{\ph{0}\,\be\,\nu\,\la}(p\!+\!s,q,r)\cr
&+S^{0\B\B\A}_{\ph{0}\,\mu\,\al\,\si}(p,q\!+\!r,s)K_{p+s,\al\be}
 \hS^{\B\B\A}_{\be\,\nu\,\la}(p\!+\!s,q,r)
+\hS^{\B\B\A\A}_{\,\mu\,\nu\,\la\,\al}(p,q,r,s)c'_s\hS_{\al\si}(s)\cr
&+\hS^{\B\B\A\A}_{\,\mu\,\nu\,\al\,\si}(p,q,r,s)c'_r\hS_{\al\la}(r)
+\hS^{\B\B\A\A}_{\,\mu\,\al\,\la\,\si}(p,q,r,s)K_{q,\al\be}
 \hS_{\be\nu}(q)\cr
&+\hS^{\B\B\A\A}_{\,\al\,\nu\,\la\,\si}(p,q,r,s)K_{p,\al\be}\hS_{\be\mu}(p)
+\hS^{\B\B\A}_{\,\mu\,\nu\,\al}(p,q,r\!+\!s)c'_\la(r;p\!+\!q,s)
 \hS_{\al\si}(s)\cr
&+\hS^{\B\B\A}_{\,\mu\,\al\,\si}(p,q\!+\!r,s)c'_{\nu,\al\be}(q;p\!+\!s,r)
 \hS_{\be\la}(r)
+\hS_{\la\si\al}(r,s,p\!+\!q)c'_{\mu,\al\be}(p;r\!+\!s,q)
 \hS^{\B\B}_{\,\be\,\nu}(q)\cr
&+\hS^{\B\B\A}_{\,\al\,\nu\,\la}(p\!+\!s,q,r)K_{\si,\al\be}(s;q\!+\!r,p)
 \hS^{\B\B}_{\,\be\,\mu}(p)
+\hS^{\B\B\A}_{\,\mu\,\nu\,\al}(p,q,r\!+\!s)c'_\si(s;r,p\!+\!q)
 \hS_{\al\la}(r)\cr
&+\hS^{\B\B\A}_{\,\mu\,\al\,\si}(p,q\!+\!r,s)K_{\la,\be\al}(r;q,p\!+\!s)
 \hS^{\B\B}_{\,\be\,\nu}(q)
+\hS_{\la\si\al}(r,s,p\!+\!q)c'_{\nu,\be\al}(q;p,r\!+\!s)
 \hS^{\B\B}_{\,\be\,\mu}(p)\cr
&+\hS^{\B\B\A}_{\,\al\,\nu\,\la}(p\!+\!s,q,r)c'_{\mu,\be\al}(p;s,q\!+\!r)
 \hS_{\be\si}(s)
+\hS^{\B\B}_{\,\mu\,\al}(p)c'_{\nu\la,\al\be}(q,r;p,s)\hS_{\be\si}(s)\cr
&+\hS_{\si\al}(s)K_{\mu\nu,\al\be}(p,q;s,r)\hS_{\be\la}(r)
+\hS_{\la\al}(r)c'_{\si\mu,\al\be}(s,p;r,q)\hS^{\B\B}_{\,\be\,\nu}(q)\cr
&+\hS^{\B\B}_{\,\nu\,\al}(q)c'_{\la\si,\al\be}(r,s;q,p)
 \hS^{\B\B}_{\,\be\,\mu}(p)
+\hS^{\B\B}_{\,\mu\,\al}(p)c'_{\nu,\si,\al\be}(q;s;p,r)\hS_{\be\la}(r)\cr
&+\hS^{\B\B}_{\,\nu\,\al}(q)c'_{\la,\mu,\al\be}(r;p;q,s)\hS_{\be\si}(s)
\Big\}\cr
&+2\delta_{\mu\si}\delta_{\nu\la}-4\delta_{\mu\la}\delta_{\nu\si}
+2\delta_{\mu\nu}\delta_{\la\si}-2{\ct'_0\over\ct^2_0}\delta_{\mu\nu}
\delta_{\la\si}+\gamma^{BBA}(\delta_{\mu\si}\delta_{\nu\la}-
\delta_{\mu\la}\delta_{\nu\si})\quad.
}}
The integrand is a straightforward translation from the diagrams, together 
with some cancellations of $L$ wine terms, in particular all $\C$-$\A$ wine
terms, as a result of drifting. Further such simplifications occur only
at the expense of expanding the seed vertices containing $\B$,
as $\hS_A+\hS_B$, to isolate the pure gauge part $\hS_A$. 
Note that, as with the four-$\A$ vertex \ymi, 
gauge invariance \gid\ strongly constrains the overall form of the
integrand and acts as a powerful consistency check.
The integration constant is dimension zero and therefore cannot generate
any new gauge invariant terms. Consequently the integration 
constant follows from expansion of \dBBA,
and the $\Lambda_1$ integral diverges in just such a way as to be cancelled
by the divergence in \gaBBA.

By drifting, all the $L$ wine terms cancel out in the $\C\C\A\A$ vertex:
\eqn\pCCAA{\eqalign{&S^{0\C\C\A\A}_{\ph{0\C\C}\,\la\,\sigma}(p,q,r,s)=
-\int_\Lambda^\infty\!\!{d\Lambda_1\over\Lambda_1^3}\Big\{
-S^{0\C\C\A}_{\ph{0\C\C}\,\al}(p,q,r\!+\!s)c'_{p+q}
 S^0_{\al\la\si}(p\!+\!q,r,s)\cr
&-{1\over\Lambda_1^2}S^{0\C\C\A}_{\ph{0\C\C}\,\si}(p,q\!+\!r,s)M_{p+s}
 S^{0\C\C\A}_{\ph{0\C\C}\,\la}(p\!+\!s,q,r)
+\hS^{\C\C\A}_{\ph{\C\C}\,\al}(p,q,r\!+\!s)c'_{p+q}
 S^0_{\al\la\si}(p\!+\!q,r,s)\cr
&+S^{0\C\C\A}_{\ph{0\C\C}\,\al}(p,q,r\!+\!s)c'_{p+q}
 \hS_{\al\la\si}(p\!+\!q,r,s)
+{1\over\Lambda_1^2}\hS^{\C\C\A}_{\ph{\C\C}\,\si}(p,q\!+\!r,s)M_{p+s}
 S^{0\C\C\A}_{\ph{0\C\C}\,\la}(p\!+\!s,q,r)\cr
&+{1\over\Lambda_1^2}S^{0\C\C\A}_{\ph{0\C\C}\,\si}(p,q\!+\!r,s)M_{p+s}
 \hS^{\C\C\A}_{\ph{\C\C}\,\la}(p\!+\!s,q,r)
+\hS^{\C\C\A\A}_{\ph{\C\C}\,\la\,\al}(p,q,r,s)c'_s\hS_{\al\si}(s)\cr
&+\hS^{\C\C\A\A}_{\ph{\C\C}\,\al\,\si}(p,q,r,s)c'_r\hS_{\al\la}(r)
+{1\over\Lambda_1^2}\hS^{\C\C\A\A}_{\ph{\C\C}\,\la\,\si}(p,q,r,s)\left[
 M_q\hS^{\C\C}(q)+M_p\hS^{\C\C}(p)\right]\cr
&+\hS^{\C\C\A}_{\ph{\C\C}\,\al}(p,q,r\!+\!s)c'_\la(r;p\!+\!q,s)
 \hS_{\al\si}(s)
+{1\over\Lambda_1^2}\hS^{\C\C\A}_{\ph{\C\C}\,\la}(s\!+\!p,q,r)
 M_\si(s;q\!+\!r,p)\hS^{\C\C}(p)\cr
&+\hS^{\C\C\A}_{\ph{\C\C}\,\al}(p,q,r\!+\!s)c'_\si(s;r,p\!+\!q)
 \hS_{\al\la}(r)
+{1\over\Lambda_1^2}\hS^{\C\C\A}_{\ph{\C\C}\,\si}(p,q\!+\!r,s)
 M_\la(r;q,p\!+\!s)\hS^{\C\C}(q)\cr
&+{1\over\Lambda_1^2}\hS^{\C\C}(q)M_{\la\si}(r,s;q,p)\hS^{\C\C}(p)
\Big\}\quad  +2{\ct'_0\over\ct^2_0}
 \left\{(2q+r)_\la(2p+s)_\si-(p^2+q^2)\delta_{\la\si}\right\}\cr
&+\gamma^{CCA}\left\{p_\la q_\si-p_\si q_\la-\delta_{\la\si}p.s
+s_\la p_\si-\delta_{\la\si}q.r+r_\si q_\la\right\}+\gamma^{CCAA}(
\delta_{\la\si}r.s-s_\la r_\si)\quad.
}}
Gauge invariance again acts as a powerful consistency
check on the integrand. The dimension two
integration constant allows for one new gauge invariant term 
$\sim \gamma^{CCAA}\C\F_{\mu\nu}^2\C$, 
otherwise gauge invariance requires the rest to be taken from \dCCA.
The quadratic divergence $2\Lambda^2_0\ct^{-1}_0\delta_{\la\si}$
is not displayed but cancels exactly such a divergence in the integral. 
Together with $\gamma^{CCA}$, our final free real dimensionless parameter 
$\gamma^{CCAA}$, mop up all logarithmic divergences in the integral.

\newsec{The $\beta$ function without gauge fixing}
The $\beta$ function is determined in essentially the same way as in 
ref. \ymi. As for $\hS$ and $S_0$, when the vertices are not labelled by 
their flavours, we now mean the pure-$\A$ vertices.
Using the renormalization condition \defg,  we have for the $\A\A$ vertices
\eqn\rencon{S_{\mu\nu}(p)+S^\sigma_{\mu\nu}(p)={2
/ g^2}\,\Delta_{\mu\nu}(p)+O(p^3)\quad,}
and thus by \twop\ and \hSex,
$$S_{\mu\nu}(p)+S^\sigma_{\mu\nu}(p)=
{1\over g^2}S^0_{\mu\nu}(p)+O(p^3)\quad.$$
By \Sloope\ and \scsiz, this implies {\sl that the $O(p^2)$ component of
all the higher loop contributions 
$S^n_{\mu\nu}(p)+S^{n\sigma}_{\mu\nu}(p)$,  must vanish}.
This greatly simplifies the $O(p^2)$ part of the $\A\A$ vertex flow in
\ergone\ -- \ergtwo,  in particular reducing them to algebraic equations.
Thus we see that
\eqn\betaone{ 
 a_1[S_0-2\hS]^\sigma_{\mu\nu}(p) =-4\beta_1\Delta_{\mu\nu}(p)+O(p^3)\quad,}
where $a_1[S_0-2\hS]^\sigma_{\mu\nu}(p)$ is the $\sigma_3\A\A$
vertex in $a_1[S_0-2\hS]$ and we have used the fact that all one-loop
vertices contain a trapped $\sigma_3$ [\cf discussion 
above \bareone]. This fixes $\beta_1$. Similarly,
at $n\ge2$ loops,  the $\beta_n$ are determined by the requirement that  
$$a_1[S_{n-1}]_{\mu\nu}(p)+a_1[S_{n-1}]^\sigma_{\mu\nu}(p)=
-4\beta_n\Delta_{\mu\nu}(p)+O(p^3)\quad.$$
And non-perturbatively from \ERG\ and \Sloope,
$$a_1[g^2S-2\hS]_{\mu\nu}(p)+a_1[g^2S-2\hS]^\sigma_{\mu\nu}(p)=
-{4\over g^3}\beta(g)\Delta_{\mu\nu}(p)+O(p^3)\quad.$$

\subsec{One loop $\beta$ function without gauge fixing}
The 
diagrams giving the LHS of \betaone\ take the same form as in ref. \ymi\
(apart from the fact that the large $N$ limit has been taken):
\nfig\fbeta{fig. 14 of ymi}
$$
{2\over N}\ \vcenter{\epsfxsize=0.085\hsize\epsfbox{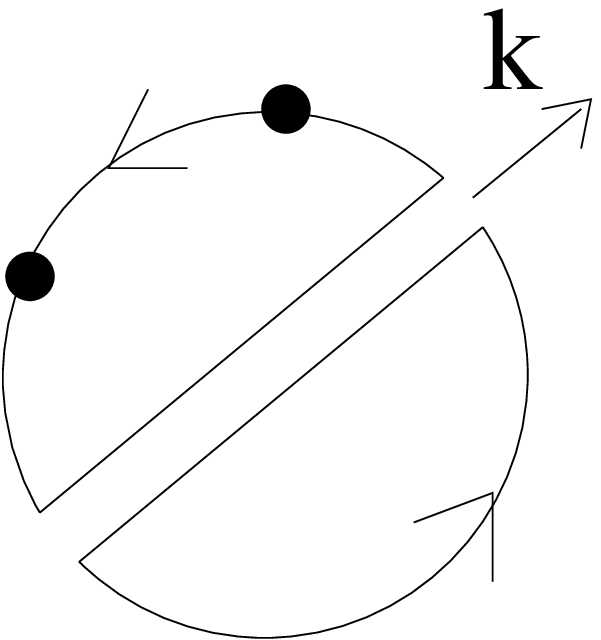}}
+{2\over N}\ \vcenter{\epsfxsize=0.085\hsize\epsfbox{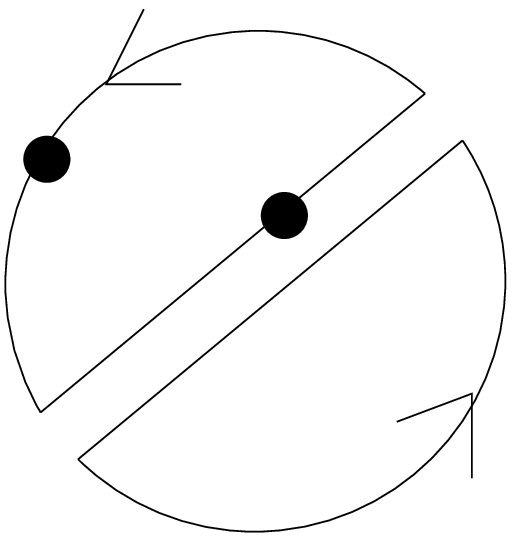}}
+{2\over N}\ \vcenter{\epsfxsize=0.085\hsize\epsfbox{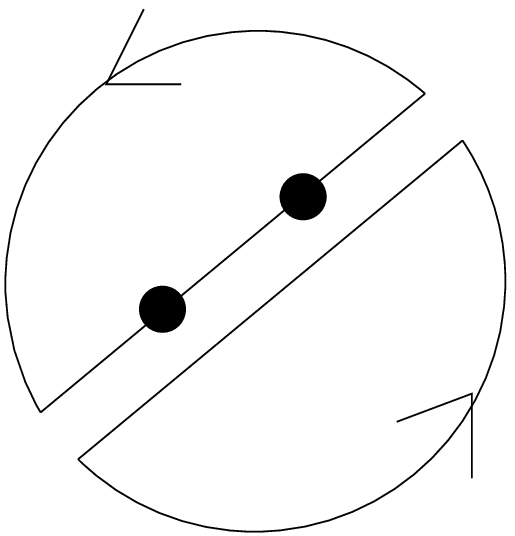}}\quad,
$$
\centerline{ 
{\bf Fig.13.} The one-loop two-point diagrams.}
where, as in ref.\ymi, we let the circle stand for $\Sigma_0=S_0-2\hS$.
Note that
from \twop\ and \ergone, these diagrams also sum to
$\Lambda\partial S^1_{\mu\nu}(p)/\partial\Lambda-2\beta_1 \hS_{\mu\nu}(p)$,
and thus after $\Lambda$ integration, yield the one-loop $\A\A$ vertex. 

Translating as described in sec. 4, since there
is no tree-level vertex with a single $\C$, 
we see that the wines in \fbeta\ attach via
two $\A$s, two $\B$s or two $\C$s.
Expanding these attachments
as in \fsig, the terms that survive have a $\sigma_3$ pair
at the top or bottom of the wine (\cf the discussion below \bpsplit\ --
the terms with either both $\sigma_3$ pairs,
or neither $\sigma_3$ pair,
vanish by the supertrace mechanism).
Thus we obtain:\foot{Note that the wine vertices appear reflected in \fbeta\
compared to sec. 4.}
\eqn\ol{\eqalign{ &a_1[\Sigma_0]^\sigma_{\mu\nu}(p) =
{2\over\Lambda^2}\int\!{d^Dk\over(2\pi)^D}\Bigg\{
c'_{k,\al\be}\Sigma^0_{\al\be\mu\nu}(-k,k,p,-p)\cr
&+c'_{\mu,\al\be}(p;k\!-\!p,-k)\Sigma^0_{\al\be\nu}(p\!-\!k,k,-p)
+c'_{\mu\nu,\al\be}(p,-p;k,-k)\Sigma^0_{\al\be}(k)\cr
&+{1\over\Lambda^2}M_k\Sigma^{0\C\C\A\A}_{\ph{0\C\C}\,\mu\,\nu}(-k,k,p,-p)
+{1\over\Lambda^2}M_\mu(p;k\!-\!p,-k)\Sigma^{0\C\C\A}_{\ph{0\C\C}\,\nu}
  (p\!-\!k,k,-p)\cr
&+{1\over\Lambda^2}M_{\mu\nu}(p,-p;k,-k)\Sigma^{0\C\C}(k)
-K_{k,\al\be}\Sigma^{0\B\B\A\A}_{\ph{0}\,\al\,\be\,\mu\,\nu}(-k,k,p,-p)\cr
&-K_{\mu,\al\be}(p;k\!-\!p,-k)\Sigma^{0\B\B\A}_{\ph{0}\,\al\,\be\,\nu}
  (p\!-\!k,k,-p)
-K_{\mu\nu,\al\be}(p,-p;k,-k)\Sigma^{0\B\B}_{\ph{0}\,\al\,\be}(k)
\Bigg\}\quad,
}}
where we have noted that, by Lorentz invariance, the 
$p_\mu\leftrightarrow-p_\nu$ transposition just yields a factor 2 \ymi.

The momentum integral is finite as required,
for appropriately chosen covariantization and cutoff functions. We 
use \pourchoice\ (evaluated in sec. 5) and keep $c$ and $\ct$ 
general except for the requirements $c(0)=1$ and $\ct(0)>0$ 
discussed in sec. 2, and
\eqn\powlaw{c(x)\propto x^{-r} \ins11{and} \ct(x)\propto x^{-\rt}  }
for large $x$. Physically, we are interested in $D=4$ where $g$ is 
marginal, and $\beta_1$ is universal. We set 
\eqn\rr{r>\rt>1.}
This is helpful for calculation, but more stringent than necessary,
since we show in the next section that physical one-loop 
contributions are finite for $r>\rt>0$.

Note that although \ol\ is thus finite, it is the sum of integrals
with cancelling divergences and therefore still 
ambiguous, a generic problem with Pauli-Villars regularisation \warr.
We will keep $D$ general and take the limit $D\to4$ only at the end
of the calculation. This amounts to dimensional preregularisation \warr,
and allows us to discard those parts which are
surface terms in any dimension $D$, but keep those parts which become
surface terms only in $D=4$ dimensions; the independence of $\beta_1$
on the choice of cutoff functions, \ie its universality, 
arises from being expressed {\sl entirely} through such latter terms.
For an explicit demonstration of these subtleties, see the example 
in ref. \alg. Further comments can be found later in this section
and in the conclusions.

By \twop, in the two-point vertices of \ol\ we have $\Sigma_0\equiv-\hS$. 
Applying \gid, the first term of \ol\ becomes
$$c'_{k,\al\be}\Sigma^0_{\al\be\mu\nu}(-k,k,p,-p)\equiv
c'_k\Sigma^0_{\al\al\mu\nu}(-k,k,p,-p)
+L_k\left\{\hS_{\mu\nu}(p)-\hS_{\mu\nu}(k+p)\right\}\quad,$$
where we use the above comment, and the fact that $L_kk_\al$ vanishes
by Lorentz invariance of $k$ integral. The expression for
$\Sigma^0_{\al\al\mu\nu}(-k,k,p,-p)$ derived from the last section, may
be further simplified by using Lorentz invariance (in fact $k\leftrightarrow
-k$, $\mu\leftrightarrow\nu$ \etc) and the coincident line identities
\coids\ in a manner already described in ref. \ymi. After these
manipulations, we simply substituted all the expressions derived earlier,
and expanded to $O(p^2)$ to extract $\beta_1$ via
\betaone. The large number of resulting terms were handled with care
by algebraic computing. (We used FORM.) This calculation is
described below. 

It seems likely that we 
could have proceeded more intelligently: after all, $\beta_1$ should be
universal not only to choices of $c$ and $\ct$ but also to the
choices of wine vertex and seed action $\hS$, \ie covariantization 
(which, recall, need not even be the same in each wine).
Therefore the result should fall out without the need for all these
expressions to be explicit. We leave such investigations for the
future.

A formula for
the gauge dependent part falls out simply as follows, and acts as a check
on the calculation \ymi. Using \gid\ and \wga, collecting terms using
Lorentz invariance under $k\leftrightarrow-k$, and again using \twop,
\eqn\gilch{\eqalign{p^\mu p^\nu a_1[\Sigma_0]^\sigma_{\mu\nu}(p) &=
{2\over\Lambda^2}\int\!{d^Dk\over(2\pi)^D}\Big\{ c'_{k,\al\be}
\hS_{\alpha\be}(k)+{1\over\Lambda^2}M_k\hS^{\C\C}(k)
-K_{k,\al\be}\hS^{\B\B}_{\,\al\,\be}(k)\cr
&-c'_{k+p,\al\be}\hS_{\al\be}(k\!+\!p)
-{1\over\Lambda^2}M_{k+p}\hS^{\C\C}(k\!+\!p)
+K_{k+p,\al\be}\hS^{\B\B}_{\,\al\,\be}(k\!+\!p)
\Big\}\quad.}}
By a shifting momentum $k\mapsto k+p$ in the first three terms we see that 
the gauge dependent part is zero as it must be.
However, expanding \gilch\ to order $p^4$ using \hstwo, and as in the
ensuing calculation, leaving the radial part of the $k$ integral to last,
we obtain
\eqn\gad{\eqalign{a_1[\Sigma_0]^\sigma_{\mu\nu}(p) &=\cr
-4(D-1)&\Omega_D\Lambda^{D-4}
\left\{\Lambda^2\delta_{\mu\nu}\int^\infty_0\!\!\!\!\!\!dx{d\over dx}G_0
+p_\mu p_\nu\int^\infty_0\!\!\!\!\!\!dx{d\over dx}G_L
\right\} -4\beta_1\Delta_{\mu\nu}(p)+O(p^3)\cr
{\rm where}\qquad G_0 &={x^{D/2}F'\over D(D-1)}
\quad,\quad G_L={1\over D(D+2)(D-1)}\left(x^{D/2+1}F''\right)'\cr
{\rm and}\qquad F &=(D-1)(c'-K)x/c-(DK+xL)/\ct+(x/\ct+\sigma)M
\quad.
}}
Here $\Omega_D=2/[\Gamma(D/2)(4\pi)^{D/2}]$ is the solid angle of a 
$(D-1)$-sphere divided by $(2\pi)^D$, and as before prime is 
differentiation with respect to its argument (here $x$). The transverse
part follows from \betaone. One readily verifies, \cf (9.4), that
with $r>\rt>0$, $F\sim1/x^f$ as $x\to\infty$ with $f>1$, and thus the
$\delta_{\mu\nu}$ and $p_\mu p_\nu$ contributions to \gad\ do indeed 
vanish.

We sketch the main steps of the remaining calculation. 
After simplifying \ol\ as discussed earlier, and
substituting for all $S^0$ and $\hS$ four-point vertices, we simplify
the special-momenta terms by recognizing that both 
$S^{0\X\X\A}_{\ph{0}a_1a_2\la}(q,-q,0)$
and $\hS^{\X\X\A}_{a_1a_2\la}(q,-q,0)$ collapse to 
${\partial\over\partial q^\la}\hS^{\X\X}_{a_1a_2}(q)$, 
as follows from gauge invariance.\foot{\cf the notation of \sexp; use 
\gid\ on $s^\la S_{\mu\nu\la}(q,r,s)$ and expand for small $s$.}
After substituting for all remaining vertices, the wine expressions
\wone, \wtwo\ and \wtwop\ are substituted, together with the 
special-momentum case \wtwosp. No other special-momenta terms 
requiring careful limits, arise. 

We collected terms under 
the substitution $k\leftrightarrow-k$, and
expanded to overall order $p^2$, which is now straightforward. 
The angular part of the $k$ integration is achieved through the 
following equivalences under the integral
$$(k.p)^2k_\mu k_\nu\equiv {k^4\over D(D+2)}\left(p^2\delta_{\mu\nu}
 +2p_\mu p_\nu\right)\ins11{and}
k_\mu\k_\nu\equiv {k^2 \over D}\delta_{\mu\nu}\quad,$$
as follows from considering Lorentz invariance (of in the first case
$k_\al k_\be k_\mu k_\nu$). The result is written as a linear combination
of $\delta_{\mu\nu}$, $p_\mu p_\nu$ and $\Delta_{\mu\nu}(p)$.

As well as the radial $k$ integral, there are a number of 
$\Lambda$ integrals to evaluate. The most involved of these are the 
integrals for the classical four-point vertices since these contain 
in turn, some inner 
$\Lambda$ integrals for classical three-point vertices.
Starting with these innermost $\Lambda$ integrals, we substitute for 
$K$, $L$ and $M$ using \KLM, and expand the derivatives until they
are all expressed directly in terms of differentiated cutoff functions,
\ie $c^{(n)}(x)$ and $\ct^{(m)}(x)$;
note that these always appear in the numerator of the integrands.

The resulting expression is very large, yet it must boil down to the one
number $\beta_1$. The game then is to manipulate it in
such a way as to successively cancel out nearly all the terms.
To do this, after any differentiation or integration
we express the resulting terms in a unique algebraic form:
in this way any algebraic cancellations take place automatically. 
Denominators involve positive integer powers of
$b(x)=1/(x\ct+c)$ and $f(x)=1/(x+\sigma c)$ ($x=k^2/\Lambda^2$ 
being effectively the integration variable). Iteratively, every appearance
of $xb$ is replaced by $(1-cb)/\ct$, and every appearance
of $b/x$ by $(1/x-\ct b)/c$. These relations thus eliminate
$x$ or/and $b$ in $x^mb^n$, in favour of $c$ and $\ct$.
A number of other ways of bringing such terms to a 
canonical form also exist, but this one has
the advantage that it does not exchange terms which behave well as
$x\to0$ or $\infty$, with terms that individually do not. 
Similarly we tidy $f$ terms via iterating
$xf=1-\sigma\ct f$ and $f/x=(1/x-f)/(\sigma\ct)$. Finally terms with both
$b$ and $f$ in but no explicit power of $x$, are brought to canonical 
form by iterating the substitutions $\ct^2bf=(cbf+\ct b-f)/\sigma$ and
$cbf/\ct=\sigma\ct bf+f/\ct-b$. 

We now integrate by parts the inner integrals
as follows. We take first the terms with a $c^{(n)}(x)$ or 
$\ct^{(n)}(x)$ where the derivative $n$ is the largest that appears.
(For the inner integrals the maximum derivative
is $n=4$; overall, it is $n=5$.) We then integrate by
parts to systematically reduce this maximum $n$. If such $\ct^{(n)}$ terms
also contain some positive power $m$ of $\ct^{(n-1)}$, we eliminate 
$\ct^{(n)}$ by first writing
\eqn\ctelim{\ct^{(n)}\left(\ct^{(n-1)}\right)^m={1\over m+1}{d\over dx}
\left(\ct^{(n-1)}\right)^{m+1}}
and similarly for $c$. If a $c^{(n-1)}$ multiplies
the terms in \ctelim, we still integrate by parts, thereby exchanging
a $\ct^{(n)}$ for a $c^{(n)}$ -- which we take to be the canonical order. 
Obstructions to further reduction of the maximum derivative $n$ thus 
occur if there are terms with a product of more than one $n$-derivative
factor [\ie $c^{(n)}$ or $\ct^{(n)}$] or if the product $c^{(n)}\ct^{(n-1)}$
appears. Although such terms have the potential to exist they all cancel
away as we iterate $n$ down to $n=1$. 

Incidentally, as mentioned in sec. 7, we also derived $\beta_1$ 
by keeping an explicit initial $\Lambda_0$. Integrating by parts then 
often yields negative powers of $\Lambda_0$ from the $\Lambda=\Lambda_0$ 
boundary (if necessary by Taylor expansion in 
$1/\Lambda_0$). At this stage it is straightforward to
estimate the maximum divergence of  factors containing further 
$\Lambda$ integrals, as a power of $\Lambda_0$ (or as $\ln\Lambda_0$), 
and thus we
drop all terms that go overall as a negative power of $\Lambda_0$;
in fact in this way we find that at most a single
power of $1/\Lambda_0^2$ need be kept. 

When there are only first derivatives left the strategy for a unique
simplification route is more involved
because differentiating $b$ ($f$) produces $\ct'$ and $c'$ ($\ct'$).
We solved this with the following algorithm.
If the integrand contains a single $c'$, 
and otherwise only $c$ and $x$ then the final derivative
can be eliminated similarly to \ctelim\ (introducing $\ln c$ if $m=-1$). 
A similar strategy applies for $\ct'$, except that we allow a power
of $c$ -- thus exchanging $\ct'$ terms for those with $c'$, our canonical
choice. 
Next, if there is no $c'$, one $\ct'$ and a factor of $b^m$ where $m\ge2$
(there are such terms up to $m=5$) then we write:
$$b^m\ct'={1\over m-1}{\ct\over c}{d\over dx} b^{m-1}+b^m(\ct+c')/c
+b^{m-1}\ct'/c\qquad (m\ge2),$$
integrate by parts the explicit $d/dx$, and iterate until there are no
more such terms. 
Next,  if there is no $c'$, no $b$, one $\ct'$ and a factor of 
$f^m$ where $m\ge2$ we write
$$f^m\ct'=-{1\over\sigma}\left(f^m+{1\over m-1}{d\over dx}f^{m-1}\right)
\qquad (m\ge2),$$
integrate by parts the explicit $d/dx$, and iterate until there are no
more such terms. After all this is done (mindful always to map
the results to a unique algebraic form as above) no further simplification
is possible on these integrals.

For the $\delta_{\mu\nu}$ and $p_\mu p_\nu$ terms, these manipulations
are sufficient to evaluate all the inner $\Lambda$ integrals and then
the outer integrals; we checked that the remaining integral, the 
radial $k$ integral, indeed takes the form \gad\ as expected. (In fact
we also checked that this is so, independent of \KLM, again as expected.)

For the $\Delta_{\mu\nu}(p)$ terms however, there are quite a few
inner $\Lambda$ integrals with at most singly differentiated cutoff
functions, that cannot be further simplified. At this stage we integrate 
by parts the outer $\Lambda$ integral (and single $\Lambda$ integrals), 
and simplify as much as possible, proceeding
in a similar manner to above. (We need also
to eliminate $c'$ in $c' \ln c=(c\ln c-c)'$ \etc)
This causes some of the remaining
inner $\Lambda$ integrals to disappear by differentiation.

Finally, we convert the measure in the $k$ integral to
\eqn\kint{\Lambda^D\Omega_D\int_0^\infty\!\!\!\!\!\!dx\,x^{D/2-1}\quad,}
where $x=k^2/\Lambda^2$, and integrate
this by parts, again in the manner sketched above. 

In the case that
we simply set $D=4$ from the beginning we find that {\sl all} the 
remaining integrals are thus evaluated, converting the 
expression to a set of boundary terms. In this way, the remaining 
calculation consists of a set of limits: $x\to0$ and $x\to\infty$ from
the boundaries of the radial $k$ integral (and first $x_0:=k^2/\Lambda_0^2
\to0$ if the explicit initial $\Lambda_0$ was kept). These limits may
themselves involve integrals, however at this stage these integrals
are easily done by substituting the appropriate asymptotic behaviour
for the integrands as $x\to0$ or $\infty$.
Indeed, just by power-counting, the vast majority
of the remaining terms simply vanish. Let us give just one example.
In order to be finite under \rr, at one stage
we have to collect together some separately divergent terms. Together
these yield, using \betaone, a contribution to $\beta_1$ of 
$$\displaylines{{1\over(4\pi)^2}\lim_{y\to\infty}\Bigg\{8yc(y)\ct(y)\left(
\int^y\!\!\!\!\!dx\,[1/x-b(x)\ct(x)]/c(x)\right)^2\hfill\cr
\hfill-[15y\ct(y)+3y^2\ct'(y)]\int^y\!\!\!\!\!dx\,[1/x
-b(x)\ct(x)]/c(x)\Bigg\}\quad.\cr}$$
In here the lower limit of the integrals corresponds to $\Lambda=\Lambda_0$
and is evaluated first. It
is either explicitly finite by cancellation from other parts or
the divergences are 
discarded in the manner already described in sec. 7.
Under the ultraviolet limit $y\to\infty$, the first term vanishes and
the second yields
$${1\over(4\pi)^2}\left({3\over\rt-1}-12\right),$$
as follows from \powlaw. 

After all the limits are evaluated in the $D=4$ calculation,
we obtain an expression for $\beta_1$ which is
independent of all the extra parameters $\gamma$ introduced in the
previous section,\foot{Ref. \alg\ illustrates this by showing how the
$\left(\gamma^{BBA}\right)^2$ term disappears.}
and all the details of $c$ and $\ct$ except that there is
still a dependence on the powers $r$ and $\rt$:
$$\beta_1={1\over(4\pi)^2}\left(16{\rt\over r}+{8\over r}+{1\over 1-\rt}
-r^2-\rt^2+r\rt-r+{\rt\over2}-{205\over6}\right)\quad.$$
Nevertheless the dependence on $r$ and $\rt$ means that
it is therefore not universal -- a clear signal that we can attach no
physical significance to the result. Indeed, we have already seen \alg\
that the result will alter if different momentum routings are chosen
(\eg shifting $k$ by $\pm p$) in different parts of \ol.

We return to step \kint,
and leave $D$ general, setting $D=4-\epsilon$.
Integrating by parts and expanding in small $\epsilon$,
we now obtain some extra remainder terms which are
$\epsilon$ times integrals that diverge as $D\to4$. (As in dimensional
regularisation terms with any higher power of $\epsilon$ are easily seen
to vanish at one loop as $\epsilon\to0$.) 
We do not expect it to make any difference
whether $\epsilon$ is taken positive or negative since the 
integral is finite for any small $\epsilon$ providing we are allowed to shift
momenta and discard the resulting surface terms. Rigorously however, 
we choose $D<4$ since then all surface terms vanish for
$D$ small enough, after which the result may defined by analytic continuation
to all $D$. In the present case any value of $D<4$ is sufficient.
Throwing away 
those terms that clearly vanish for $D<4$, and restoring the factor
$N$ absorbed by the large $N$ change of variables \ymi,
we find \alg:
\eqn\atlast{\eqalign{\beta_1 &=
{N\over (4\pi)^2}{\epsilon\over2}\Bigg[
\int^\infty_0\!\!\!\!\!\!dx\ x^{\epsilon/2-1}\left\{ 2+4c^2-6c\right\}
+\int^\infty_0\!\!\!\!\!\!dx\ x^{\epsilon/2}\left\{{59\over2}{\ct\over
x\ct+c}-{59\over6}{1\over x+\sigma\ct}\right\}\cr
&\ +\int^\infty_0\!\!\!\!\!\!dx\ x^{\epsilon/2}\,
\ct \int^x_0\!\!\!\!dx_1\left\{ {20\,\ct\over (x_1\ct+c)^2}
-{2\over\ct(x_1+\sigma\ct)^2}+{24\,\ct'\over\ct(x_1\ct+c)}
-{6\,\ct'\over\ct^2(x_1+\sigma\ct)}\right\}\Bigg]\cr
&\to-{11\over3} {N\over(4\pi)^2}\ins11{as}\epsilon\to0,}}
the famous one-loop result for $SU(N)$ Yang-Mills,
completely independent of the $\gamma$s and $c$ and $\ct$, 
this time including $r$ and $\rt$.

({\it N.B.} We display in \atlast, as an example,
the penultimate terms in
a way closely similar to those that
we computed.  There is no sense however of a 
canonical choice here: they are
easily altered by adding terms whose differences obviously vanish
as $D\to4^-$. Only the final limit is invariant.)

\newsec{One loop finiteness}

For the following
proof, it is helpful to constrain the form of the covariantization and 
to specify the
general form of the asymptotic behaviour of the cutoff functions.
We specialize to covariantizations that are minimal
(in the sense of appendix A of ref. \ymi), such as but not exclusively 
\pourchoice, and to
cutoff functions where the scale is set by the momentum if it is much
greater than $\Lambda$ which implies here that they decay as a power for 
large momenta (\cf again appendix A \ymi), \ie are of form 
\powlaw\  
for large $x$.\foot{\ie in the sense that $x^rc(x)$ and $x^\rt\ct(x)$ 
attain finite non-zero limits as $x\to\infty$.}
The conditions that $c$ and $\ct$ decay,  and \ctlim, imply
\eqn\conds{r>0\ins11, \rt>0 \ins11{and} r>\rt-1\quad.}
In this section we prove the following theorem, which relies also on
the more stringent conditions: \eqna\condr
$$\eqalignno{\Delta r &>{D-2\over2} &\condr a\cr
\noalign{\hbox{where we set $\Delta r:=r-\rt$, and}}
\rt &>{D-2\over2}\quad. &\condr b\cr}$$
\proclaim Theorem 1. Conditions \conds\ and \condr{}\ are necessary
and sufficient (at $N=\infty$) to ensure that
the one-loop corrections in the
flow equation \pRG, with any number of
external $\A$ fields and no external $\B$ or $\C$ fields,
are finite.\par

Note that by replacing $D$ in \condr{}\ by $D+\Delta$ we can thus
obtain for the one-loop momentum integral over $k$,
any degree of convergence $\sim k^{-\Delta}$  that we desire.
In fact, as we discuss later, for those
one-loop corrections other than the vacuum amplitude (\ie for those
corrections containing at least one external $\A$), we can relax \condr{}\
to \eqn\condrlx{\Delta r>{D-4\over2}\ins11{and}\rt >{D-4\over2}\quad.}
In particular in four dimensions, $r>\rt>0$ is sufficient for our purposes.

Clearly the cases considered are sufficient for the computation of
physics to one loop. For compactness we will refer to  contributions
without external $\B$s or $\C$s as {\sl BC bereft}. Note that the 
supertraces will generically still contain trapped $\sigma_3$s as well as 
external $\A$ fields. It may be possible to lift the restriction 
on no external $\C$s, which is made here for convenience. On the other
hand we show at the end of this section that in the present formulation,
there are one-loop corrections
with external $\B$s that diverge, and there are divergent diagrams beyond
one loop.


Before launching into the proof proper, we sketch the main reasons for
finiteness. This in turn helps to explain why the flow equation \pRG\
takes the form it does. See also ref. \alg. Again, it is helpful 
to refer to the
three terms in \pseed\ as $\hS=\hS_A+\hS_B+\hS_C$, respectively.
It is also helpful to refer to the different terms in \pRG\ by their
covariantized kernels, 
\ie as the attachment of the associated
wines $\{c'\}$, $\{L\}$, $\{K-c'\}_\A$ and $\{M-L\}$.
(Thus here we will not use the fact that the 
covariantization is linear \eg $\{M-L\}=\{M\}-\{L\}$ which implies 
the $\C$-$\C$ attachment is 
actually via $\{M\}$).

Apart from ensuring that the high energy behaviour of the Feynman rules
is as expected by the $SU(N|N)$ properties (this is discussed in the
next three paragraphs below), there are three main mechanisms for
finiteness which are essentially very simple:  One is the supertrace
mechanism that we have already discussed in sec. 2.  Another is
`drifting' \alg:  many potentially distastrously divergent
contributions following from attaching $L$ in \pRG\ via the $\AA$
differential in $\l$ on or near $\hS_A$, disappear by gauge invariance.
These two mechanisms are sufficient to ameliorate the problems
\oneslip\ caused by covariantizing the higher derivatives.  The final
mechanism for finiteness is simply the existence of the higher
derivatives themselves. As we will see the proof of finiteness is
nevertheless sufficiently involved to make a more mathematical style
convenient. We expect that much simpler proofs will be possible in a
manifestly local $SU(N|N)$ framework.

We choose the kernels \KLM\ so that the kinetic terms (bilinears in the
fields) in $\hS$ and $S$ can coincide as in \twop, and thus lead to
the simplifications we saw in sec. 7, mimicking ref. \ymi.  This is
all the more desirable in the PV sector because we want the Fermi-Bose
cancellations that regularise diagrams with $\hS_{\mu\nu}$ in, to work
also on those diagrams with $S_{\mu\nu}$.  
In fact the problem would be
worse than this because when the two-point functions do not coincide,
and thus the cancellations \ymi\  described in sec. 7.2 do not take place,
the higher point functions of $S$ pick up integration factors that can
radically alter their ultraviolet properties. With the cancellations
described in sec. 7.2, diagrammatic contributions may be immediately
integrated with respect to $\Lambda$, which means in particular that
their ultraviolet properties follow more or less straightforwardly
from the component vertices,
just as they do in the usual application of Feynman rules.
(Their leading ultraviolet properties 
are the same as those of the integrand providing that the coefficient 
of the leading ultraviolet behaviour of the integrand also converges
when integrated over $\Lambda$.)

At the same time however, we need the kernels themselves to have the
right normalisations to lead to cancellations at high momentum just as
the propagators do (as in ref. \alg, or via the spontaneously broken
$SU(N|N)$ interpretation).  Since the kernels are essentially the
$\Lambda$ differentials of these propagators, we need care in choosing
the powers of $\Lambda$ multiplying these propagators, equivalently the
powers of $\Lambda$ in front of the $\hS$ two-point functions in
\pseed, or again equivalently, the natural dimensions of the fields.
This requires the assignment of mass dimension one for $\B$ and the
less conventional choice of dimension zero for $\C$.

With these choices we have as required that the transverse
$\B$'s kernel $K$ and $\A$'s kernel $c'$ coincide at high energy,
and the longitudinal $\B$'s kernel $L$ and
$\C$'s kernel $M$ coincide at high energy. Indeed
from \conds\ and \KLM,
\eqn\lgx{ \eqalign{
K(x) &= c'(x)  \left\{1+O\left(1/x^{1+\Delta r}\right) \right\} \cr
L(x) &= (x\ct)'/x \left\{1+O\left(1/x^{1+\Delta r}\right) \right\} \cr
M(x) &= (x\ct)'/x \left\{1+O\left(1/x^{1+\rt}\right) \right\}
}}
This high momentum restoration of the unbroken $SU(N|N)$
clearly improves the ultraviolet behaviour of the last two terms
in \pRG. Meanwhile, quantum corrections following
from differentiating the full $\AA$, will
cancel by the supertrace mechanism
\psplit, unless a $\sigma_3$ is trapped in between.
A slightly more subtle supertrace mechanism works between $\B$ and $\C$
differentials of $\BB$ (reflecting the fact that in essence,
$\B$ ate ${\bf D}$).

As we have already mentioned, this supertrace mechanism
together with the drifting property cure the problems associated with
covariantizing the higher derivatives. Indeed, as we now show,
the conditions \condr{}\ are anyway required
to ensure that there are sufficient higher derivatives to regularise,
independently of the problems caused by covariantization.
(Note \conds, which is also required, has already been discussed in sec. 2.)

\proclaim Lemma 2. \condr{a}\ and \condr{b}\ are necessary for
ultraviolet finiteness.\par
\midinsert
$$
\epsfxsize=0.1\hsize\epsfbox{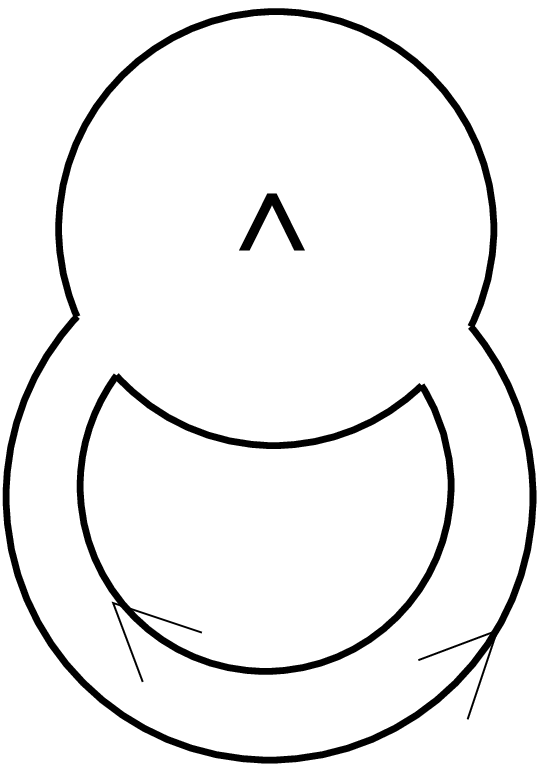}
$$
\centerline{{\bf Fig.14.}  Simple one loop diagram.} 
\endinsert

\proof\ Consider the vacuum contribution
to the simple one-loop vacuum diagram $a_1[\hS_C]$,
as illustrated in \fig\fsimp{Simple one-loop diag see p14 again},
where the wine is $L$.
Clearly there is no supertrace cancellation since there is
no analogous $\B$ term in \pseed.
By \powlaw\ and \lgx,
this contribution is UV finite if and only if \condr b\ holds.

Similarly, consider $a_1[\hS_B]$ where the kernel is $c'$ which we attach
to the lone $\B$ in the $\BB$s. Again there is no supertrace cancellation,
because there is no corresponding longitudinal $\A$ term in \pseed.
By \powlaw, this contribution is UV finite if and only if \condr a\
holds. $\lform$

Note that \condr{a}\ and \condr{b}\ are strictly necessary in the above
examples only for the vacuum contributions. Discarding vacuum energy contributions, the
first non-trivial contribution
in \fsimp\ must be the two-point contribution
since the one-point contribution vanishes by $\tr A^i=0$ (as well as many
other reasons \ymi).
By secs. 4 and 5.5 (or more generally appendix A of ref. \ymi), 
wherever the two points are placed, the large
momentum behaviour is improved by a power of two. Therefore in \condr{}
we can replace $D$ by $D-2$, obtaining \condrlx. By keeping careful
track of where the external points must be, we can prove these more
relaxed conditions for the general contributions, but we will not pursue
it further here:
the important point we wish to establish is that conditions {\sl do}
exist that guarantee finiteness for all physically relevant
one-loop graphs.

We have already shown that there are no terms of the form of \ftail.
Therefore a one loop graph is formed only by attaching wines to
{\sl other} wines, or lobes $\hS$, as shown in
\fig\foneloop{general form of one loop graph}.
\midinsert
$$
\epsfxsize=0.4\hsize\epsfbox{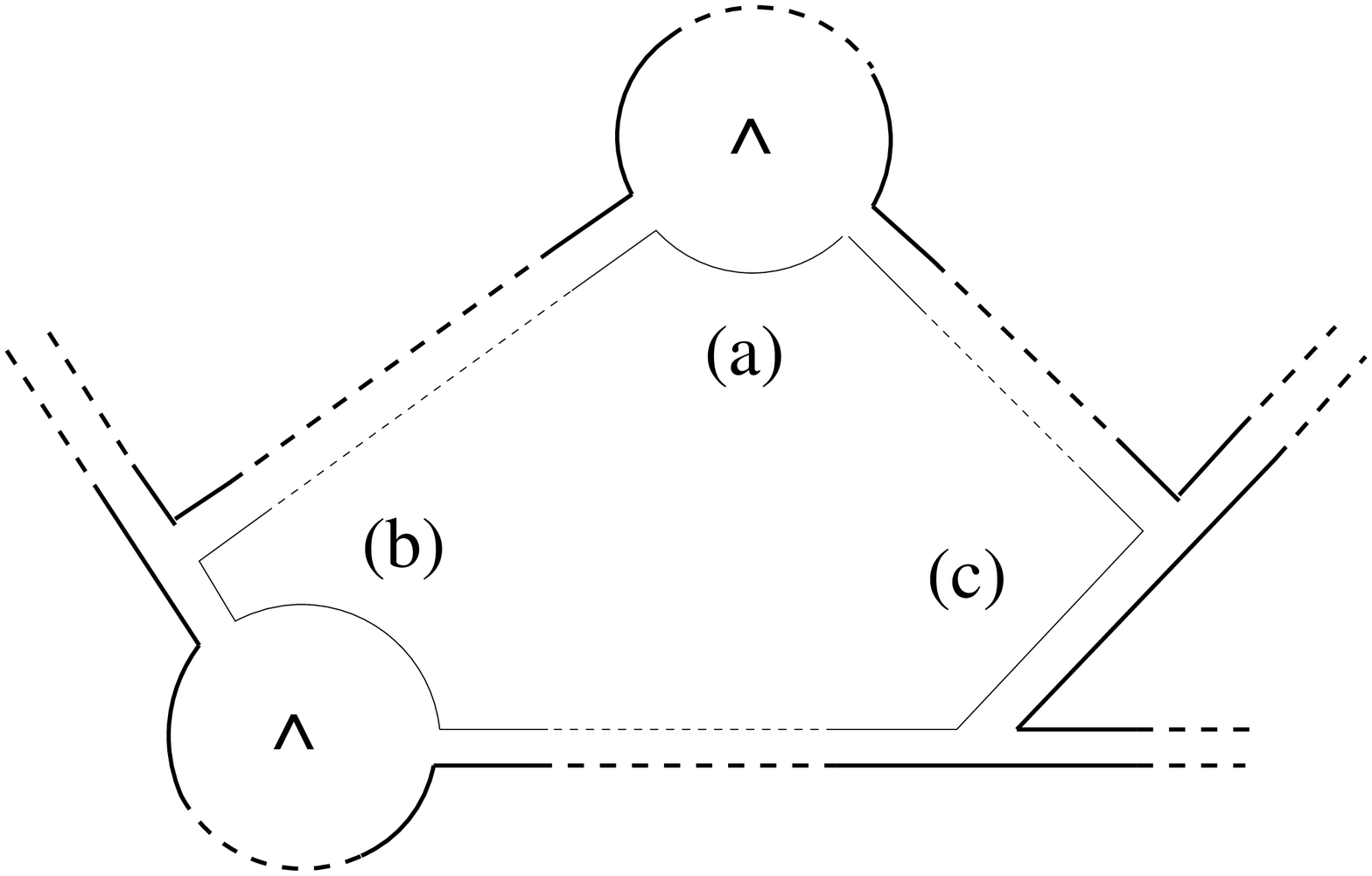}
$$
\centerline{{\bf Fig.15.}  General form of the one loop diagram.} 
\endinsert
\noindent Here we have used the fact that
one of the two superloops must be field free to survive the large $N$ limit.
Without loss of generality, we take it to be the inner Wilson loop.
It follows that the inner loop also grows no tree corrections,
because tree corrections have at least one lobe with
only one wine attached, and these lobes must have at least one external
field because there are no $\hS$ one-point vertices.

As illustrated in \foneloop, the one loop contribution
will have one or more of the following features:
\item{(a)} The loop momentum $k$ is carried directly into and
out of a lobe by wines. The lobe $\hS$ can have any number of external fields
on the outer Wilson loop and any number of tree corrections branching off it.
\item{(b)} As in (a) except that
one or both wines carrying $k$, first attach to another wine,
the rest of which is attached to some tree correction.
\item{(c)} The loop momentum passes through three wines (in one
of several ways) with no intermediate lobe.

At first sight it is easy to construct a divergent one-loop diagram:
we use the $\hS_A$ terms for the lobes and $\{L\}$ for the wines. By
\conds\ or Lemma 2, the $k^{-2\rt}$ is in general no way sufficient
to soften the divergent $k^{2r}$ behaviour in $\hS_A$. However,
the $L$ attachment must be made via the
$\nabla_\mu\cdot\delta/\delta\AA_\mu$ part of $l$, there being no
$\C$ terms in $\hS_A$ or the wines (where
attachment may be made as in \foneloop b). Gauge invariance then implies
drifting which as we will see, entirely cancels all such dangerous
contributions (in BC bereft one-loop diagrams).

Thus, consider an $\{L\}$ wine attaching to a pure $\AA$
section of a superloop. The attachment takes the form \eqnn\drift
$$\eqalignno{\int\!\!d^D\!w\ \str\!\!\left[\X(w)\nabla_\mu\cdot{\delta\over
\delta\AA_\mu(w)}\right]\,\Phi_{xy}
&= -\int\!\!d^D\!w\ \str\!\!\left[\nabla_\mu\cdot\X(w){\delta\over
\delta\AA_\mu(w)}\right]\,\Phi_{xy}\cr
&=i\left\{\Phi_{xy}\,\X(y)-\X(x)\,\Phi_{xy}\right\}\quad, &\drift}$$
where $\X(w)$ contains $\{L\}$ and all that it attaches to at its other
end, $\Phi_{xy}[\AA]$'
is the pure gauge section, with ends at $x$ and $y$,
and the last line follows since $\Phi_{xy}$ transforms homogeneously
under gauge transformations \gatr. The interpretation in terms of
drifting, or sliding,
is now clear from \fig\fdrift{drifting or sliding L. Interpretation
of \drift.}.
\midinsert
$$
\epsfxsize=0.6\hsize\epsfbox{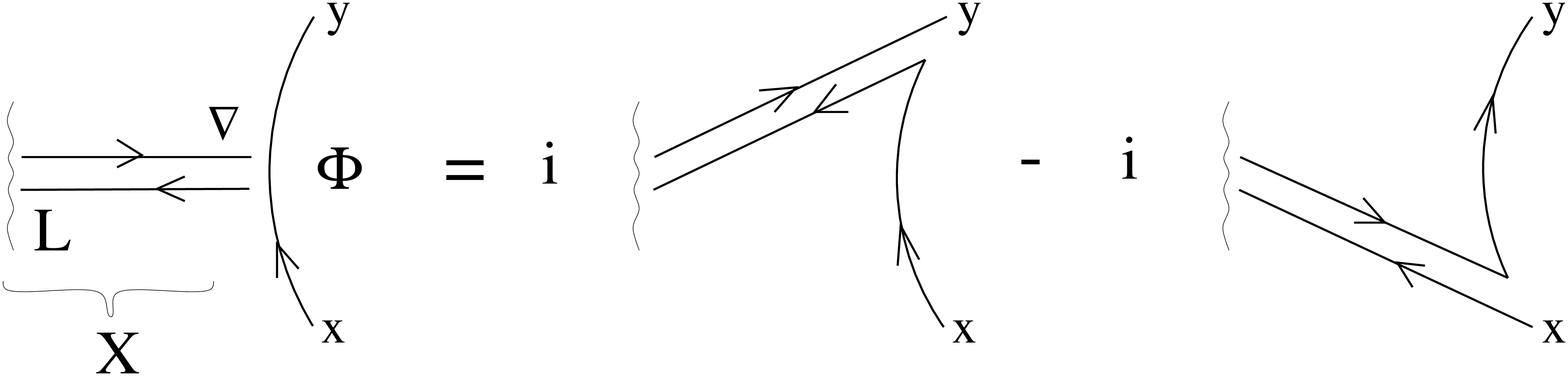}
$$
\centerline{{\bf Fig.16.}  Drifting or sliding $\{L\}$.} 
\endinsert

The obstructions at the ends of $\Phi_{xy}$ are $\sigma_3$s,
$\C$s, explicit $\B$s,\foot{which may be regarded as arising from $\sigma_3$s
by writing $\B=\di-\, \AA$, \cf \proj} or `remaindered' $\B$s which
arise in the following sense. The pure gauge section may
depend only on $\A$ rather than the full supermatrix $\AA$,
as in  $\{K-c'\}_\A$ of \pRG. Drifting still occurs by the first
gauge transformation relation \gatr, but because $\{L\}$ attaches
with the covariant derivative $\nabla=\d-i\B$, we obtain
remainder terms with a $\B$ at the join. Equivalently we can
write the pure $\A$ section as $\Phi_{xy}[\AA-\B]$, so that
$\Phi_{xy}$ expands into a $\Phi_{xy}[\AA]$
plus a series of smaller pure $\AA$ sections with remainder $\B$s
at one or both ends.

If we replace the section $\Phi_{xy}[\AA]$ in \drift\ by a full
pure gauge Wilson loop $\phi[\AA]$,\foot{Recall that we can think of
these pure gauge contributions as fluctuating Wilson loops, integrated
over with some suitable measure \ymi.} then since this is gauge invariant,
we find that attaching $\{L\}$ just results in zero. This is equally clear
from \fdrift\ by sowing the two ends of $\Phi_{xy}$ together, \ie
by setting $x=y$ and taking a supertrace.

$\{L\}$ can drift of course across a
join between two sections, say $\alpha$ and $\beta$,
in a composite Wilson loop (\eg drifting off a wine and onto
a lobe). We can see this either by treating the two sections together
as one pure gauge section $\Phi$ in \drift, or more carefully as follows.
We can add together the diagrams with $\{L\}$ attached to
$\alpha$, and with $\{L\}$ attached to
$\beta$. By \drift\ and \fdrift, the drift to the join from $\alpha$
cancels the drift to the join from $\beta$.

Note however: we are making an assumption when sliding $\{L\}$ across a
join, that the new diagram actually exists, \ie can be constructed
from \pRG. In \fig\fspecial{Special case
cf. p14, L touches an $\hS_B$} we show an exception. 
\midinsert
$$
\epsfxsize=0.25\hsize\epsfbox{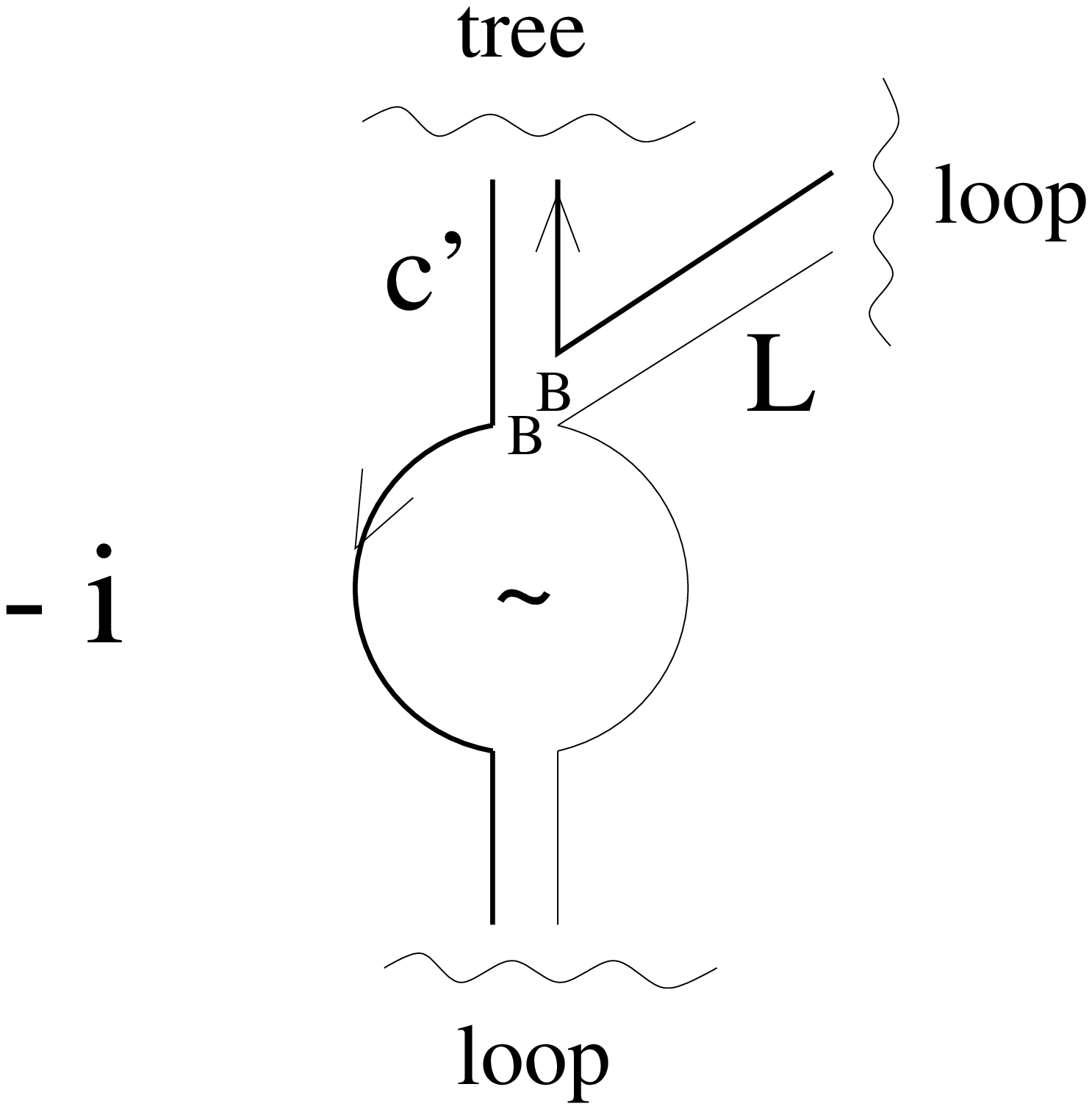}
$$
\centerline{{\bf Fig.17.}  A diagram with no external $\B$s
where $\{L\}$ cannot slide across the join.} 
\endinsert
\noindent We take the diagram to have no external $\B$s. The lobe is
$\hS_B$ and the $\{c'\}$ wine attaches by differentiating the lone $\B_\mu$ in
$\BB_\mu$.
The $\{L\}$ attaches to $\{c'\}$, absorbing the $\B$ that has to be there
by fermion number \fermin, and then slides back to the join.
The corresponding diagram with $\{L\}$
attached to $\hS_B$, but all other factors the same,
does not exist however: since $\{c'\}$ and above is $\B$-free,
in the corresponding diagram $\{c'\}$
must attach by differentiating an $\A_\mu$ in the same place as the lone
$\B_\mu$ as illustrated in \fig\fnonexist{non-existing cancelling diag
see p14}, but there is no such $\A_\mu$ term. 
\midinsert
$$
\epsfxsize=0.25\hsize\epsfbox{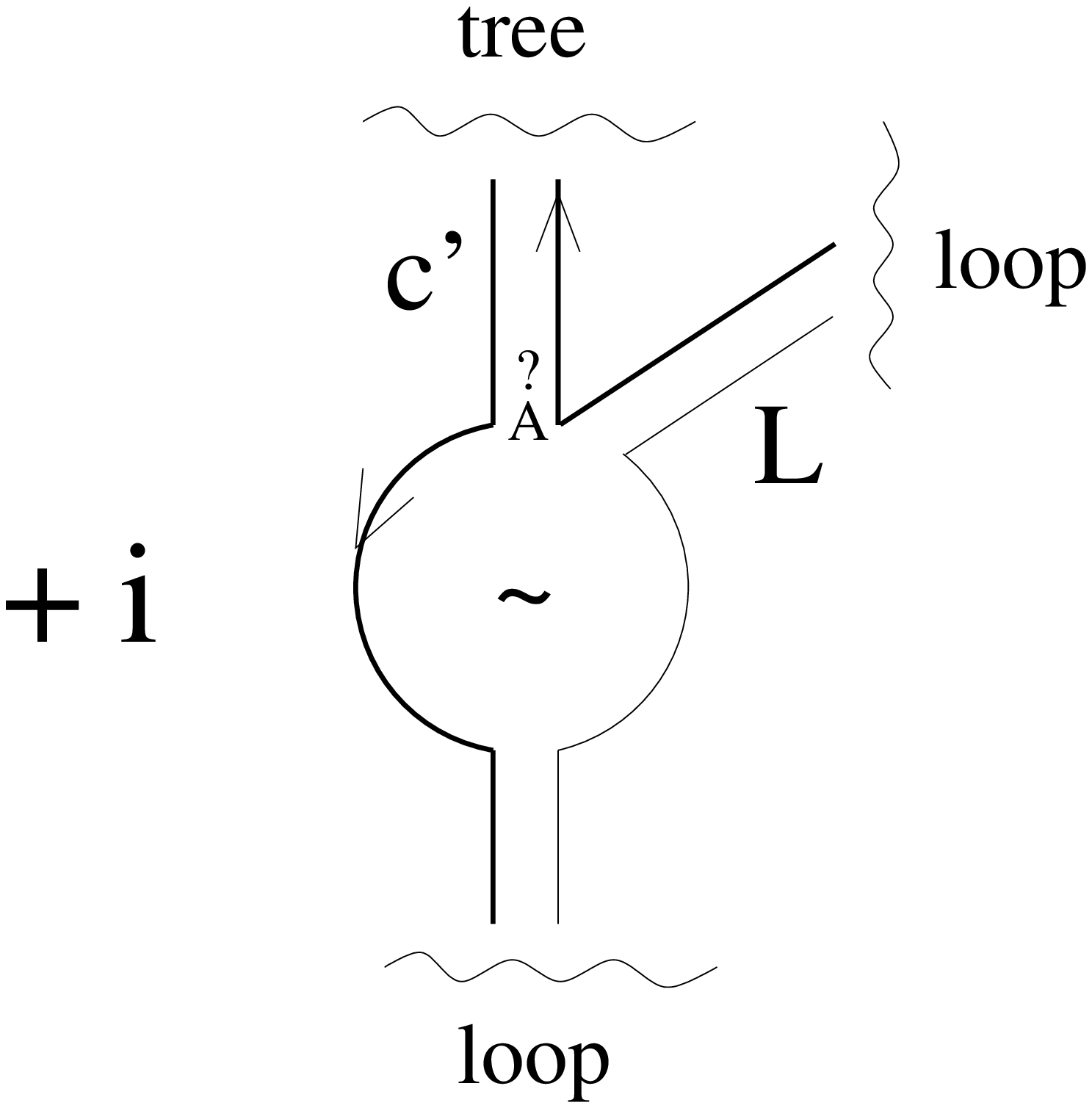}
$$
\centerline{{\bf Fig.18.}  A diagram with no external $\B$s 
and drifted $\{L\}$, that does not exist.} 
\endinsert
\noindent (Such a longitudinal $\A_\mu$
term is not allowed by gauge invariance.)
Fortunately, as we see in this example, firstly the problem only arises
where a bosonic attachment must change into a fermionic one (or \vv)
as a consequence of sliding a $\delta/\delta\B$ across the join, and
secondly the problem is always flagged by the appearance of $\sigma_3$s
at the join expressing the fact that the attachment is only via a partial
supermatrix as in \fsig.

Collecting the observations above, we thus obtain the
following lemma.

\proclaim Lemma 3.
\item{(a)} $\{L\}$ wines drift to the ends of
pure $\AA$ sections of the superloop.
\item{(b)} $\{L\}$ wines cannot attach to pure $\AA$
Wilson loops.
\item{(c)} $\{L\}$ wines drift to the ends of
pure $\A$ sections up to terms involving remaindered $\B$s.
\item{(d)} $\{L\}$ wines
attach to pure $\A$ Wilson loops, in particular BC bereft
tree-level contributions, only if accompanied by remaindered $\B$s.\par

\proof\  We only need note that in (d)
a (BC bereft) tree-level contribution contains no $\sigma_3$s since they
occur in pairs and thus
may be commuted past $\A$s and combined together until they all
disappear (\cf \fsig and sec. 7), and $\{L\}$ has to attach by a bosonic 
differential since there are only $\A$s to differentiate, so  diagrams
exist for $\{L\}$ attaching at all points on the tree level contribution.
$\lform$

\proclaim Lemma 4. There are no tree-level contributions with just
one $\B$ or one $\C$.\par

\proof\ We stress that this holds, whatever the number of $\A$s.
That there are no one-point $\B$ diagrams is obvious,
from the fermion number conservation \fermin\ already mentioned. If
an external $\C$ appears it must do so in $\hS_B+\hS_C$. But then
together with any number of $\A$s, it accompanies in the $\hS$ lobe,
either two $\B$s which at tree level cannot be
removed by further attachments (again by fermion number), or another
$\C$. This other $\C$ can be propagated to other similar lobes by $\{L\}$
or $\{L-M\}$ in \pRG\ where it may be turned into pairs of $\B$s
(which again translate to external $\B$s),
but it cannot disappear in any other way except apparently
by attaching $\{L\}$
where we use the $\C$ differential in $\l$,
and at the other end we use the
$\nabla_\mu\cdot\delta/\delta\AA_\mu$ differential. However, this latter
must attach to a BC bereft tree-level contribution
and produce no remainder $\B$s, which is
excluded by Lemma 3(d). $\lform$

\proclaim Corollary 4. (a) BC bereft trees and BC bereft tree corrections
are made only of lobes $\hS_A$ and wines $\{c'\}$.
(b) In a BC bereft one-loop diagram
$\{K-c'\}_\A$, $\{M-L\}$ and $\{L\}$ must appear entirely within the loop
\ie such that the loop momentum $k$ goes in one end of the wine and out
the other, while $\hS_B$ ($\hS_C$) must have the loop momentum enter and 
leave via the lone $\B$ or the $\C$ in
$\BB$ ($\C$).\par

\proof\ (a)
For a BC bereft tree or BC bereft tree correction,
if there existed $\hS_B$ or $\hS_C$, then breaking open the diagram
by removing the wine connection to the $\BB$ or $\C$ respectively,
results in a tree with just one $\B$ or $\C$ in violation of Lemma 4.
Similarly if the BC bereft tree (correction)
contained $\{K-c'\}_\A$, $\{M-L\}$ or $\{L\}$,
then breaking open the diagram at the (tree)
end of one of these wines we obtain a
tree with one $\B$ and no $\C$s, one $\C$ and no $\B$s,
and either one $\C$ and no $\B$s or neither $\B$s nor $\C$s, respectively.
All these possibilities are excluded by Lemmas 4 and 3(d).

(b) For the wines this is just a restatement of (a).  For
$\hS_B$ and $\hS_C$, if the loop momentum did not enter and leave
in this way we would have to have an external $\B$ or $\C$
since by Lemma 4, they
cannot be absorbed by attaching a tree correction. $\lform$

\midinsert
$$
\epsfxsize=0.25\hsize\epsfbox{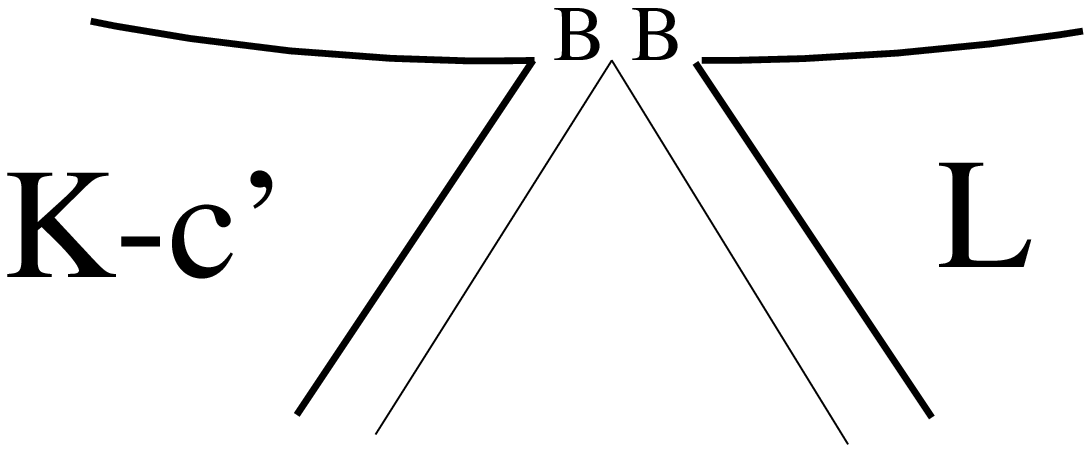}
$$
\centerline{{\bf Fig.19.} An $\{L\}$ attached to $\{c'\}$ or $\hS_A$
which has drifted to touch a $\{K-c'\}_A$ wine.}
\endinsert
\proclaim Proposition 5. In a BC bereft one-loop diagram, $\{L\}$ can
only attach either:
(a) directly to the lone $\B$ or $\C$ in $\BB$ of $\hS_B$, 
or directly to the $\C$ in $\hS_C$,
(b) to $\{c'\}$ as in \fspecial, or
(c) to $\{c'\}$ or $\hS_A$ as in \fig\ffinite{cd. p14 notes for paper,
L attaches to K-c' with blob on top}.\par

\proof\ By Corollary 4(b), if $\{L\}$ attaches to a lobe it either does
so as in (a), or attaches to $\hS_A$. However,
it will drift off $\hS_A$ unless it meets an obstruction which may be regarded as being
caused $\sigma_3$ (\cf discussion
above Lemma 3). This in turn would arise only from
a wine attached to $\hS_A$ by a partial supermatrix derivative, uniquely
specifying the attachment of $\{K-c'\}_\A$ and hence \ffinite.
Otherwise $\{L\}$ must attach to a wine, which by Corollary 4(b)
and \foneloop, must be $\{c'\}$. If $\{L\}$ is not to drift off $\{c'\}$,
it must meet an obstruction which can only be due to
either a partial supermatrix
derivative attachment, again giving uniquely $\{K-c'\}_\A$ as in
\ffinite, or $\{c'\}$ attaching to a partial supermatrix, which
uniquely specifies \fspecial. (By Corollary 4,
$\{c'\}$ cannot attach to $\{K-c'\}_\A$
since one end of the latter would be outside the momentum loop.) $\lform$

Note that (since $\{L\}$ is entirely inside the loop)
if $\{L\}$ fails to find any obstruction then it will slide up
its own tail as in \ftail, which we saw in sec. 2 gives zero. Equivalently,
if it fails to find an obstruction then there are no trapped $\sigma_3$s,
but since $\{L\}$ attaches by full supermatrix differentials, the result
vanishes by the supertrace mechanism.

In \ffinite, by Corollary 4(b), the loop momentum enters and
leaves via $\{L\}$ and $\{K-c'\}_\A$. Since these
two wines meet at the same point, the pure gauge vertex above
(\viz $\hS_A$ or $\{c'\}$) carries no loop momentum. Similarly
in \fspecial, by Corollary 4(b),
the loop momentum enters and leaves through $\{L\}$
and $\hS_B$, and $\{c'\}$ carries no loop momentum. We see that
drifting has entirely removed the dangerous combinations of $\{L\}$
and $\hS_A$ discussed above \drift, as promised.

By \foneloop, there are at least as many wines carrying loop
momentum as there are lobes carrying loop momentum. The insertion
of extra wines as in \foneloop b\ and \foneloop c, cannot lessen
the superficial\foot{\ie as established by power counting
and not taking into account possible cancellations} convergence. 
To be more precise, note that in the case where there
is only one lobe carrying loop momentum (for example as in \fsimp),
there is always at least one `internal'
wine, in the sense of an internal propagator,
\ie a wine carrying loop momentum $k$ from end to end.
In a one-loop diagram with more than one $k$ carrying lobe,
there is always at least one such internal wine
between any adjacent pair of such lobes. 
These internal wines are therefore at least equal in number to
the lobes carrying $k$. The extra wines, over and above these
internal wines, would be
only partially within the loop, but by \endtopj\ and \bysidee, 
for large momentum $k$ they contribute at worst $\sim k^0$. 

Consider a $k$-carrying lobe and all the wines on one side
of the lobe that $k$ passes through until the next $k$-carrying lobe
is reached (or the same lobe is reached, if there is only one $k$-carrying
lobe in the loop). Such a combination forms a dressed internal tadpole,
which we will refer to simply as a tadpole. Note that such a tadpole
thus always includes at least one internal wine.
Every one-loop diagram can be split up (in one of two ways) into a set 
of disjoint tadpoles as illustrated in \fig\ftad{One loop diag split
into tadpoles}.
\midinsert
$$
\epsfxsize=0.3\hsize\epsfbox{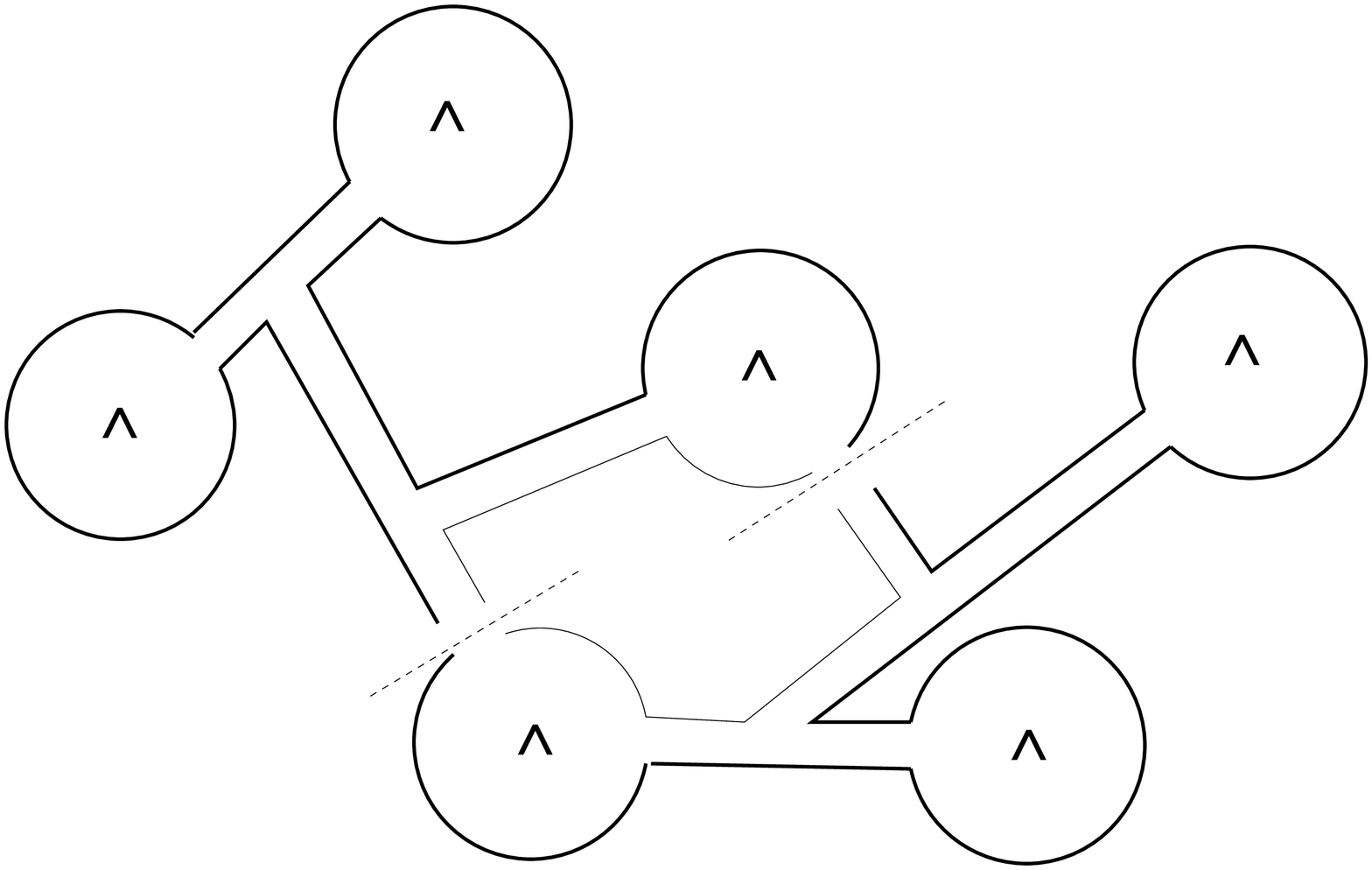}
$$
\centerline{{\bf Fig.20.} An example one-loop diagram split into
two tadpoles by the dashed lines.}
\endinsert

For a given tadpole, and large $k$, 
let $k^{2p}$ be the power contributed by the 
$k$-carrying lobe, and let
$k^{-2w}$ be the total power of $k$ contributed by the wines
(again as established by power counting).
We now use \powlaw\ and the results of sec. 5.5 (more generally appendix A of \ymi), 
to compute upper
bounds for the tadpole's superficial degree of divergence $2(p-w)$,
for each choice of lobe and choices of internal wine.

\item{(i)} Suppose that the 
$k$-carrying lobe is $\hS_A$, then the maximum $p$ that
this can contribute is $p=r+1$. 
\itemitem{(ia)} $\{M-L\}$ cannot be
an internal wine in this tadpole
since it joins directly to $\hS_B+\hS_C$
by \pRG\ and Corollary 4(b). 
\itemitem{(ib)} If $\{K-c'\}_\A$ appears as an
internal wine then by \lgx, 
\eqn\kc{p-w\le-1-\Delta r\quad.} 
\itemitem{(ic)} If $\{L\}$ is an internal wine in this tadpole
then so is $\{K-c'\}_\A$ by Proposition 5, \ffinite\ and Corollary 4(b).
({\it N.B.} As noted above, the vertex shown above
$\{K-c'\}_\A$ and $\{L\}$ in \ffinite\ {\sl is not}
carrying $k$.) Therefore \kc\ at least applies.
\itemitem{(id)} Finally $\{c'\}$ could be an internal wine, 
thus contributing a minimum of $w=r+1$, and hence $p-w\le0$.

\item{(ii)} Now suppose that the $k$-carrying lobe is $\hS_B$. 
This will contribute a maximum $p=\rt$ if attached to the loop by the
$\B$s, or $p=\rt+1$ if attached to the loop via the $\C$s. 
\itemitem{(iia)} If $\{c'\}$
is an internal wine in this tadpole then by Corollary 4(b)
(and Lemma 4), the $\hS_B$ must 
attach via the $\B$s and $p-w\le-1-\Delta r$. 
\itemitem{(iib)} Similarly if $\{K-c'\}_\A$
is an internal wine, then by \lgx, $p-w\le-2-2\Delta r$.
\itemitem{(iic)} If $\{M-L\}$ appears as an internal wine
then this attaches to a $\C$ in $\hS_B$, and 
by \lgx, $p-w\le-1-{\rm min}(\rt,\Delta r)$.
\itemitem{(iid)} Finally if $\{L\}$ is an internal wine 
in this tadpole then
either it can attach directly to the $\C$s in which case $p-w\le0$,
or attach via the $\B$s in which case again $p-w\le0$.

\item{(iii)} Finally, suppose that the $k$-carrying lobe
is $\hS_\C$. This gives $p=0$. The least convergent internal wine is 
$\{L\}$, thus we have that whatever internal wines there are,
$p-w\le-1-\rt$.

Collecting these observations together we have

\proclaim Proposition 6. In a BC bereft one-loop diagram,
all tadpoles have $p-w\le0$.\par
\proof\ The result follows trivially from the above observations on using
the second two conditions of \conds. $\lform$

\proclaim Proposition 7. Any BC bereft one-loop diagram containing 
a $\{K-c'\}_\A$
wine, or $\{M-L\}$ wine, 
or $\hS_B$ and an internal $\{c'\}$, or $\hS_C$, is finite.\par
\proof\ If $\{K-c'\}_\A$ appears in the diagram, it does so as an
internal wine by Corollary 4(b). Splitting the diagram into tadpoles,
we have by  (ib), (iib) and (iii) above, and Proposition 6, that the
degree of divergence of the diagram is no worse than $D-2-2\Delta r$, 
and thus the diagram is finite by \condr{a}.

Similarly if $\{M-L\}$ appears, it also must appear as an internal wine.
Then by (ia), (iic), (iii) and Proposition 6, the degree of divergence
of the diagram is no worse than $D-2-2{\rm min}(\rt,\Delta r)$,
and thus by \condr{}\ the diagram is finite.

If $\hS_C$ appears, it must do so as a $k$-carrying lobe by Corollary
4(b). Splitting the diagram into tadpoles, we see from Proposition 6 and
(iii) above, that the diagram is finite by \condr{b}.

Finally, if $\hS_B$ appears, it must do so as a $k$-carrying
lobe by Corollary 4(b). If an internal $\{c'\}$ also appears in the loop
then either it or another $\{c'\}$ 
may incorporated in a tadpole with the $\hS_B$, in which case the
diagram is finite by Proposition 6, (iia) and \condr{a}, or else the internal
$\{c'\}$s must only lie between adjacent 
$k$-carrying $\hS_A$s. In this case
the $\hS_B$ can be
incorporated in a tadpole with $\{K-c'\}_\A$, there being no other way
to attach $\hS_B$ to the $k$-carrying $\hS_A$s, as follows by 
Corollary 4 and Proposition 5.
But we have already shown that such a diagram is finite. $\lform$

By Proposition 7, and Corollary 4 and Proposition 5,
we are left only to prove the convergence
of diagrams either (a) consisting only of $\hS_A$ and $\{c'\}$, or (b)
where all the $k$-carrying wines are $\{L\}$s attaching to $\hS_B$s,
either directly to the $\BB$s, or as in \fspecial. We complete the 
proof of Theorem 1, by showing that
both these types vanish by the supertrace mechanism. As we will see,
this is obvious for case (a). In case (b) the supertrace mechanism 
has been obscured by spontaneous symmetry breaking, the unitary gauge,
and the `missing' $\sigma_3$ described below \bb.

\proclaim Proposition 8. Diagrams of type (a) or (b) above, vanish
identically.\par

In case (a) the attachments are all made with full supermatrix 
differentials. There are thus no $\sigma_3$s generated and
the internal Wilson loop in \foneloop\ is $\str 1=0$.

In constructing the diagram in case (b), if an $\{L\}$ is attached
to a full $\BB$ at tree level (\ie before the loop is closed) then the
attachment is of the form
\eqn\abb{\str\!\int\!\!d^D\!w\ \X\left({\delta\over\delta\C}
+\nabla_\mu\!\cdot\!{\delta\over\delta\AA_\mu}\right)\ \
\str\!\int\!\!d^D\!x\ \left(\B_\nu+\nabla_\nu\!\cdot\C\right)
\Y_\nu\quad,}
where we have written out $\l$ by \pRG\ and the
$\BB_\nu$ it attacks, using \bb. Here
$\X$ and $\Y_\nu$ stand for the rest of 
the trees on either side. By Proposition 5, we are interested only
in the case where the $\AA$ differential attacks the lone $\B_\nu$.
Thus using \proj\ and \bpsow,
the part of the above we are interested in, evaluates to
\eqn\lbb{-\,\str\!\!\int\!\!d^D\!x\,\left(\nabla_\nu\!\cdot\!\X\,\di-\!\Y_\nu
+\di+\X\,\nabla_\nu\!\cdot\!\Y_\nu\right)\,=\,
{1\over2}\ \str\!\!\int\!\!d^D\!x\, \left(
\nabla_\nu\!\cdot\!\X\,\,\sigma_3\!\Y_\nu\sigma_3-
\sigma_3\X\sigma_3 \,\nabla_\nu\!\cdot\!\Y_\nu\right).}
We see that we have exactly one $\sigma_3$ trapped on each side of the join
as in the second term of \fsig. As in \fsig, the same local process takes
place on supersplitting, thus
if the $\{L\}$ attachment
to $\BB$ closes the loop, 
a closely similar calculation shows that exactly
one $\sigma_3$ gets trapped in each Wilson loop.

For every occurence of \fspecial, there is a corresponding diagram
with attachments as in \fig\fveryspecial{p15 ..}, where $\{c'\}$ {\sl first}
attaches to the covariant derivative in $\BB_\nu$, \ie to the term
$i\,\C\AA_\nu$ (by \gatr, \bb\ and
the ordering implied in \fveryspecial).\foot{Contributions \fspecial\ 
and \fveryspecial\ 
were discussed as fig. 24 of ref. \alg.} 
\midinsert
$$
\epsfxsize=0.2\hsize\epsfbox{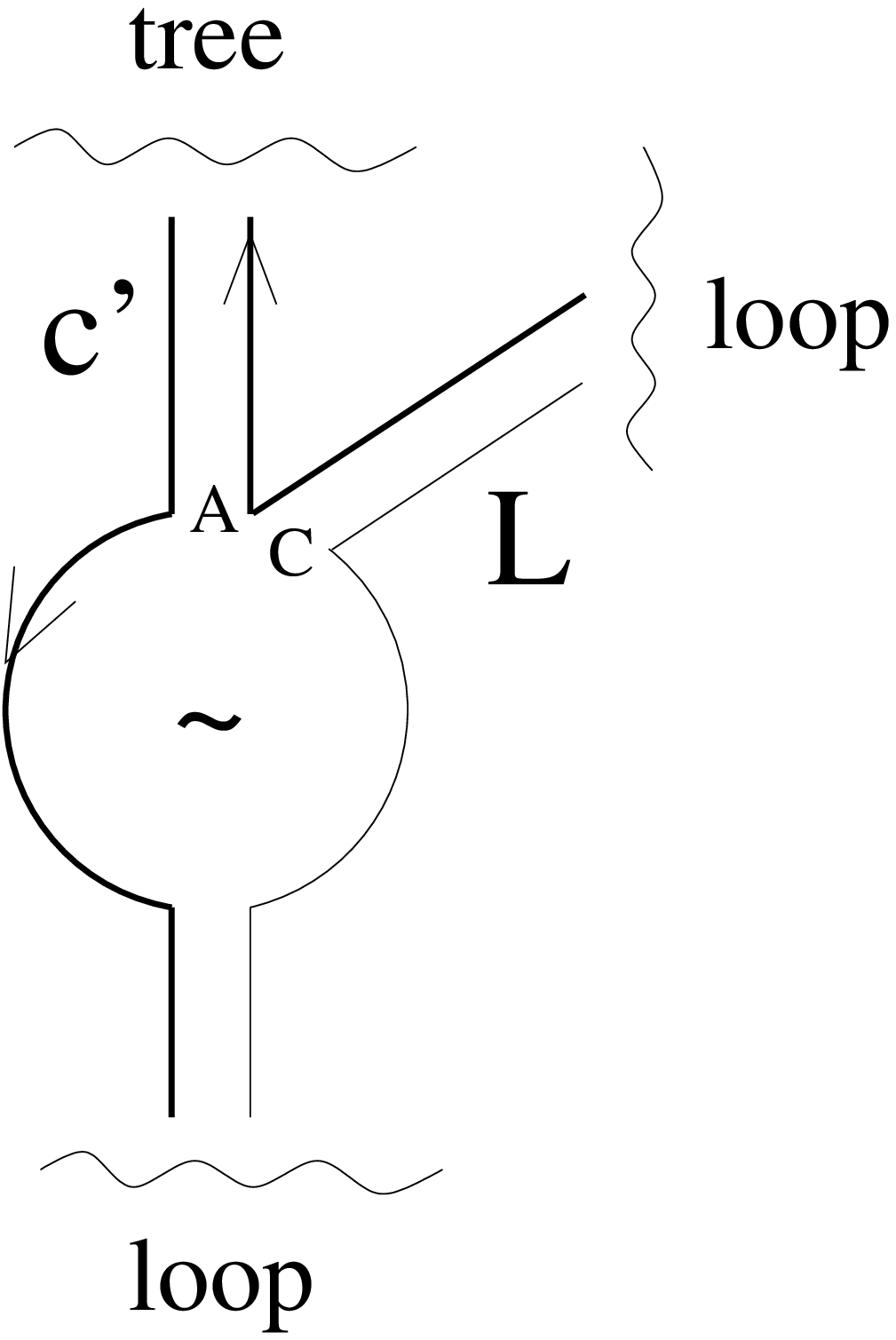}
$$
\centerline{{\bf Fig.21.} The partner diagram for Fig.17. }
\endinsert
\noindent Consider these attachments made
at tree level, with $\{L\}$ and everything it attaches to at its other
end being $\X(w)$, and the rest of $\hS_B$ and its attachments 
being $\Y_\nu(x)$,
as in \abb, and $\{c'\}$ and all it attaches to being $\ZZ_\mu$.
By the description above Lemma 3, \fspecial\ is given by 
$-i\,\str\!\!\int\!\!d^D\!x\,\,\X\ZZ_\nu\,\di-\!\Y_\nu$.
On the other hand by the description directly
above, \fveryspecial\ is given by
$i\,\str\!\!\int\!\!d^D\!x\,\X\,\di+\left(\ZZ_\nu\Y_\nu\right)$.
Recognizing that, by Corollary 4, for BC bereft diagrams $\di+\!\ZZ=\ZZ$,
these contributions sum to
$$i\,\str\!\!\int\!\!d^D\!x\,\X\,\ZZ_\nu\,\sigma_3\!\Y_\nu\sigma_3\quad.$$
As in \lbb, we have exactly one $\sigma_3$ trapped on either side
of the join. Also as above, a similar calculation for the case where \fspecial\
or \fveryspecial\ closes the loop, shows that exactly one $\sigma_3$ gets 
trapped in each Wilson loop in this case. Therefore we see
that in diagrams of type (b), an even number of $\sigma_3$s appear in the
inner Wilson loop (one for every corner or join), which thus  
again furnishes $\str 1=0$. $\lform$

\subsec{Divergent diagrams}
The above concludes the proof of finiteness of BC bereft one-loop diagrams.
We now show by example where the difficulties lie in extending the 
ultraviolet finiteness of the present
`unitary gauge' formulation beyond this class, in 
particular we provide examples of 
divergent one-loop diagrams with external $\B$s,
and divergent two-loop diagrams. 

The real difficulties appear to arise through the restriction to $\A$
of $\{K-c'\}_\A$ in \pRG.
Thus we readily obtain one-loop divergent diagrams with remaindered $\B$s
(\cf the Lemma 3 and the discussion before it) for example
\fig\frB{remainder B divergent diag p16}. Note that since the attachment
of $L$ to $\{K-c'\}_\A$ can only be made via $\A$, and the other end
can only attach via $\B$ of $\BB$ (by fermion number conservation given
that from \pRG, $\{K-c'\}_\A$ attaches via $\B$) there is no 
corresponding diagram that could cancel the divergence via the 
supertrace mechanism. Therefore some trapped $\sigma_3$s remain,
after expanding as in \fsig. Noting \endton\ and the Feynman rules,
one readily shows that \frB\ is quadratically divergent in $D=4$
dimensions (the integrand $\sim 1/k^2$ for large loop momentum $k$).
\midinsert
$$
\epsfxsize=0.3\hsize\epsfbox{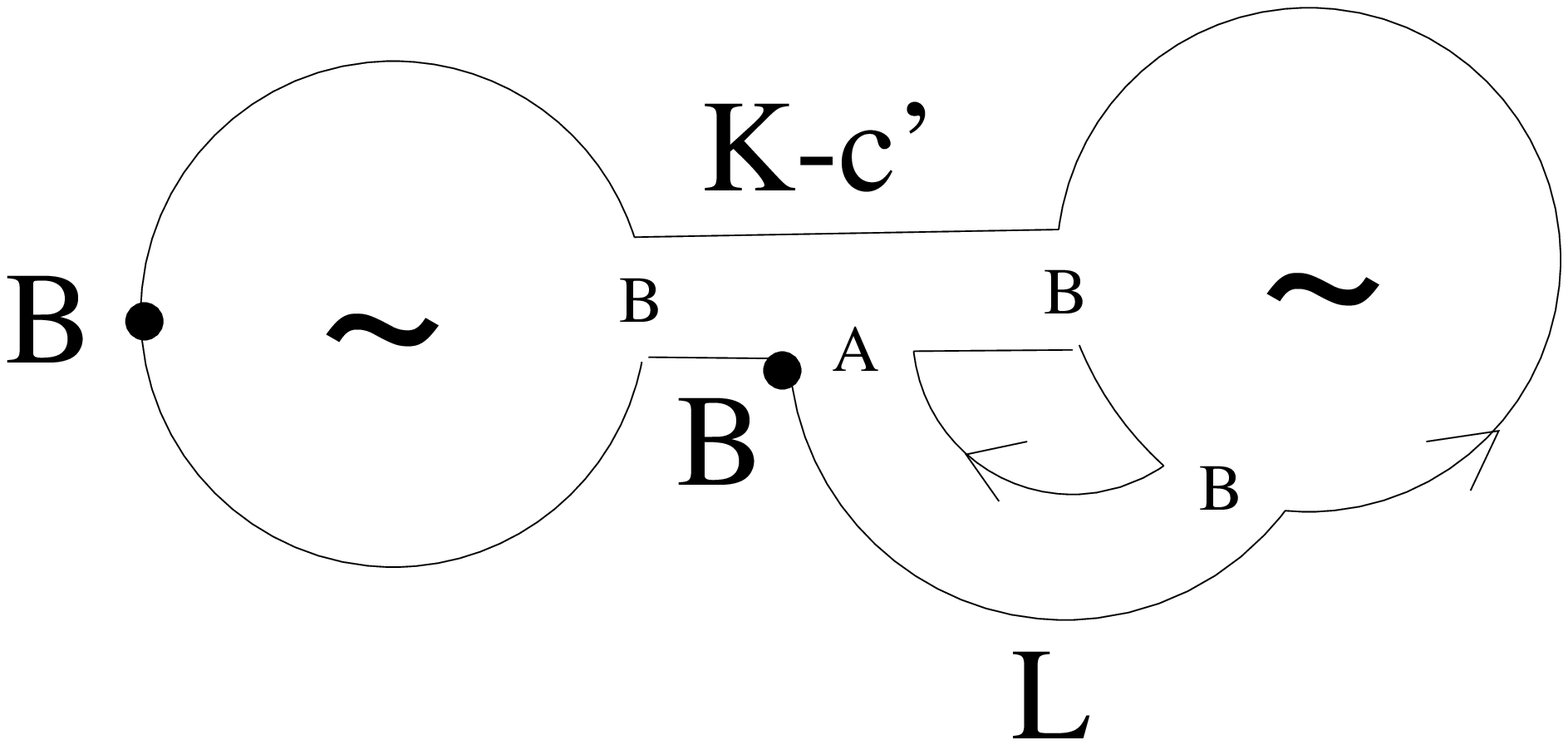}
$$
\centerline{{\bf Fig.22.} Divergent diagram with a remaindered $\B$. }
\endinsert

We see that the $\A$ restriction destroys the supertrace mechanism as
well as drifting. Another example is that of \fig\fB{c' example see
p16}.  Since the $c'$ attachment has to be made via $\A$s there is no
supertrace cancellation, and from the $\hS_A$ part of the RH lobe, we
have again that the integrand $\sim1/k^2$.  (Note that the external
$\B$s are forced via the $\B$ differential of the $\{K-c'\}_\A$
attachment.)
\midinsert
$$
\epsfxsize=0.3\hsize\epsfbox{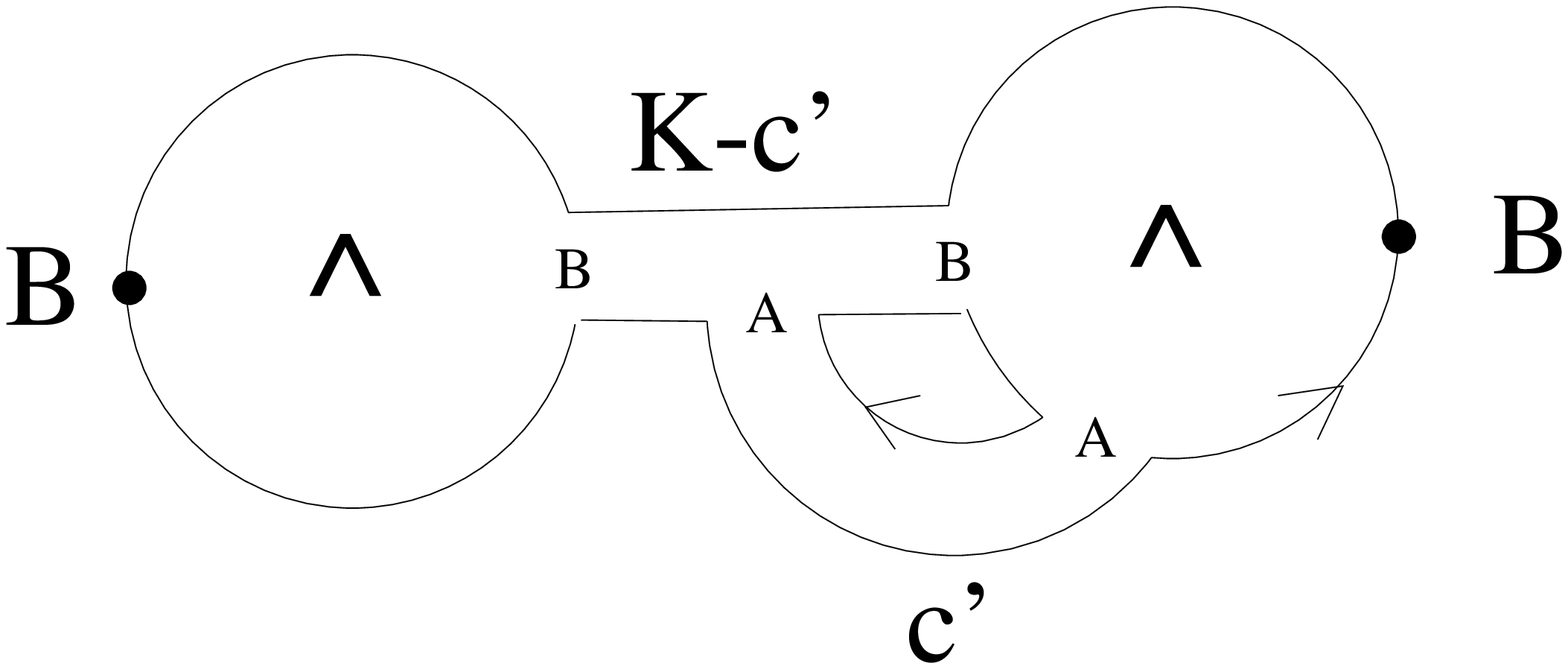}
$$
\centerline{{\bf Fig.23.} Divergent diagram through restriction of supertrace
mechanism.}
\endinsert

The problem is particularly severe beyond one loop. It is easy to give
severe cases with external $\B$s, but now also BC bereft diagrams are
affected, as illustrated in
\fig\ftwoloop{p16}. 
Here $\{K-c'\}_\A$ is attached to $\hS_A$ which is
finite, but then a two-loop diagram is made by attaching $L$. $L$s
ends will drift to the joins between $\{K-c'\}_\A$ and $\hS_A$ and
cancel the diagrams where $L$ attaches and drifts from 
the other side of the join.
However in the case that $L$ attaches via $\B$ this cancellation is
imperfect because the corresponding diagram does not exist for
$L$ attaching to $\{K-c'\}$. (Recall the arguments above Lemma 3.
Equivalently we can start by noting that when $L$ attaches exclusively
to $\hS_A$ the result vanishes by the supertrace mechanism.) 
The loop momentum routing round $L$ through $\hS_A$ gives a
violently divergent integral with an integrand 
that {\sl diverges} as $k^{2\Delta r}$.
\midinsert
$$
\epsfxsize=0.16\hsize\epsfbox{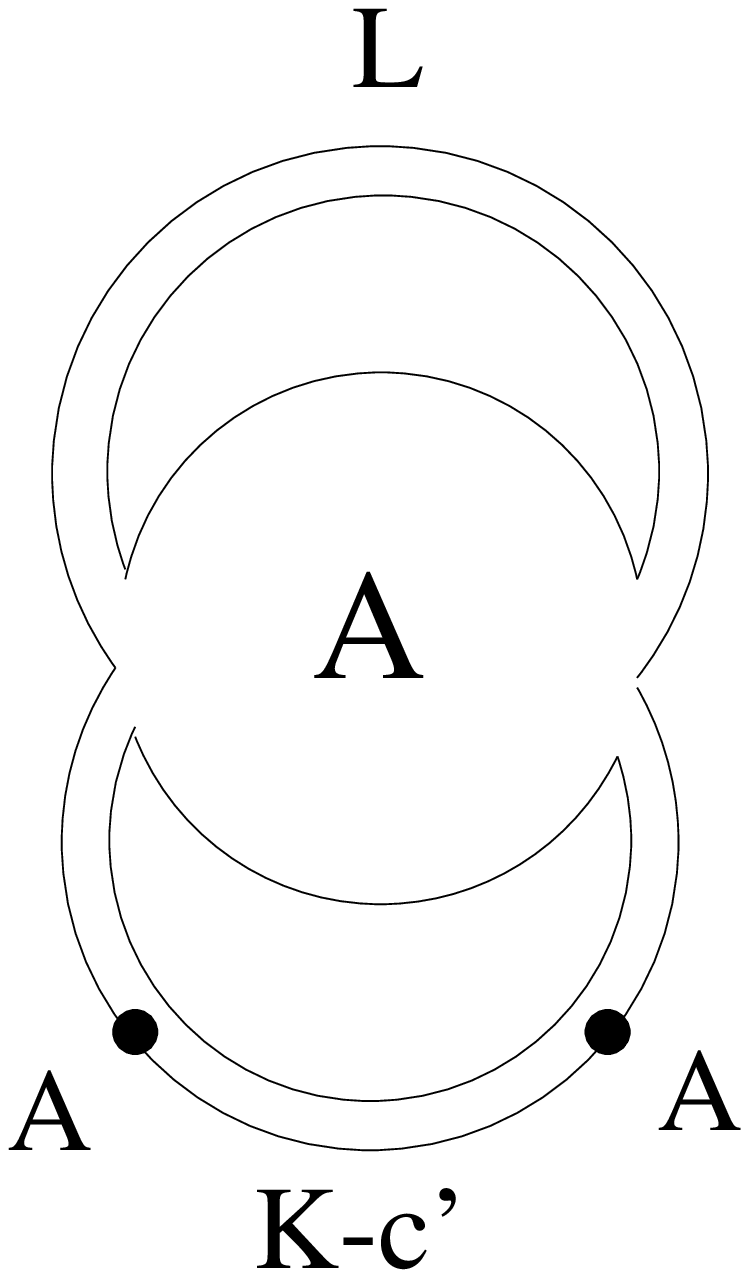}
$$
\centerline{{\bf Fig.24.} Divergent BC bereft two-loop diagram. }
\endinsert

We cannot remove the $\A$ restriction however since the $c'$ in \fspecial\
would then become $K$ and would not then cancel 
against \fveryspecial\ as in Proposition 8 and ref. \alg,
resulting in quadratically divergent one-loop diagrams 
(integrand $\sim1/k^2$) even in the BC bereft sector \alg.

Also if we removed the $\A$ restriction, $\{K-c'\}$ would bite its
own tail as in \ftail, and since it does so from \pRG\ via
the partial supermatrix $\B$, there is no supertrace cancellation in
the inner Wilson loop, leading by \endtonn, 
to an unregulated integrand $\sim1/k^2$. (Here we note that the leading
divergence would superficially arise from \endton, but this vanishes
under Lorentz invariance of the $k$ integral.)
However this problem might need a separate solution as we explain below.

Actually, tail-biting is a generic problem beyond one loop or for finite
$N$. Recall that $\{c'\}$ did not bite its own tail because the inner Wilson
loop in \ftail\
had to be field free in the large $N$ limit, thus giving $\str 1=0$.
But at finite $N$, $\AA$s can appear on the inner loop and thus
$\str 1$ does not arise. Such a term could still be logarithmically
divergent in $D=4$ dimensions (integrand $\sim1/k^4$). 

At two-loops the subtleties of the large $N$ limit of the flow equation
itself \ymi\ may allow such divergent tail biting diagrams even at
$N=\infty$ as illustrated in \fig\ftwobite{p16}, providing
neither field free Wilson loop vanishes by the supertrace mechanism.
\midinsert
$$
\epsfxsize=0.2\hsize\epsfbox{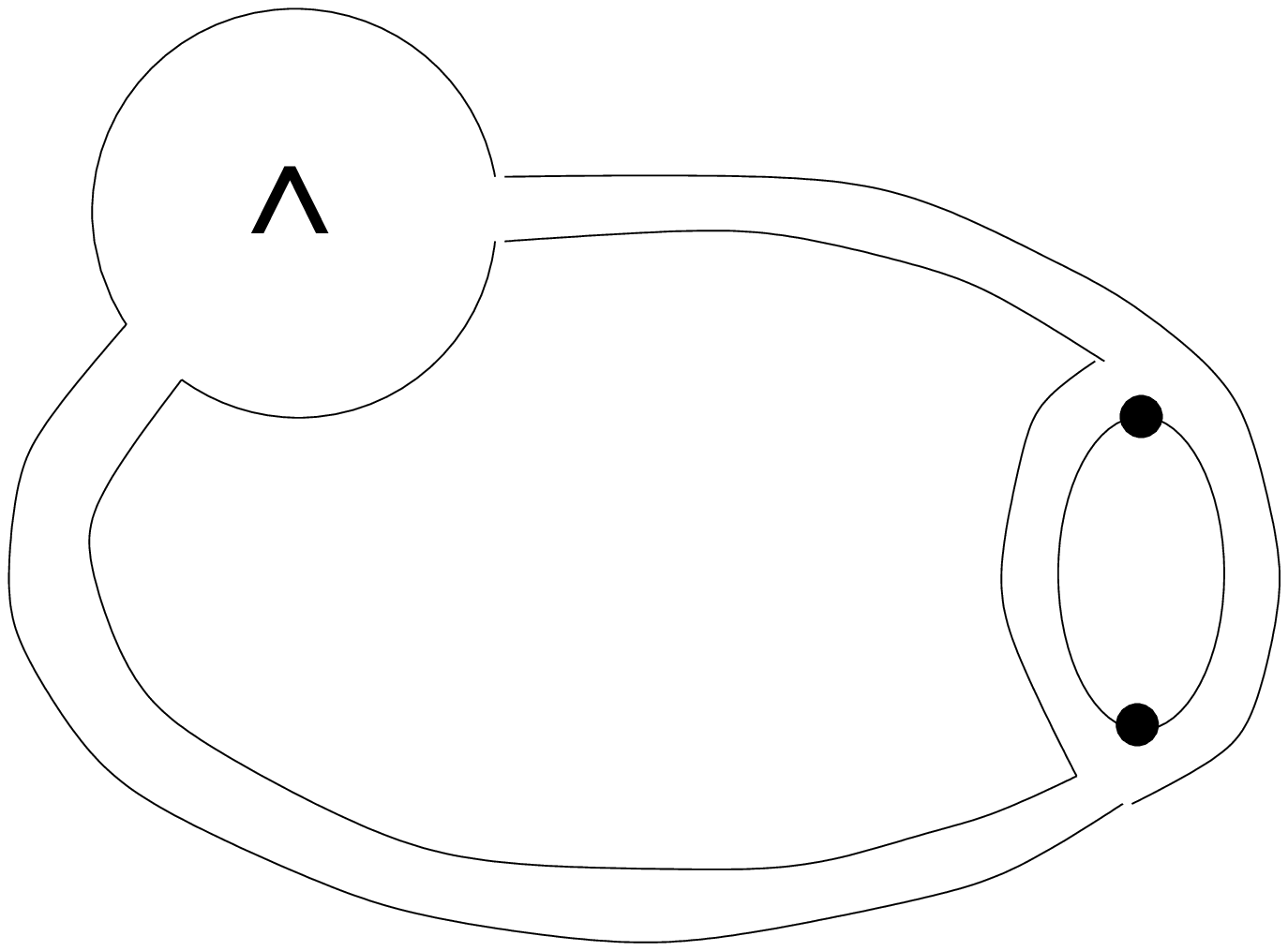}
$$
\centerline{{\bf Fig.25.} An $N=\infty$ divergent two-loop tail-biting
diagram. }
\endinsert

For these reasons, 
extending these ideas to finite $N$ and beyond one-loop seem to require
returning to more geometric covariantization (\eg straight line wines)
and regulating the wines as already described in \alg\ymi\ so as to
eliminate diagrams of form \ftail. 

As we have seen, apart from this, divergences appear to
arise only because of the restriction to $\A$ in $\{K-c'\}_\A$ which is
necessary to ensure that the hidden supertrace mechanism described in 
Proposition 8 works at least in BC bereft one-loop diagrams. We expect
that in a finite manifestly local $SU(N|N)$ approach such problems will be
absent.


\newsec{Conclusions}
We have presented a methodology which for the first time allows manifestly
gauge invariant continuum calculations to be performed. No gauge fixing
or ghosts are required, and the full power and beauty of local gauge
invariance is directly incorporated. 
We have demonstrated just a few 
of the consequences:
the quantum gauge field does not renormalize --
only the gauge coupling requires renormalization;
possible counterterms are strongly constrained, easily
and elegantly determined; the one-loop $\beta$ function 
falls out in a manifestly gauge invariant manner; the gauge invariant 
continuum Wilsonian effective action is for the first time precisely
defined.

It is important to recognize that the crucial problem  solved in
constructing  a successful manifestly gauge invariant calculus 
in quantum field theory,
is the passage to the continuum limit, equivalently renormalization. 
This is achieved by at the same time solving
the long standing problem of combining gauge
invariance with the exact RG. (In the exact RG, the continuum limit is almost
automatic; in the gauge sector we
only need to require a finite coupling constant $g(\Lambda)$  \YKIS.) 

However, in order to write down such a gauge invariant exact RG,
one must first solve the problem 
of finding a continuum gauge invariant physical cutoff. As we
have seen from this paper and ref. \alg, this is in a sense uniquely given by 
spontaneously broken $SU(N|N)$ gauge theory 
with covariant higher derivatives. This method of regularization
stands separately and is interesting and useful in its own
right. It will be discussed more fully in ref. \sunn. We also comment
further below.

Our method should be compared with previous approaches to the
exact RG in gauge theory which implement a physical cutoff 
at the expense of not only fixing, but breaking, 
the gauge invariance at least at intermediate stages
\ref\prev{C. Becchi, in {\sl Elementary Particles, Field Theory and Statistical
Mechanics}, (Parma, 1993), hep-th/9607188;\
M. Bonini {\it et al}, Nucl. Phys. B409 (1993) 441, B418 (1994) 81, B421 (1994)
81, B437 (1995) 163;\
U. Ellwanger, Phys. Lett. 335B (1994) 364;\
U. Ellwanger {\it et al}, Z. Phys. C69 (1996) 687;\
M. Reuter and C. Wetterich, Nucl. Phys. B417 (1994) 181, B427 (1994) 291;\
M. D'Attanasio and 
T.R. Morris, Phys. Lett. B378 (1996) 213;\
K-I. Aoki {\it et al}, Prog. Theor. Phys. 97 (1997) 479, 102 (1999) 1151,
Phys. Rev. D61 (2000) 045008;\ K-I. Kubota and H. Terao, Prog. 
Theor. Phys. 102 (1999) 1163;\
M. Pernici {\it et al}, Nucl. Phys. B520 (1998) 469;\
D.F. Litim and J.M. Pawlowski, Phys. Lett. B435 (1998) 181;\ 
M. Simionato, hep-th/9809004, hep-th/9810117, hep-th/0005083.}, and it
should be compared from this point of view with stochastic quantization
(the manifestly gauge invariant form of which fails to regulate in the
gauge orbit directions) \ref\zinn{See \eg the discussion
in J. Zinn-Justin, {\it Quantum Field Theory and Critical Phenomena} (1993)
Clarendon Press, Oxford.} and the famous loop-space, \aka Migdal-Makeenko 
or Dyson-Schwinger,  approach \ref\mig{A. Migdal, 
Ann. Phys. 109 (1977) 365;\
Yu.M. Makeenko and A.A. Migdal, Nucl. Phys. B188 
(1981) 269.}\nref\LN{A.A. Migdal, Phys. Rep. 102 (1983) 199.}\nref\polbo{
A.M. Polyakov, {\sl Gauge fields and Strings}
(Harwood, 1987).}--\ref\others{V.A. Kazakov and I.K. Kostov, Nucl. Phys.
B176 (1980) 199; V.A. Kazakov, Nucl. Phys. B179 (1981) 283.}
as already indicated in refs. \alg\ymi, where it is unknown how to 
formulate it at other than the bare level \polbo.  At finite $N$,
Mandelstam's relations \ref\mand{S. Mandelstam, Phys. 
Rev. 175 (1968) 1580.}\LN, which encode the overcomplete nature
of the Wilson loop representation, also significantly complicate this
approach since both equations are relations between intersecting loops.  

Here this overcounting has no
great consequence,
amounting to a harmless overparametrization of the 
effective action \alg; intersecting loops do not here carry 
any special significance. 
We stress that while it is possible and we believe useful to express
the pure gauge sector and pure gauge sections geometrically
in terms of Wilson loops and lines respectively, it is not necessary
for our formulation. Nor need the apparent non-locality be any more than
usual. Indeed if we choose  $c^{-1}$ and $\ct^{-1}$ to be
polynomials, then the seed action $\hS$ is just
a sum over a finite number of ultralocal operators
(\ie vertices polynomial in their momenta) as would be normal
for a bare action.\foot{Despite some similarities $\hS$ is a renormalized
action however.} The effective action $S$ is
necessarily only quasilocal (\ie Taylor expandable in momenta) as is the
case for any Wilsonian effective action. 

We should emphasise some subtleties in our construction.
Even the inclusion of Pauli-Villars fields as part of the effective
cutoff in an exact RG is novel,
and has to be done carefully so that flow to lower energies
amounts to {\sl lowering} their
masses (rather than raising them as naturally occurs for a relevant
perturbation such as a mass)
and thus corresponds to integrating out. We saw that at the {\sl classical
level} the solutions involve integrals over the effective cutoff 
$\Lambda$
which diverge, requiring integration constant `counterterms'
or in effect a finiteness prescription for these $\Lambda$ integrals.

The Pauli-Villars contributions lead to finiteness in the momentum integrals,
by subtracting separately
divergent contributions. The answers are finite --
but well defined only after applying and
removing a preregulator. (In our case we took $D<4$ and only let $D\to4$
at the end of the calculation.)
Only the symmetries respected in this 
procedure are guaranteed preserved.
It is important in a regularisation such as this,
that appears to generalise so
straightforwardly to in particular anomalous gauge theories, that there
are subtleties in its construction. 
For non-anomalous theories, an imperfect pre-regulator
can be repaired by adding symmetry-breaking operators
that vanish on removing the pre-regulator,
but for anomalous theories this procedure will fail.

From below \bb, we also recall the wrong sign action for $A^2_\mu$. 
As we noted below \supersc, and
in ref. \alg, this unphysical gauge field decouples from the physical
gauge field $A^1_\mu$ at energies much less than $\Lambda$, but  
problems could potentially
arise for $A^1$ if only at the \nonp\ level. At first sight
the wrong-sign $A^2$ action leads to instability, 
however the fact that the kinetic term has also the wrong sign
results instead in unitarity violations, whose effect in the physical sector
should disappear in the limit that the overall cutoff is removed \sunn.

We now understand the wrong sign $A^2$ action as an 
inevitable consequence of the underlying 
$SU(N|N)$ gauge theory. 
The idea of regularising via such a gauge theory 
is another novel development in this paper. (For some previous
applications of $SU(N|N)$ gauge theory see refs.\foot{The author thanks John Tighe for bringing these and ref. \bars\ to his attention.} 
\ref\earlysunn{Y. Ne'eman, Phys. Lett. 81B (1979) 190;\
H.C. Lee and Z.Y. Zhu, Phys. Rev. D44 (1991) 942;\ 
C-Y Lee, D.S. Hwang, Y. Ne'eman, 
J. Math. Phys. 37 (1996) 3725.}.)
Note that geometrically this corresponds to realising a superspace in the
fibres of the principal bundle rather than in the base space as 
would correspond  to the usual spacetime supersymmetry. 

As we saw in sec. 6.9,
an intriguing property, in fact inherent to $SU(N|N)$, 
is a duality that exchanges
the r\^ole of $A^1$ and $A^2$ and thus also in a sense changes $g^2$ to $-g^2$.
The present `unitary gauge formulation' closely imitates a spontaneously 
broken $SU(N|N)$ theory but actually differs in some small details
(as discussed in secs. 2 and 6.6), and 
it is not clear to us that this duality survives spontaneous breaking
in a true local $SU(N|N)$ invariant formulation. We must also be mindful
that the present formulation only regulates to one loop with any number
of external gauge fields. Nevertheless, it is intriguing to 
note that the duality symmetry,
the local $SU(N|N)$ and the global fermionic $U(1)$, do not commute
with each other. In view of the intimate connection of $g^2$ with
$\sigma_3$ as suggested by duality,
which may be taken to imply that $\sigma_3$ appears with
the same power as $g^2$,
it seems that we should regard the coupling constant as the
space-time independent field $g^2\sigma_3$. The action of local
$SU(N|N)$ then forces us to consider it 
a space-time dependent field. 
Perhaps these ideas hold the key to 
develop this duality at a deeper level.

We saw however that the duality is at least obscured when we come to
renormalize. To renormalize the $A^2$ sector we should introduce a
separate coupling $g_2$ for $A^2$. (We expect this to be true
also in the complete $SU(N|N)$ theory.) A very intriguing scenario
now arises as is clear from \bareone. The wrong sign action for $A^2$
is equivalent to the wrong sign $g^2_2$, and means that for negative
$\beta_1$, $g_2$ is not asymptotically free but trivial. As
$\Lambda_0\to\infty$, the continuum limit for the physical Yang-Mills
field $A^1$ is reached by $g_0\to0$.  But in this limit $g_2$ vanishes,
and thus it would appear that for dynamical reasons the unphysical
$A_2$ sector entirely decouples and loses all interactions in the continuum
limit.

Finally, let us remark that the formalism presented here appears
to open many doors, and suggests many further avenues of exploration.
We note that
there is considerable freedom: the use of completeness relations for
the generators, coincident lines, power law cutoff functions, or the same
covariantization for the different kernels, is not 
necessary. Other interactions in $\hS$ are possible, and other ways of
arranging the covariantization and regularisation
may be considered along similar lines 
\ref\hirano{S. Hirano, Phys. Rev. D61 (2000) 125011.}.
This may help to understand more deeply holographic RG flow in the
AdS/CFT correspondence and string theory \ref\li{Miao Li, hep-th/0001193}.
We have already touched on the issues of
extending the regularisation to finite $N$ \sunn, of
separately developing the $SU(N|N)$
regularisation \sunn, of formulating a manifestly
$SU(N|N)$ invariant exact RG and generalisation to other groups.
We do not directly compute  correlators of gauge fields,
the integration over modes being indirect
through in effect iterating infinitessimal
gauge invariant changes of variables \alg.
Nevertheless we expect that gauge invariant 
correlators can be computed by introducing sources for the
appropriate gauge invariant operators and absorbing these
sources as space-time dependent couplings. As $\Lambda\to0$, all modes
are integrated out and the partition function $\Z$ should then be just 
proportional to ${\rm e}^{-S}$. Differentiating with respect to the sources
should then allow all the physics to be extracted.  We should also
understand if/how gauge {\sl variant} operators are gauge averaged over.
Other  future directions include higher loop calculations
such as $\beta_2$, investigating the Gribov problem in the
continuum -- perhaps through
a limit as gauge fixing is removed, incorporation of matter fields in other
representations; $U(1)$ gauge theory \eg QED, 
through $U(1|1)$, should be much simpler in many
respects since the gauge field kernels do not need covariantizing 
\alg\ref\ui{T.R. Morris, Phys. Lett. B357 (1995) 225.}.
We have already touched on anomalies in this framework but these deserve further
investigation. This formalism should allow investigations 
of instantons, renormalons and other controlled
\nonp\ effects in a manifestly gauge invariant way.
It looks possible to generalise this formalism to local coordinate invariance
and thus a \nonp\ continuum framework for quantum gravity and supergravity.
Of course the exact RG is tailor made to investigate 
Seiberg-Witten methods \sw\ and super Yang-Mills theory
more generally at a deeper level. And last but by no means least, we
are hopeful that fully \nonp\ approximation methods can be developed 
to allow
accurate analytic continuum calculations in realistic gauge theories,
\eg $SU(3)$ and QCD. Such approximation methods can involve for example
the large $N$ (colour) limit \ymi\alg. Many other potentially powerful
approximation methods are also available within the exact RG
framework \alg\YKIS\ref\deriv{T.R. Morris, Phys. Lett. B329 (1994) 241;\
T.R. Morris and J.F. Tighe, JHEP 08 (1999) 7.}
\ref\revs{D.U. Jungnickel and C. Wetterich,
in {\it The Exact Renormalization Group}, Eds Krasnitz {\it et al},
World Sci (1999) 41, and hep-ph/9902316;\
T.R. Morris, in {\it New Developments 
in Quantum Field Theory},
NATO ASI series 366, (Plenum Press, 1998), 
hep-th/9709100, in {\it RG96}, Int. J. Mod. Phys. B12 (1998) 1343, hep-th/9610012.}.

\bigbreak\bigskip\bigskip\bigskip\bigskip
\centerline{{\bf Acknowledgements}}\nobreak
The author wishes to thank Jonathon Evans,
Hugh Osborn and John Tighe for perceptive comments,
PPARC for financial support through an Advanced Fellowship and PPARC grant
GR/K55738, and Ken Barnes for moral support.

\listrefs

\end